\documentclass[twocolumn,tighten, times]{aastex631}

\usepackage{graphicx}   
\usepackage{epsfig}
\usepackage{psfig}
\usepackage{color}

\usepackage{natbib} 
\usepackage{multirow}

\newcommand{\gs}{\mathrel{\raise0.35ex\hbox{$\scriptstyle >$}\kern-0.6em
\lower0.40ex\hbox{{$\scriptstyle \sim$}}}}
\newcommand{\ls}{\mathrel{\raise0.35ex\hbox{$\scriptstyle <$}\kern-0.6em
\lower0.40ex\hbox{{$\scriptstyle \sim$}}}}

\shorttitle{JWST's PEARLS: SMA and JWST demonstrate diversity in SMGs}
\shortauthors{Smail et al.}

\graphicspath{{./}{figs/}}

\begin{document}

\title{Hidden giants in JWST's PEARLS:\\ An ultra-massive $z=$\,4.26 sub-millimeter galaxy that is invisible to HST}

\author[0000-0003-3037-257X]{Ian Smail}
\affiliation{Centre for Extragalactic Astronomy, Department of Physics,
  Durham University, South Road, Durham DH1 3LE, UK}

\author[0000-0003-4748-0681]{ Ugn\.{e} Dudzevi{\v{c}}i{\={u}}t{\.{e}}} 
\affiliation{Max-Planck-Institut f\"{u}r Astronomy, K\"{o}nigstuhl 17, 69117
  Heidelberg, Germany}

\author[0000-0003-0685-3621]{Mark Gurwell} 
\affiliation{Center for Astrophysics \textbar\ Harvard \& Smithsonian, 60 
  Garden St., Cambridge, MA 02138, USA}

\author[0000-0002-0670-0708]{Giovanni G.\ Fazio} 
\affiliation{Center for Astrophysics \textbar\ Harvard \& Smithsonian, 60 
Garden St., Cambridge, MA 02138, USA}

\author[0000-0002-9895-5758]{S.\ P.\ Willner} 
\affiliation{Center for Astrophysics \textbar\ Harvard \& Smithsonian, 60 
  Garden St., Cambridge, MA 02138, USA}

\author[0000-0003-1192-5837]{A.\ M.\ Swinbank} 
\affiliation{Centre for Extragalactic Astronomy, Department of Physics,
  Durham University, South Road, Durham DH1 3LE, UK}

\author[0009-0009-7284-5049]{Vinodiran Arumugam} 
\affiliation{Institut de Radioastronomie Millim\'etrique,
300 rue de la Piscine, Domaine Universitaire,
38406 Saint Martin d'H\`eres, France}

\author[0000-0002-7265-7920]{Jake Summers} 
\affiliation{School of Earth and Space Exploration, Arizona State University,
  Tempe, AZ 85287-1404, USA}

\author[0000-0003-3329-1337]{Seth H.\ Cohen} 
\affiliation{School of Earth and Space Exploration, Arizona State University,
  Tempe, AZ 85287-1404, USA}

\author[0000-0003-1268-5230]{Rolf A.\ Jansen} 
\affiliation{School of Earth and Space Exploration, Arizona State University,
  Tempe, AZ 85287-1404, USA}

\author[0000-0001-8156-6281]{Rogier A.\ Windhorst} 
\affiliation{School of Earth and Space Exploration, Arizona State University,
  Tempe, AZ 85287-1404, USA}

\author[0000-0002-7876-4321]{Ashish Meena} 
\affiliation{Physics Department, Ben-Gurion University of the Negev, P.O. Box
  653, Beer-Sheva 8410501, Israel}

\author[0000-0002-0350-4488]{Adi Zitrin} 
\affiliation{Physics Department, Ben-Gurion University of the Negev, P.O. Box
  653, Beer-Sheva 8410501, Israel}

\author[0000-0002-6131-9539]{William C.\ Keel} 
\affiliation{Department of Physics and Astronomy, University of Alabama, Box 870324, Tuscaloosa, AL\,35404, USA}


\author[0000-0001-7410-7669]{Dan Coe} 
\affiliation{Space Telescope Science Institute, 3700 San Martin Drive, Baltimore, MD 21218, USA}
\altaffiliation{Association of Universities for Research in Astronomy for the European Space Agency, STScI, Baltimore, MD 21218, USA}
\altaffiliation{Center for Astrophysical Sciences, Department of Physics and Astronomy, The Johns Hopkins University, 3400 N.\ Charles St. Baltimore, MD 21218, USA}

\author[0000-0003-1949-7638]{Christopher J.\ Conselice} 
\affiliation{Jodrell Bank Centre for Astrophysics, Alan Turing Building,
University of Manchester, Oxford Road, Manchester M13 9PL, UK}

\author[0000-0002-9816-1931]{Jordan C.\ J.\ D'Silva} 
\affiliation{International Centre for Radio Astronomy Research and 
International Space Centre,  University of Western Australia, M468,
35 Stirling Hwy, Crawley, WA 6009, Australia}
\altaffiliation{ARC Centre of Excellence for All Sky Astrophysics in 3 Dimensions, Australia}

\author[0000-0001-9491-7327]{Simon P.\ Driver} 
\affiliation{International Centre for Radio Astronomy Research and 
International Space Centre,  University of Western Australia, M468,
35 Stirling Hwy, Crawley, WA 6009, Australia}

\author[0000-0003-1625-8009]{Brenda Frye} 
\affiliation{Steward Observatory, University of Arizona, 933 N.\ Cherry Ave,
Tucson, AZ, 85721-0009, USA}

\author[0000-0001-9440-8872]{Norman A.\ Grogin} 
\affiliation{Space Telescope Science Institute,
3700 San Martin Drive, Baltimore, MD 21218, USA}

\author[0000-0002-6610-2048]{Anton M.\ Koekemoer} 
\affiliation{Space Telescope Science Institute,
3700 San Martin Drive, Baltimore, MD 21218, USA}

\author[0000-0001-6434-7845]{Madeline A.\ Marshall} 
\affiliation{National Research Council of Canada, Herzberg Astronomy \&
Astrophysics Research Centre, 5071 West Saanich Road, Victoria, BC V9E 2E7,
Canada}
\altaffiliation{ARC Centre of Excellence for All Sky Astrophysics in 3 Dimensions, Australia}

\author[0000-0001-6342-9662]{Mario Nonino} 
\affiliation{INAF-Osservatorio Astronomico di Trieste, Via Bazzoni 2, 34124
Trieste, Italy}

\author[0000-0003-3382-5941]{Nor Pirzkal} 
\affiliation{Space Telescope Science Institute,
3700 San Martin Drive, Baltimore, MD 21218, USA}

\author[0000-0003-0429-3579]{Aaron Robotham} 
\affiliation{International Centre for Radio Astronomy Research  and 
International Space Centre,  University of Western Australia, M468,
35 Stirling Hwy, Crawley, WA 6009, Australia}

\author[0000-0001-7016-5220]{Michael J.\ Rutkowski} 
\affiliation{Minnesota State University-Mankato,  Telescope Science Institute,
TN141, Mankato MN 56001, USA}

\author[0000-0003-0894-1588]{Russell E.\ Ryan, Jr.} 
\affiliation{Space Telescope Science Institute,
3700 San Martin Drive, Baltimore, MD 21218, USA}

\author[0000-0001-9052-9837]{Scott Tompkins} 
\affiliation{School of Earth and Space Exploration, Arizona State University,
Tempe, AZ 85287-1404, USA}

\author[0000-0001-9262-9997]{Christopher N.\ A.\ Willmer} 
\affiliation{Steward Observatory, University of Arizona,
933 N.\ Cherry Ave, Tucson, AZ, 85721-0009, USA}

\author[0000-0001-7592-7714]{Haojing Yan} 
\affiliation{Department of Physics and Astronomy, University of Missouri,
Columbia, MO 65211, USA}


\author[0000-0002-5807-4411]{Thomas J.\ Broadhurst}  
\affiliation{University of the Basque Country UPV/EHU, Department of Theoretical Physics, Bilbao, E-48080, Spain}
\altaffiliation{DIPC, Basque Country UPV/EHU, San Sebastian, E-48080, Spain}
\altaffiliation{Ikerbasque, Basque Foundation for Science, Bilbao, E-48011, Spain}

\author[0000-0003-0202-0534]{Cheng Cheng}  
\affiliation{Chinese Academy of Sciences South America Center for Astronomy, National Astronomical Observatories, CAS, Beijing, 100101, China}

\author[0000-0001-9065-3926]{Jose M.\ Diego}  
\affiliation{Instituto de Fisica de Cantabria, Avda.\ Los Castros, Universidad de Cantabria, 39005 Santander, Spain}

\author[0000-0001-9394-6732]{Patrick Kamieneski}  
\affiliation{School of Earth and Space Exploration, Arizona State University, Tempe, AZ 85287-1404, USA}

\author[0000-0001-7095-7543]{Min Yun}  
\affiliation{University of Massachusetts Department of Astronomy, 710 North Pleasant Street Amherst, MA 01003-9305, USA}


\begin{abstract}
We present a multi-wavelength analysis using SMA, JCMT, NOEMA, JWST, HST, and SST of two dusty strongly star-forming galaxies, 850.1 and 850.2,  seen through  the massive cluster lens A\,1489.   These SMA-located sources both lie at $z$\,$=$\,4.26 and have  bright dust continuum emission, but 850.2 is a UV-detected Lyman-break galaxy, while 850.1 is undetected at $\ls$\,2\,$\mu$m, even with deep {JWST}/NIRCam observations.   We investigate their stellar, ISM, and dynamical properties, including a pixel-level SED analysis to derive sub-kpc-resolution stellar-mass and $A_V$ maps. We find that 850.1 is one of the most massive and highly obscured, $A_V$\,$\sim$\,5, galaxies known at $z$\,$>$\,4 with $M_\ast$\,$\sim $\,10$^{11.8}$\,M$_\odot$ (likely forming at $z$\,$>$\,6), and 850.2 is one of the least massive and least obscured, $A_V$\,$\sim$\,1, members of the $z$\,$>$\,4 dusty star-forming population.  The diversity of these two dust-mass-selected galaxies  illustrates the  incompleteness of  galaxy surveys  at $z$\,$\gs$\,3--4 based on imaging  at $\ls$\,2\,$\mu$m, the longest wavelengths feasible from  {HST} or the ground.   The resolved mass map of  850.1 shows a  compact stellar mass distribution, $R^{\rm mass}_{\rm e}$\,$\sim$\,1\,kpc, but  its expected evolution  to $z$\,$\sim$\,1.5 and then $z$\,$\sim$\,0 matches both the properties of massive, quiescent galaxies at $z$\,$\sim$\,1.5 and  ultra-massive early-type galaxies at $z$\,$\sim$\,0.   We suggest that 850.1 is the central galaxy of a group in which 850.2 is a satellite that will likely merge in the near future.   The stellar morphology of 850.1  shows arms and a linear bar feature which we link to the active dynamical environment it resides within.   
\end{abstract}


\section{Introduction}

Obscuration by dust has been a complicating factor in  understanding  the properties of high-redshift galaxies for the last $\sim$\,30\,years.  The first studies of $z$\,$\gs$\,3 star-forming  galaxy samples were selected through their restframe ultraviolet (UV) emission  \citep{Steidel93}.  However, the subsequent identification in the sub-millimeter waveband of highly dust-obscured galaxies at similar redshifts \citep[e.g.,][]{Ivison98} demonstrated a much wider diversity in the properties of galaxies at high redshifts than the just the subset detectable in the restframe UV \citep{Ivison00}.   The potential for entire classes of dusty, massive galaxies (detectable in the sub-/millimeter or the mid-infrared with {Spitzer}/IRAC) to be missed from UV, visible- and near-infrared-selected samples,  variously termed optically/near-infrared faint/dark (or less precisely ``{HST}-dark''),  has been a concern for studies attempting to construct stellar-mass-limited galaxy surveys at $z$\,$>$\,1 \citep[e.g.,][]{Dey99,Caputi12,Simpson14,Wang19}.  With the launch of {JWST}, there has been a renewed appreciation of the significant role of dust in defining the observed visible to near-infrared properties of  high-redshift  galaxy populations \citep[e.g.,][]{Barrufet23,Bisigello23,Kokorev23,Magnelli23,PG23}.

The potential significance of missing optically-undetected populations was underlined by the analysis of large samples of dust-obscured galaxies. These populations are generally optically faint, but the majority of star formation in the Universe at $z$\,$\ls$\,4--5 was  estimated to  occur  in these  obscured systems  \citep[e.g.,][]{Dudzeviciute20,Bouwens20}.  It was the rapid evolution in this dust-obscured activity that created the peak in  the star-formation rate density at $z$\,$\sim$\,2 \citep[e.g.,][]{Magnelli13}.  Discovering  the physical processes driving this dust-obscured activity is thus crucial for understanding   galaxy evolution.   The most massive and obscured examples of this high redshift, dust-rich population are known as sub-millimeter galaxies (SMGs). They are found from surveys at $\sim$\,1\,mm that select galaxies based  on  cool dust mass out to $z$\,$\sim $\,6 \citep[e.g.,][]{Simpson14,Brisbin17}. These galaxies are proposed to evolve into the cores of  massive galaxies at $z$\,$\sim$\,0 \citep[e.g.,][]{Eales99,Dudzeviciute20,Amvrosiadis23}.

Current studies of the properties of dust-obscured galaxies are hampered by both their complex dust obscuration and  high redshift.  In particular, their restframe UV-to-visible morphologies (available from {HST}) may not provide true indicators of their structures, necessary to determine the role of interactions and mergers in driving their evolution \citep[e.g.,][]{Gillman23,Huang23}.   {JWST} promises to revolutionize this field by providing high-spatial-resolution imaging in the {\it restframe} near-infrared, necessary to map  the  internal variations in the  dust reddening and star formation, as well as isolating any contributions from reddened AGN emission, which have confounded previous attempts to  estimate the stellar masses of some of these systems \citep[e.g.,][]{Hainline11}.  The analysis of   resolved {JWST} restframe near-infrared imaging  will thus hopefully deliver much more reliable stellar masses \citep{Sorba18}.   In particular, $\sim$\,0\farcs1 FWHM NIRCam  imaging will yield more robust stellar morphologies  on kpc scales that may indicate what   drives the growth of the population of massive, gas- and metal-rich galaxies   responsible for the increasing obscuration of star formation at $z$\,$\ls$\,4--5.

%
%
\setcounter{figure}{0}
\begin{figure*}[ht!]
\centerline{\includegraphics[width=5.0in]{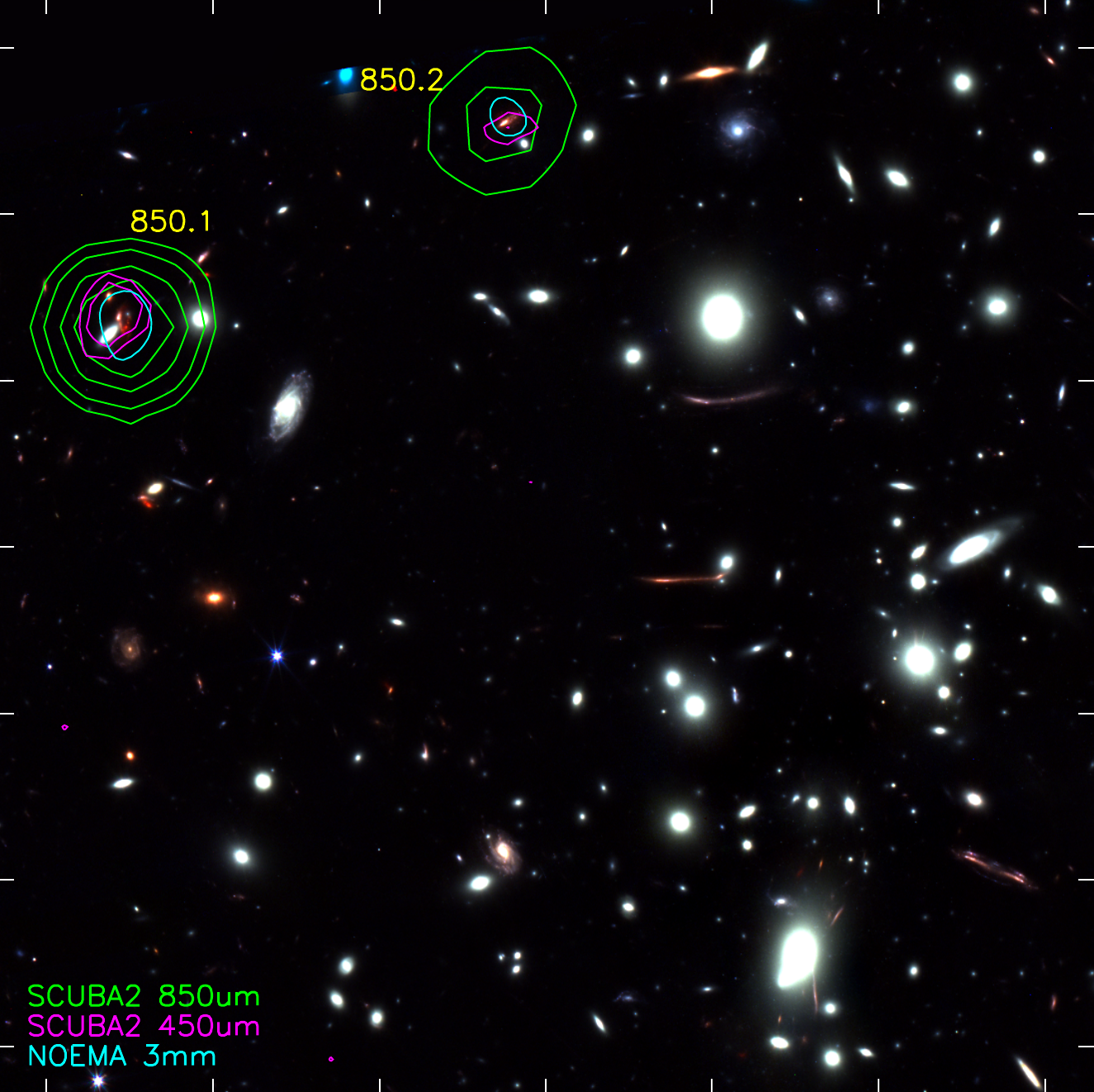}}
  \caption{\small 
    A three-color {JWST} NIRCam image of a 100$''\times$\,100$''$   region (North top, East to the left) around the $z$\,$=$\,0.35 cluster A\,1489 (where F090W+F150W is  shown as blue, F200W+F277W is green and F356W+F444W is red).   Signal-to-noise contours are shown, representing the SCUBA-2 850\,$\mu$m emission (starting at 5$\sigma$, in 5$\sigma$ increments, 14\farcs5 FWHM beam), the SCUBA-2 450\,$\mu$m emission (3$\sigma$ and 4$\sigma$, 7\farcs5 FWHM beam), and  the NOEMA 3-mm continuum emission (10$\sigma$, $\sim$\,5$''$ FWHM beam).   850.1 and 850.2 are the two bright  sources visible at 450\,$\mu$m, 850\,$\mu$m, and 3\,mm to the north east of the cluster core, which is identified by several bright elliptical galaxies, as well as associated strong lensing features.   The counterparts to these sources are unambiguously identified using the higher resolution SMA 880\,$\mu$m observations shown in Figure~2.
}
\end{figure*}

To  exploit the unique {JWST} imaging from the Prime Extragalactic Areas for Reionization and Lensing Science  (PEARLS) Guaranteed Time program \citep{Windhorst23}, we collated the  sub-millimeter imaging from the SCUBA-2 camera on JCMT of  all the PEARLS fields visible from Mauna Kea,  using the JCMT archive at CADC.  The  analysis of a 30-minute  exposure of the cluster A\,1489 ($z$\,$=$\,0.35) revealed two previously unknown, bright 850-$\mu$m sources, SMM\,J121223.0+273351 and SMM\,J121220.4+273410 (hereafter 850.1 and 850.2, respectively) with $S_{850\rm \mu m}$\,$\sim$\,60 and 20\,mJy.   This snapshot observation was so shallow that it would not normally be expected to yield any significant detections, so the presence of two bright sources was a surprise.   Initial analysis of these sources suggested that they were likely to be behind  the cluster and hence   high-redshift, dusty star-forming galaxies gravitationally magnified by the massive foreground cluster \citep{Zitrin20}.

Gravitational lensing has been used as a tool to aid  the study of dusty galaxies at high redshifts since their discovery.    The first example of a high-redshift dust-obscured galaxy, FSC\,10214 at $z$\,$=$\,2.286 \citep{RowanRobinson91}, was uncovered by virtue of the gravitational magnification caused by a serendipitously positioned foreground galaxy.   Lensing by clusters and individual galaxies has been exploited ever since to study high-redshift dusty star-forming galaxies  \citep[e.g.,][]{Smail97,Swinbank10,Ciesla20,Fujimoto23},   as lensing  both boosts the apparent  fluxes and magnifies sources across all wavelengths, making it easier to study these otherwise very faint galaxies.   

Follow up of the two A\,1489 sub-millimeter sources with SMA, NOEMA, and JCMT  gave precise locations and redshifts for the two bright sources and constrained their far-infrared emission.   These observations showed that the two sources were indeed  very dusty, star-forming galaxies lying $\sim$\,100\,kpc apart in a  structure at $z$\,$\sim$\,4.26.   One counterpart was undetected in the {HST} and {Spitzer} 3.6/4.5-$\mu$m imaging of this field, while the other corresponds to a UV-bright Lyman-break galaxy.    They are thus examples of the two populations that dominate the star-formation-rate density in the transition from primarily  unobscured at $z$\,$\gs$\,5 to predominantly obscured at $z$\,$\ls$\,4--5, with the benefit that the lens magnification will help {JWST} resolve their internal structures.

The structure of this paper is as follows:   \S2  describes our new and archival multi-wavelength observations, their reduction and analysis, including spatially resolved  spectral energy distribution (SED) fitting.   \S3  presents our results and discusses these, and  our conclusions are given in \S4. We assume a cosmology with $\Omega_{\rm M}$\,$=$\,0.3, $\Omega_\Lambda$\,$=$\,0.7, and $H_0$\,$=$\,70\,km\,s$^{-1}$\,Mpc$^{-1}$. In this cosmology the luminosity distance to  $z$\,$=$\,4.26 is 39.4\,Gpc and 1$''$ corresponds to 6.9\,kpc before correcting for lens magnification. All quoted magnitudes are on the AB system, and uncertainties on median values were derived from bootstrap resampling. The quoted stellar properties assume a Chabrier initial mass function (IMF).

\section{Observations, Reduction and Analysis}

\subsection{JCMT}

The initial  30-minute  SCUBA-2 observation of A\,1489 came from  program M15AI29,  taken in very good weather, $\tau_{225{\rm GHz}}$\,$ = $\,0.03, on 2015 May 6.  The  850-$\mu$m map showed two significantly detected point sources, one of which was exceptionally bright.   The presence of these unusual sources in this {JWST} survey field motivated us to obtain higher resolution continuum observations at 880\,$\mu$m with SMA as well as deeper continuum observations at 450/850$\mu$m with SCUBA-2, and spectroscopic observations covering the 3-mm band from NOEMA.

To improve the sub-millimeter maps, the cluster was observed  for program M23AP006 with SCUBA-2  simultaneously at 450\,$\mu$m and 850\,$\mu$m  (at resolutions of 7\farcs5 and 14\farcs5 FWHM, respectively) in very good weather conditions $\tau_{225{\rm GHz}} $\,$= $\,0.03--0.04 on the nights of 2023 February 15, March 21 and May 14. The total exposure time was 3.5\,hours using a standard constant-velocity daisy mapping pattern, and  the 30\,minutes of exposure taken in similar conditions from M15AI29 was added to give a total exposure time of 4\,hours in each band.

%
%
\setcounter{figure}{1}
\begin{figure*}[ht!]
\centerline{\begin{minipage}{7.0in}
     \centering
     \includegraphics[height=3.4in]{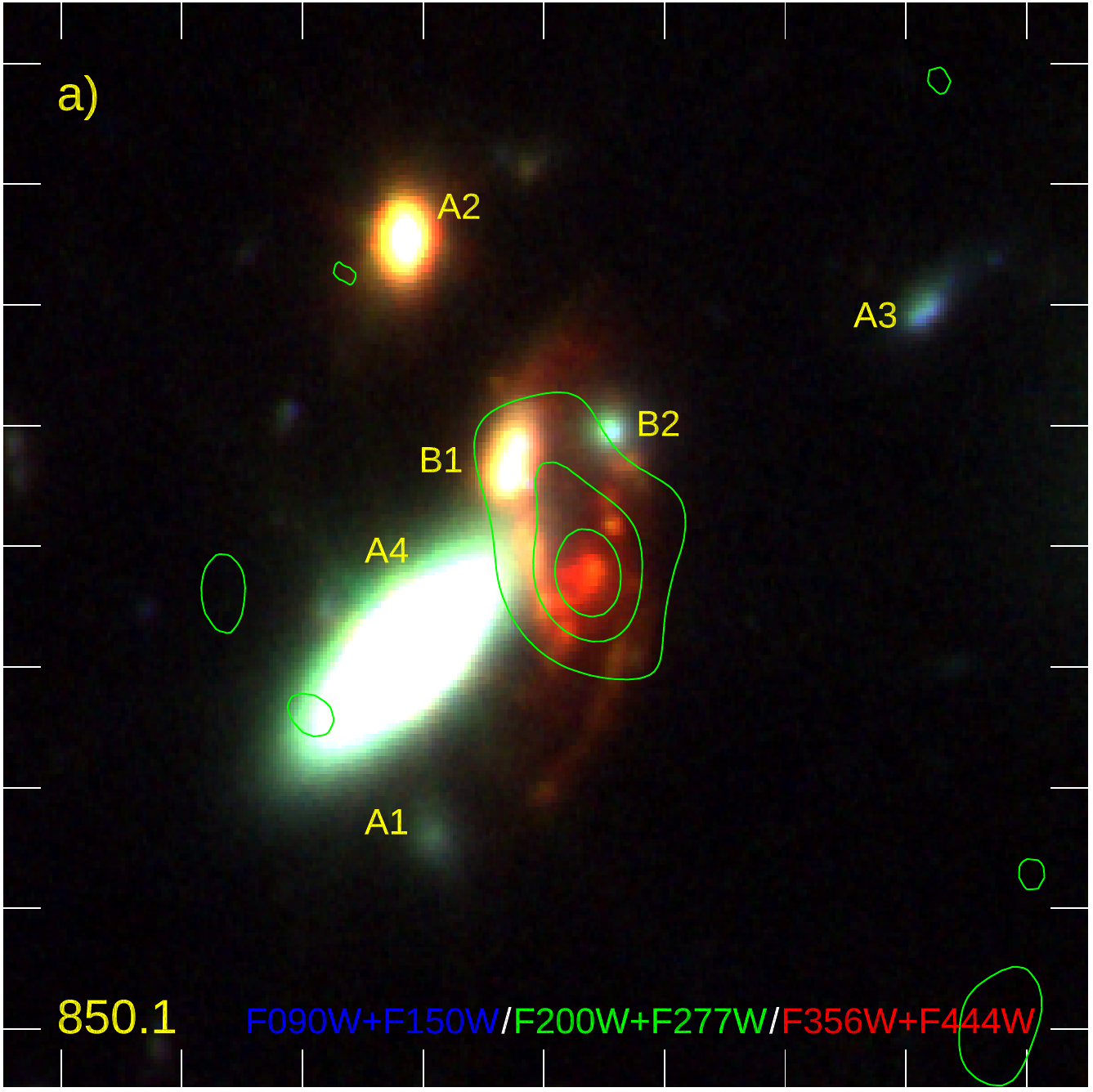}
         \hspace*{0.12in}
  \includegraphics[height=3.4in]{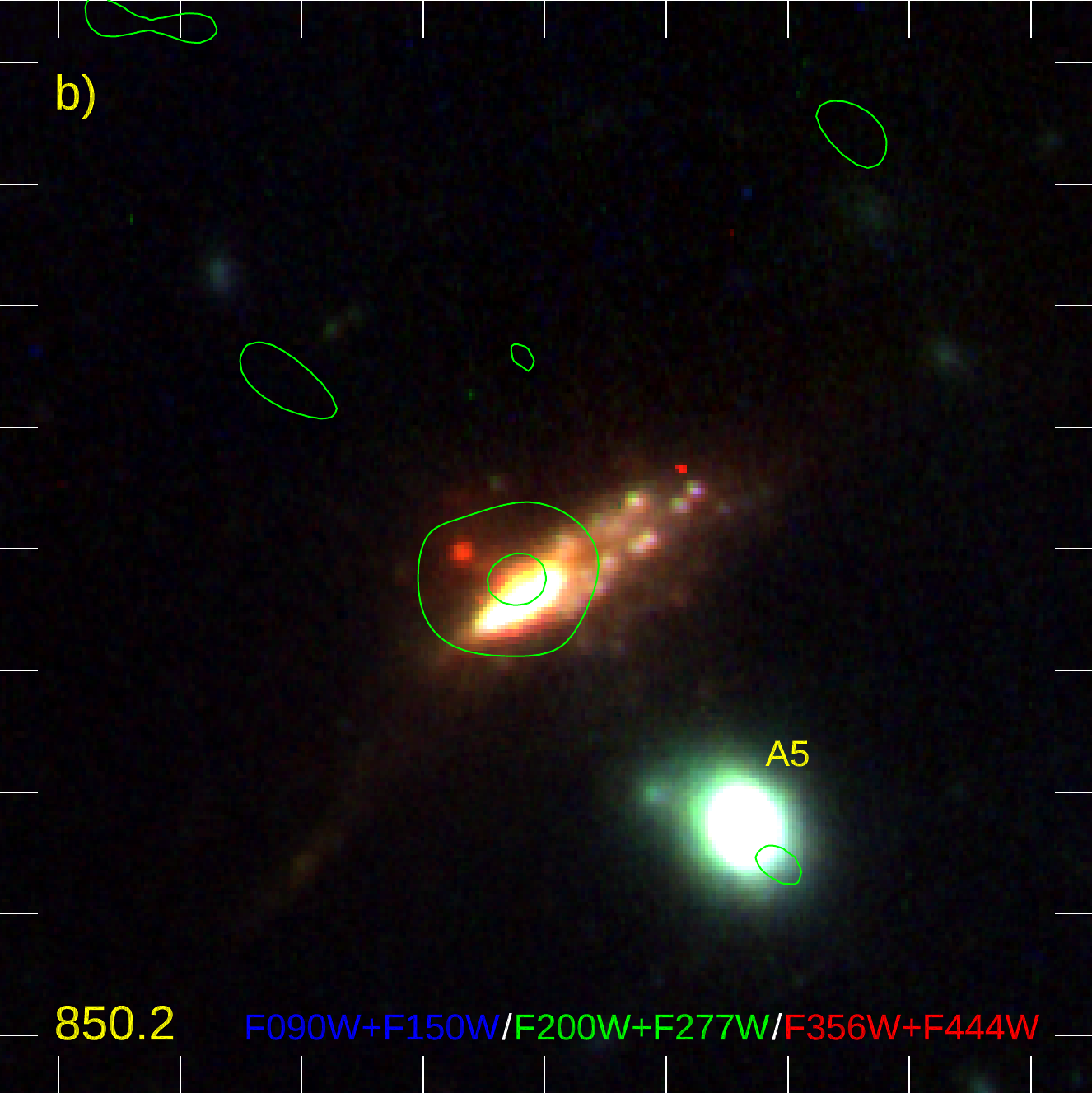}
\end{minipage}}
~\hspace*{-0.015in}
\begin{minipage}{3.4in}
   \centering
    \includegraphics[height=1.68in]{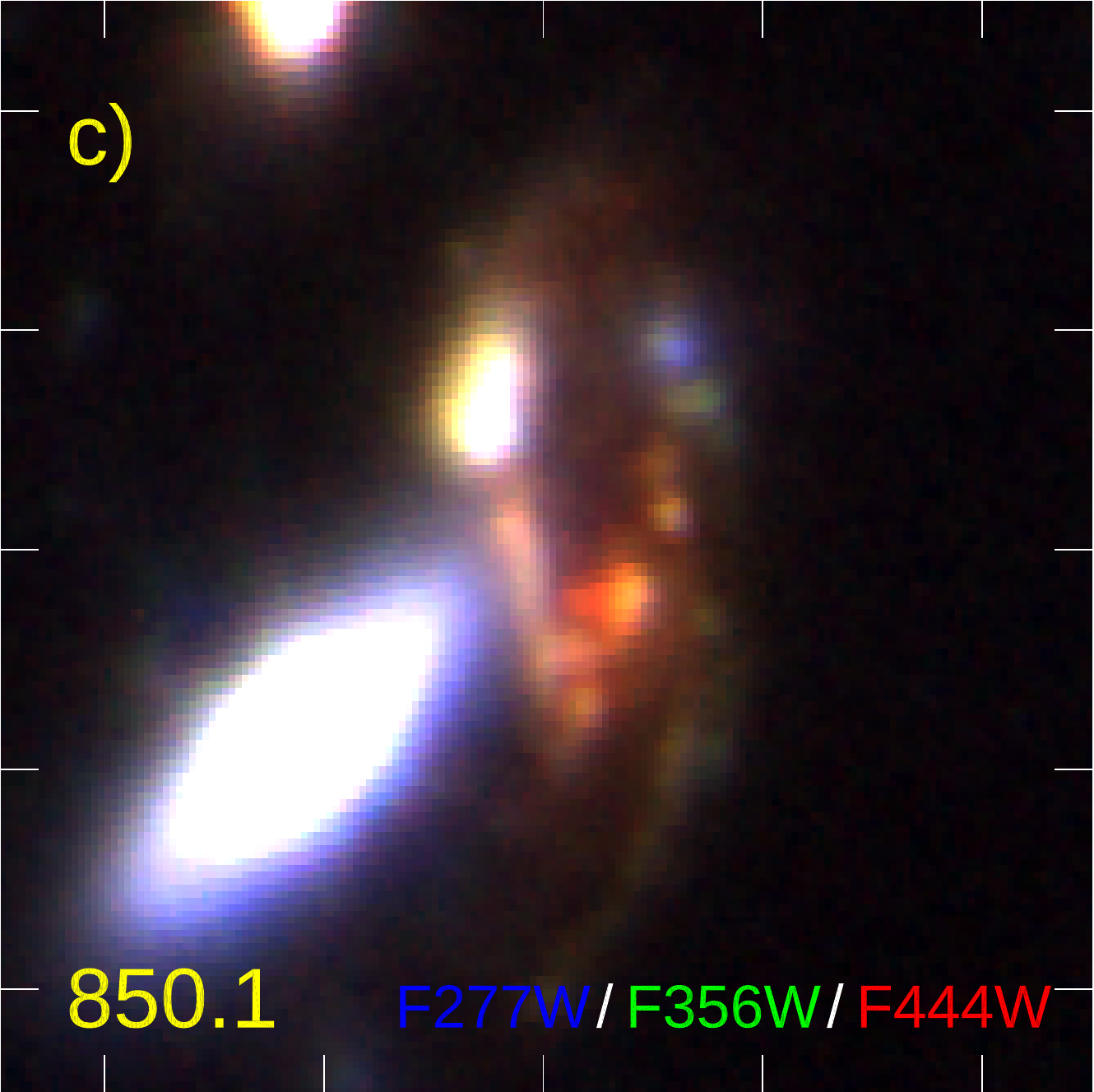}
    \includegraphics[height=1.68in]{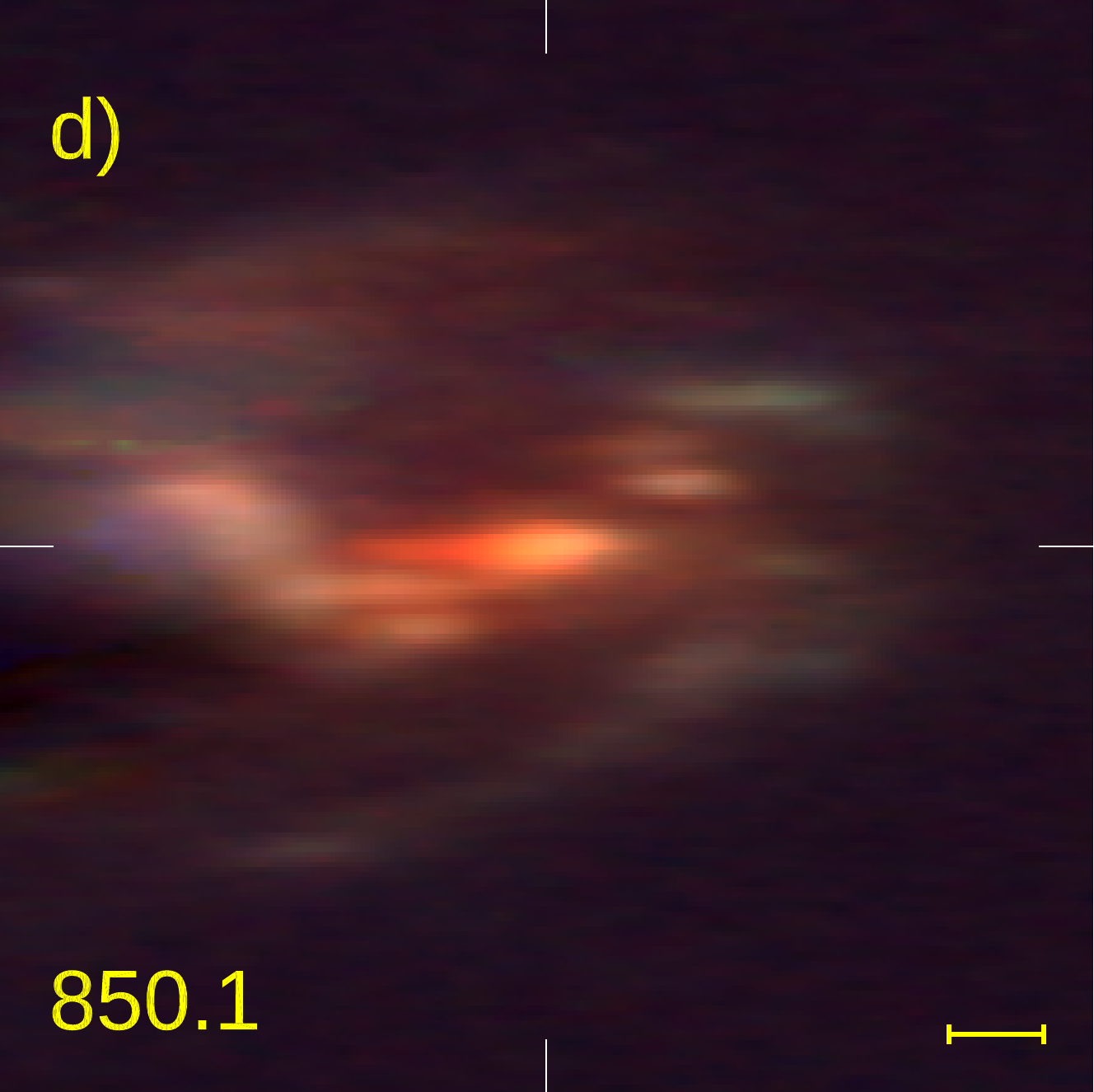}
    \end{minipage}
  \hspace*{0.08in}
  \begin{minipage}{3.4in}
    \centering
   \includegraphics[height=1.68in]{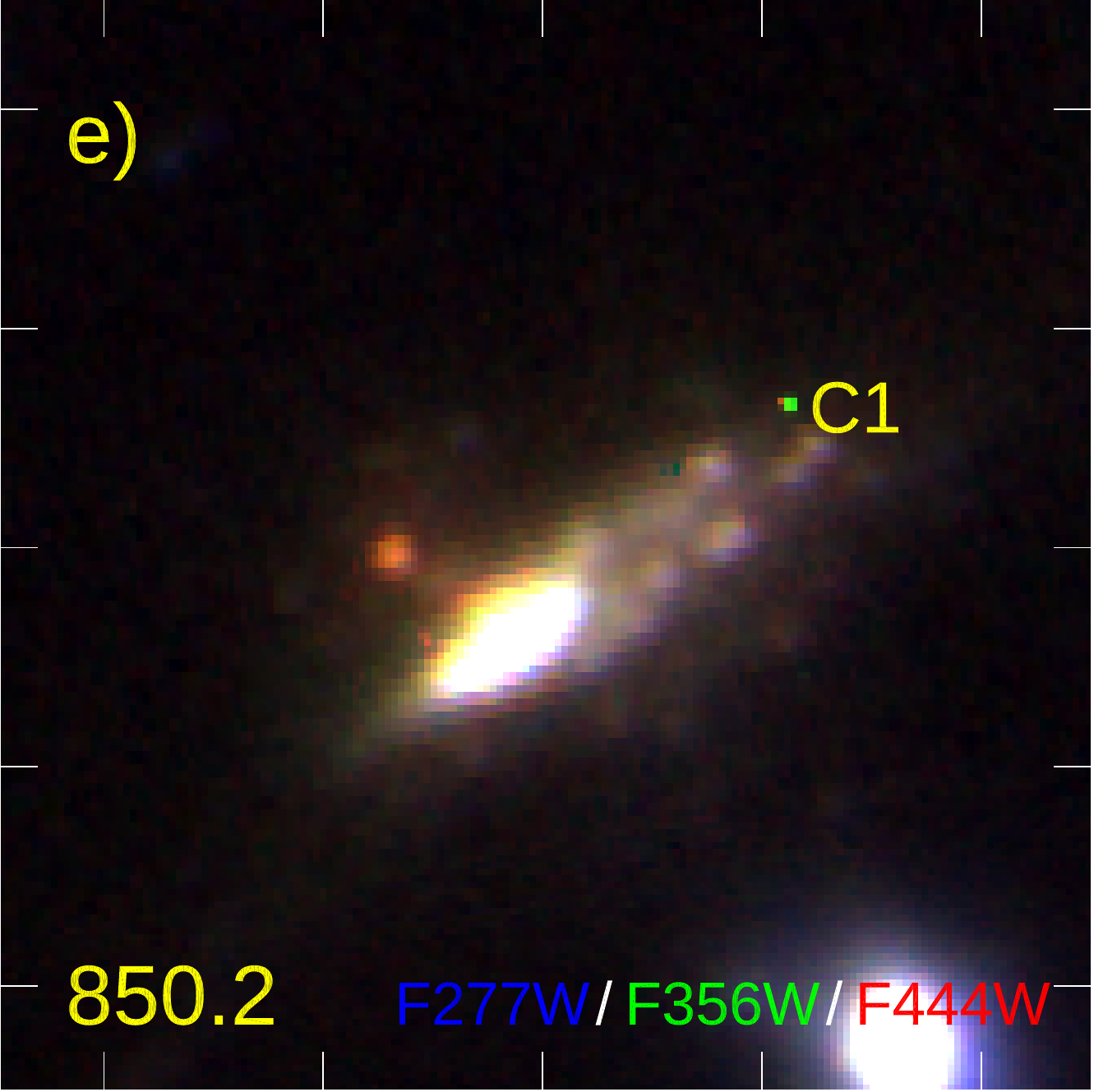}
    \includegraphics[height=1.68in]{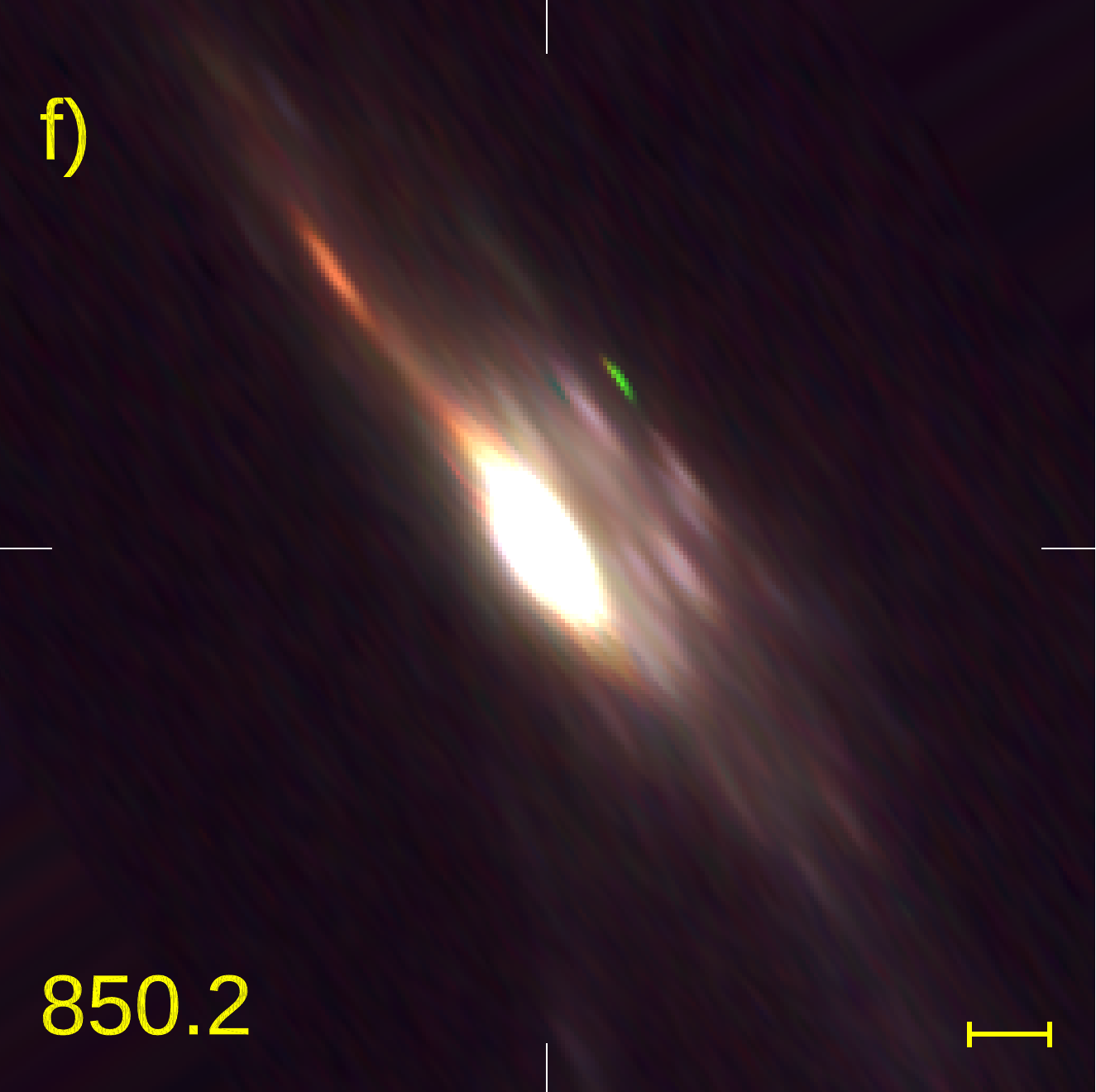}
    \end{minipage}

  \caption{\small   Three-color images of 9$''\times$\,9$''$ regions around  {\it a)} 850.1 and  {\it b)}  850.2 constructed from {JWST} NIRCam imaging, with F090W+F150W as blue, F200W+F277W as green, and F356W+F444W as red.   Panels  {\it c)} and  {\it e)}  show 5$''\times$\,5$''$ regions using F277W, F356W and F444W as blue, green, and red and the same filter combination is used in panels {\it d)} and {\it f)}, which show views of 850.1 and 850.2 ``de-sheared'' to correct for the lensing magnification (the scale bar indicates 1\,kpc).      Panels {\it a)} and {\it b)} also show the SMA 880-$\mu$m contours (starting from 2$\sigma$ in 7-$\sigma$ increments),  these unambiguously identifying the two sub-millimeter sources.  850.1 corresponds to an extremely red, highly structured source lying close to a bright, foreground  disk galaxy with several  fainter sources in close proximity. In contrast, 850.2 is more isolated and appears to be a slightly bluer, disk-like system.   The various sources and features discussed in the text are identified on each  panel: distinct sources have labels starting with “A” ,  “B” for potential subcomponents, and “C” for likely emission-line features.   The colors and photometric redshifts suggest that  ``knot'' B1 in 850.1 is  likely to lie at $z$\,$<$\,4.26, while B2 is probably a close/interacting companion seen in projection to 850.1 or a less obscured component within the galaxy.  The photometric redshift analysis also suggests that A1, A4, and A5 are likely to be cluster members at $z$\,$\sim$\,0.35, while A2 and A3 are probably behind the cluster, but foreground to 850.1 and 850.2.  }
\end{figure*}

Individual maps for each 30\,minute  observation were reduced using the Dynamic Interactive Map-Maker ({\sc dimm}) tool of the Sub-Millimetre User Reduction Facility \citep[{\sc smurf};][]{smurf} with the blank field configuration in order to detect point sources within the maps.  The  bright  source 850.1 was well-detected in  each 30\,minute exposure, so a shift-and-add method \citep{Ivison06} was applied to the data, co-aligning the positions at 850\,$\mu$m to the median position from the M23AP006 observations. The same offsets were applied to the 450\,$\mu$m maps and   this modestly improved the signal-to-noise for sources in both bands.

The individual maps were calibrated using a flux conversion factor of FCF$_{450\mu\rm m} $\,=\,491\,Jy\,beam$^{-1}$\,pW$^{-1}$ and FCF$_{850\mu \rm m} $\,=\,537\,Jy\,beam$^{-1}$\,pW$^{-1}$ \citep{Dempsey13} and then combined using inverse-variance weighting to create a final map at each wavelength. To improve point-source detection, the resulting 450\,$\mu$m and 850\,$\mu$m maps were match-filtered with their respective beams:  7\farcs5 and 14\farcs5 FWHM Gaussians.  The maps were cropped to radii of 4\farcm0, where the noise  was low and more uniform, providing coverage of all of the primary {JWST} field. The 1-$\sigma$ sensitivity in the region of 850.1/850.2 was 1.5\,mJy at 850\,$\mu$m and 21\,mJy at 450\,$\mu$m after matched filtering. The resulting 450\,$\mu$m and 850\,$\mu$m maps are shown in Figure~1.

The new maps identified a third bright sub-millimeter source to the north west of the cluster, SMM\,J121213.4+273517, with $S_{\rm 850\mu m}$\,=\,16.5\,mJy, but undetected at 450\,$\mu$m (Table~1).  Unfortunately this source is $\sim$\,2$'$ from 850.2 and lies outside the coverage of the existing {HST}, {JWST}, SMA, and NOEMA observations, and there was no obvious counterpart in the {Spitzer} imaging.   This source is not discussed further, beyond noting that it is likely to be intrinsically luminous as the lensing magnification from the cluster will be modest ($\ls$\,1.5\,$\times$) and  the 450/850-$\mu$m flux ratio suggests this source may lie at $z$\,$>$\,3, thus  it could be another member of the $z$\,$=$\,4.26 structure.

Positions and flux densities for all three 850-$\mu$m sources were derived by fitting Gaussian beams to the 850-$\mu$m and 450-$\mu$m maps.  The  resulting fluxes are given in Table~1, these were deboosted at 450\,$\mu$m following \cite{Geach17} to reflect the lower signal to noise.

%
%
\setcounter{figure}{2}
\begin{figure*}[ht]
 \centerline{ \includegraphics[width=3.5in]{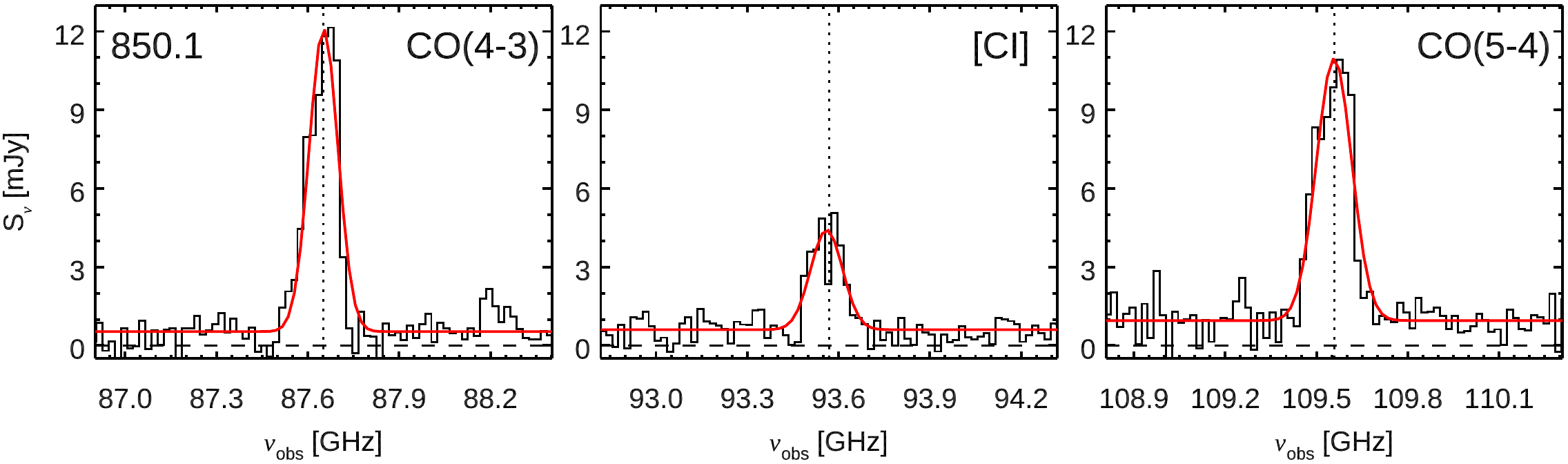}\hspace{0.1in}\includegraphics[width=3.5in]{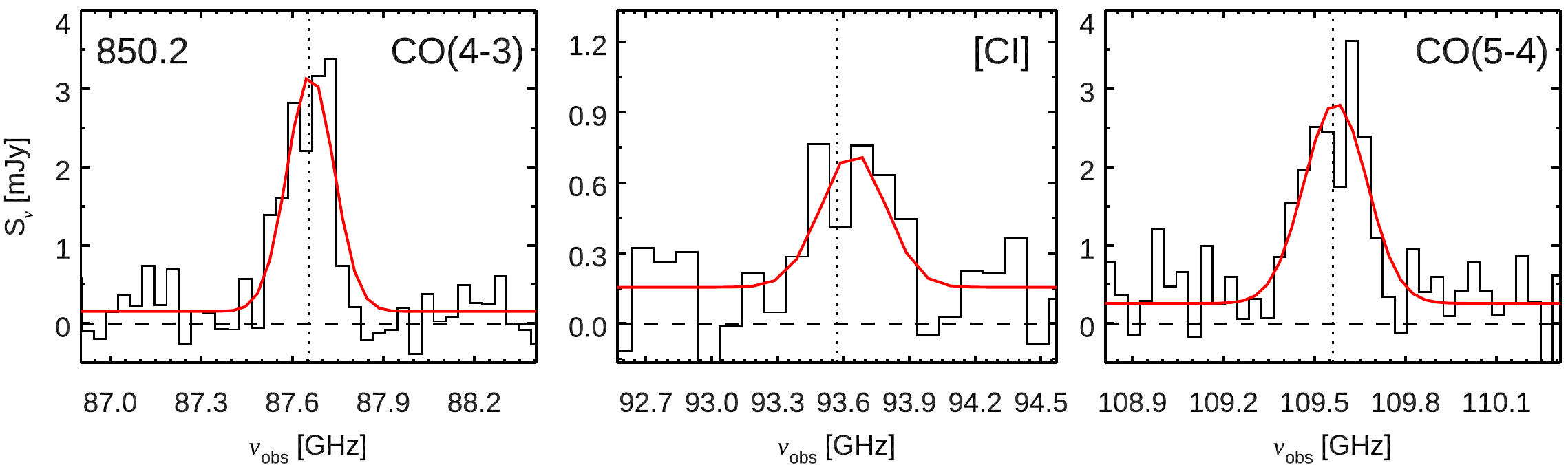} }
 \caption{\small  The NOEMA 3-mm spectra for 850.1 (left) and 850.2 (right), showing the CO(4--3), [C{\sc i}](1--0), and  CO(5--4) emission lines.  The panels are centered on the expected frequencies of the lines at the adopted redshifts for each source from Table~2 and a model fit  comprising a Gaussian and a constant continuum level  is shown for each line.  Both sources have relatively weak [C{\sc i}] emission compared to their CO(4--3) and CO(5--4).   A single Gaussian provided an adequate description of the emission for the lines in 850.1, but all three lines in 850.2 display a dip at the systemic velocity, suggesting a double-peaked profile, with the higher-frequency CO peak close to the redshift derived from the restframe UV spectroscopic features.    The spectra for 850.2 have been rebinned to 40\,MHz per channel, except for the [C{\sc i}] line which has been rebinned to 100\,MHz per channel (and as a result is shown over a slightly wider frequency window to illustrate the off-line noise).}
\end{figure*}

\subsection{SMA}

The Submillimeter Array (SMA) observed the JCMT-determined positions for 850.1 and 850.2 during a single transit between UT 6.1 and 16.3 on 2022 March 14. At the time, the SMA was operating with all eight antennas in the EXT configuration (maximum baselines of $\sim$\,220\,m). The array receiving system was tuned to an LO frequency of 346\,GHz (giving an effective band center of $\sim$\,880\,$\mu$m), providing LSB coverage from 330--342 GHz and USB coverage from 350--362\,GHz in each of two orthogonally polarized receivers, resulting in 24\,GHz of dual-polarization continuum bandwidth. The weather was good, with the precipitable water vapor column ranging from 1.5 to 2 mm and stable, low atmospheric phase variation (``seeing'').

Observations were made in a short repeated cycle, measuring the nearby phase and amplitude gain calibrator J\,1159+292 for 1.5\,minutes, then the positions of 850.1 and 850.2 for 4\,minutes each. This resulted in high quality gain calibration transfer to each target. In total, 850.1 was observed for 3.75\,hours  and 850.2 for 3.85\,hours. Passband calibration was determined from observations of J\,0854+201, J\,1159+292, and BL Lac, and the absolute flux scale was derived from observations of Ceres and MWC\,349A, taken immediately before and after the targets.

For each source, the resolution of the synthesized beam was 0\farcs85$\times$\,0\farcs63 -- around 10\,$\times$ better than from SCUBA-2 -- and 
for sources with lensing magnification of $\mu$\,$\sim$\,4--6, the circularised beam corresponds to 2.0--2.5\,kpc FWHM in the source plane.  The rms of the maps was $\sigma_{\rm 880\mu m} $\,$\sim$\,0.8\,mJy\,beam$^{-1}$ and both 850.1 and 850.2 were well detected in these sub-arcsecond resolution maps, precisely locating the sub-millimeter sources. Table~1 reports their precise positions.

850.1 was detected with a peak flux density around 19\,mJy. However, this source was clearly resolved in both dimensions  (Figure~2). Similarly, 850.2 had a peak flux density of $\sim$\,9\,mJy, although it also appeared resolved in the East$-$West direction. The vector-average amplitude versus $(u,v)$ distance plots for both sources showed increasing amplitudes on short baselines. This demonstrated that both sources were resolved. Fitting the visibility data and extrapolating these to zero spacing indicated total flux densities of $S_{\rm 880 \mu m} $\,$=$\,42.3\,$\pm$\,1.6\,mJy for 850.1 and $S_{\rm 880 \mu m} $\,$=$\,22.5\,$\pm$\,1.6\,mJy for 850.2.  Table~1 reports  these values, along with the other long-wavelength measurements.  The estimated flux density for 850.2 was in good agreement with that measured at 850-$\mu$m by SCUBA-2, but the SMA observations were $\sim$\,20\% lower for 850.1, a $\sim$\,6.5\,$\sigma$ difference.  Half of this  may be accounted for the difference in the effective band centers of the two instruments, but the remainder   suggests that there may be further extended continuum, or [C{\sc ii}] emission \citep{Smail11}, around 850.1 that was  resolved out with the sub-arcsecond SMA beam.

Using the {\sc aips} task {\sc uvfit}, elliptical Gaussian models were fitted to the $uv$ data for each source.  This derived beam-corrected FWHM  of 1\farcs00\,$\pm$\,0\farcs14\,$\times$\,0\farcs65\,$\pm$\,0\farcs14 (PA of 23\,$\pm$\,7\,degrees) for 850.1 and 1\farcs05\,$\pm$\,0\farcs16\,$\times$\,0\farcs43\,$\pm$\,0\farcs20 (PA of $-$22\,$\pm$\,8\,degrees) for 850.2.   These correspond to circularised equivalent FWHM sizes of 5.6\,$\pm$\,1.4\,kpc and 4.6\,$\pm$\,2.4\,kpc (uncorrected for lensing).

At the redshifts of 850.1 and 850.2 determined from their NOEMA 3-mm spectra (see next section), the  [C{\sc ii}]\,1900.54\,GHz line falls in the spectral coverage of the SMA observations.  The 1-$\sigma$ sensitivity of these data after binning to  $\sim$\,120\,km\,s$^{-1}$ (143\,MHz) channels was $\sim$\,25\,mJy.  However, this was insufficient to  detect [C{\sc ii}], assuming typical [C{\sc ii}] to far-infrared luminosity ratios for the sources.  A search for the line in the sources did not show any significant line emission associated with the continuum sources at the relevant frequencies.

\subsection{NOEMA}

NOEMA observations were obtained as part of the S22CX program in several partial tracks during 2022 July.  An earlier Gemini-N GMOS program (GN-2020A-Q-903, PI: Zitrin) had yielded an optical spectroscopic redshift from template-fitting of $z$\,$=$\,4.267 (Nonino, Zitrin priv.\ comm.) for the SMA-located {HST} counterpart to 850.2.  This allowed the LO tunings to be chosen  to cover the expected frequencies of the CO(4--3), CO(5--4), and [C{\sc i}] lines of this source,  in the expectation that 850.1 was likely to lie at high redshift and potentially the same redshift as 850.2.  Hence, two setups were used in Band 1 (3\,mm)  with LO frequencies of 93.5\,GHz and  101.244\,GHz.   Tthe wide bandwidth of the PolyFiX correlator means that these two setups covered a contiguous frequency range of $\sim$\,31\,GHz from 81.88--112.86\,GHz. All tracks were observed in the compact D-configuration (maximum baseline length $\sim$\,175\,m) with 10 antennas in the array for five partial tracks, nine antennas for three partial tracks, and eight antennas for one partial track.

Calibration was performed using the Grenoble Image and Line Data Analysis Software ({\sc gildas}) package. Either MWC\,349 or LkHa\,101 was observed for use as the flux calibrator; in cases where one of these sources was either not observed or the data were of poor quality, the gain calibrator 1156+295 was used as the flux calibrator, bootstrapping its flux from other tracks with reliable observations of MWC\,349 or LkHa101. Bandpass calibration was obtained from observations of 3C\,273, 3C\,279, and BL Lac. After flagging for bad visibilities, the effective on-source time for a 10-antenna array for each of 850.1 and 850.2 was 1.6\,hours from the lower frequency setup and 1.8\,hours from the higher frequency setup. Finally, dirty cubes were produced with {\sc mapping} (part of the {\sc gildas} package) using natural weighting from $uv$-tables resampled to 20\,MHz. The synthesised beam sizes for the cubes ranged between 3\farcs8 to 6\farcs1 FWHM. 

The reduction of the NOEMA observations showed 3-mm continuum detections of both sources with positions that agree within the relative astrometric uncertainty with those from SMA (Figure~1).   Spectra extracted at the positions of the two galaxies yielded three lines in each spectrum (Figure~3).     The frequencies of the brighter two lines in each source were 87.65\,GHz and 109.56\,GHz (Figure~3, Table~2)  corresponding to CO(4--3)  and CO(5--4), at 461.041\,GHz and 576.268\,GHz in the restframe, respectively. The third, fainter line in each source (at $\sim$\,93.6\,GHz) was [C{\sc i}](1--0) at 492.161\,GHz.     These low-resolution observations in NOEMA's compact configuration showed no significant evidence for extended emission in either the continuum or lines.   

Gaussian fits to all three emission lines in each source yielded redshifts of $z$\,$ = $\,4.2599\,$\pm$\,0.0001 for 850.1 and $z$\,$= $\,4.2593\,$\pm$\,0.0003 for 850.2 (Figure~3).   The latter was in reasonable agreement with the  optical measurement for 850.2.     The Gaussian fits  were then used to derive line centers and widths, which were  used to  integrate the emission to derive the intensity (after continuum subtraction) within a window of $\pm$\,FWHM around the peak. For the weak [C{\sc i}] line in 850.2 the FWHM of the CO lines was used to determine the integration window.     From these the line intensity,  center and width (expressed as an equivalent Gaussian FWHM), were measured along with their uncertainties. The second moment of the line yields FWHM-equivalent estimates that were $\sim$\,20\% smaller than the Gaussian fitted values.  850.2 shows evidence of a double-peak profile in both the CO(4--3) and CO(5--4) lines (Figure~3), and so for that source  the velocity difference between the two peaks was instead used as the best estimate of the kinematics of the source.    The line and continuum properties of both sources are given in Tables~1, 2, \& 3.    The CO line luminosities were  used to estimate  line luminosity ratios between transitions
from $J_{\rm up}$\,$=$\,$i$ to $J_{\rm up}$\,$=$\,$j$, $r_{ij}$, following \citet{Solomon05}.

%
%
\setcounter{figure}{3}
\begin{figure*}[ht!]
  \centerline{ \includegraphics[width=7.0in]{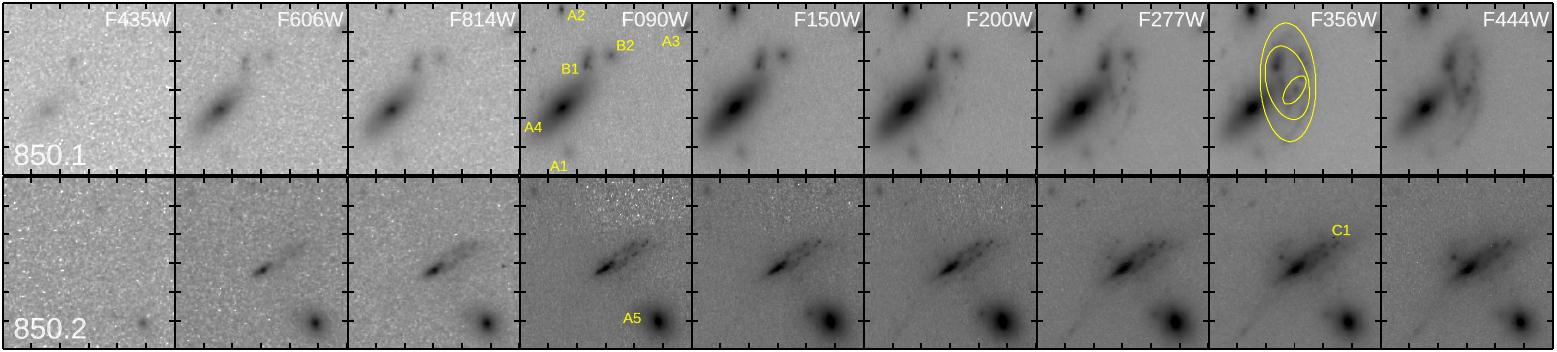} }
  
  \caption{\small 6$''\times$\,6$''$ thumbnails of 850.1 and 850.2.  From left to right the panels show the {HST} ACS F435W, F606W, and F814W imaging and  {JWST} NIRCam F090W, F150W, F200W, F277W, F356W, and F444W images.  The striking red colors of 850.1 are clear as the bulk of the emission from the galaxy, apart from the knots B1 and B2 (the former of which we suggest is foreground), is undetectable shortward of 2\,$\mu$m (the F200W band).     In contrast the bright central regions of 850.2 are detectable down to F606W (corresponding to $\sim$\,1100\AA\ in the restframe at $z$\,$=$\,4.26), but are undetected in F435W (restframe $\sim$\,830\AA) showing the presence of a strong Lyman break.    Fitted ellipses at three isophote levels are  overlaid on the F356W image of 850.1, illustrating the ``bar''-like feature seen in the central regions of the galaxy.    At  $z$\,$=$\,4.26 the 912-\AA\ Lyman break falls at the extreme red end of the F435W filter transmission, while    the Ly$\alpha$ emission line  falls in  F606W,  H$\alpha$  in F356W (strong emission from which likely explains the point-like feature C1 in 850.2, which probably corresponds to a giant [H{\sc ii}] region with an intrinsic size of $\sim$\,50--100\,pc), and the nebular emission lines [O{\sc iii}]\,4959/5007 are in F277W and [O{\sc ii}]\,3727 lies in F200W.  The various sources and features discussed in the text are identified on the F090W thumbnails.  
}
\end{figure*}

\subsection{{HST}, {JWST}, \& {SST}}

The details of the observations, reduction, and analysis of the {HST}  imaging of A\,1489  (PID: 15959) were reported in \cite{Zitrin20}.  The observations were obtained in 2020 March and comprised a total of five orbits, including one orbit ($\sim$\,1.9\,ks) in each of the F435W, F606W, and F814W filters with ACS and shorter exposures with WFC3 in the F105W, F125W, F140W, and F160W filters.   The WFC3 data were not used in our analysis as they were  shallower than the equivalent {JWST} imaging and also do not cover both targets.    As described by \cite{Zitrin20}, mosaiced images were produced for all of the {HST} filters using the calibrated exposures at a pixel scale of 0.03$''$\,pixel$^{-1}$ (matching that used for NIRCam) on an astrometric grid aligned to {Gaia} DR2.  See \cite{Zitrin20} for more details of the reduction and analysis.

{JWST} observed A\,1489 as part of GTO program 1176 (PEARLS, \citealt{Windhorst23}) targeting moderate-redshift clusters and blank fields to study galaxy evolution. A\,1489 was imaged in 2023 January  with NIRCam filters F090W, F150W,  F200W, F277W, F356W, and F444W.   Observations of F090W and F444W were taken with the MEDIUM8 readout with six groups per integration and four integrations for a total exposure time of 2.5\,ks. F150W and F356W were observed with a SHALLOW4 readout with nine groups per integration and four integrations for a total exposure of 1.9\,ks.  While F200W and F277W were observed with a SHALLOW4 readout with ten groups per integration and four integrations for a total exposure of 2.1\,ks.

Images were initially processed using the STScI {\sc calwebb} pipeline with the {\sc jwst\_0995.pmap} context. Further  processing was then applied to remove artifacts in the images. These steps were described in detail by \cite{Windhorst23}.  In brief, ``snowballs''  (resulting from energetic cosmic ray hits) were identified and masked, then ``wisps'' caused by straylight from the secondary-mirror support were corrected using filter-specific templates, next  striping due to amplifier differences was corrected by matching the background level in adjacent columns using the {\sc ProFound} code \citep{Robotham18}. Finally, detector-level offsets in the short-wavelength camera were corrected by constructing a ``super-sky'' with {\sc ProFound} and scaling each detector to that sky level.

%
%
\setcounter{figure}{4}
\begin{figure*}[ht!]
 \centerline{ \includegraphics[width=3.3in]{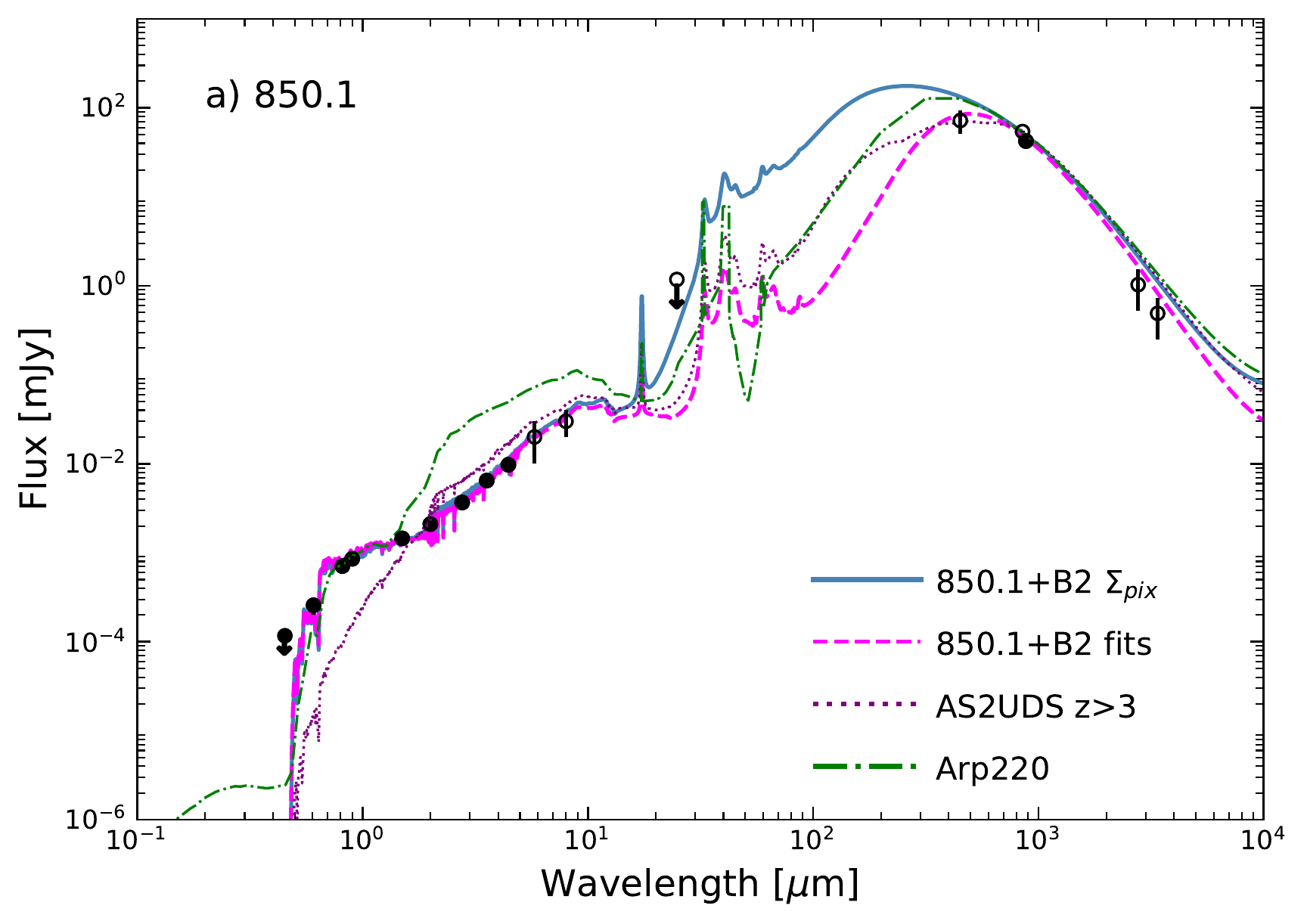}\hspace{0.2in}\includegraphics[width=3.3in]{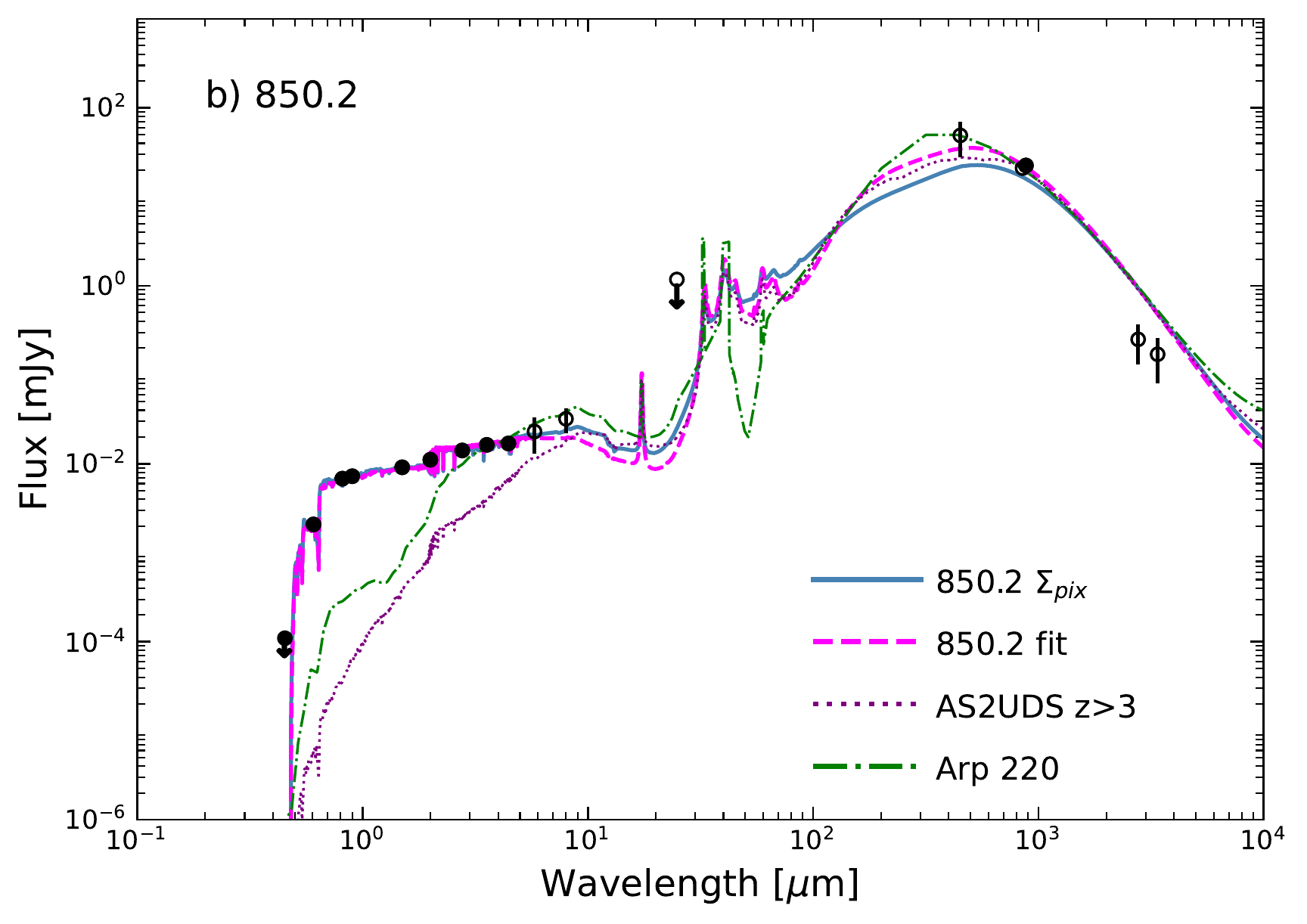} }
 \centerline{ \includegraphics[width=3.3in]{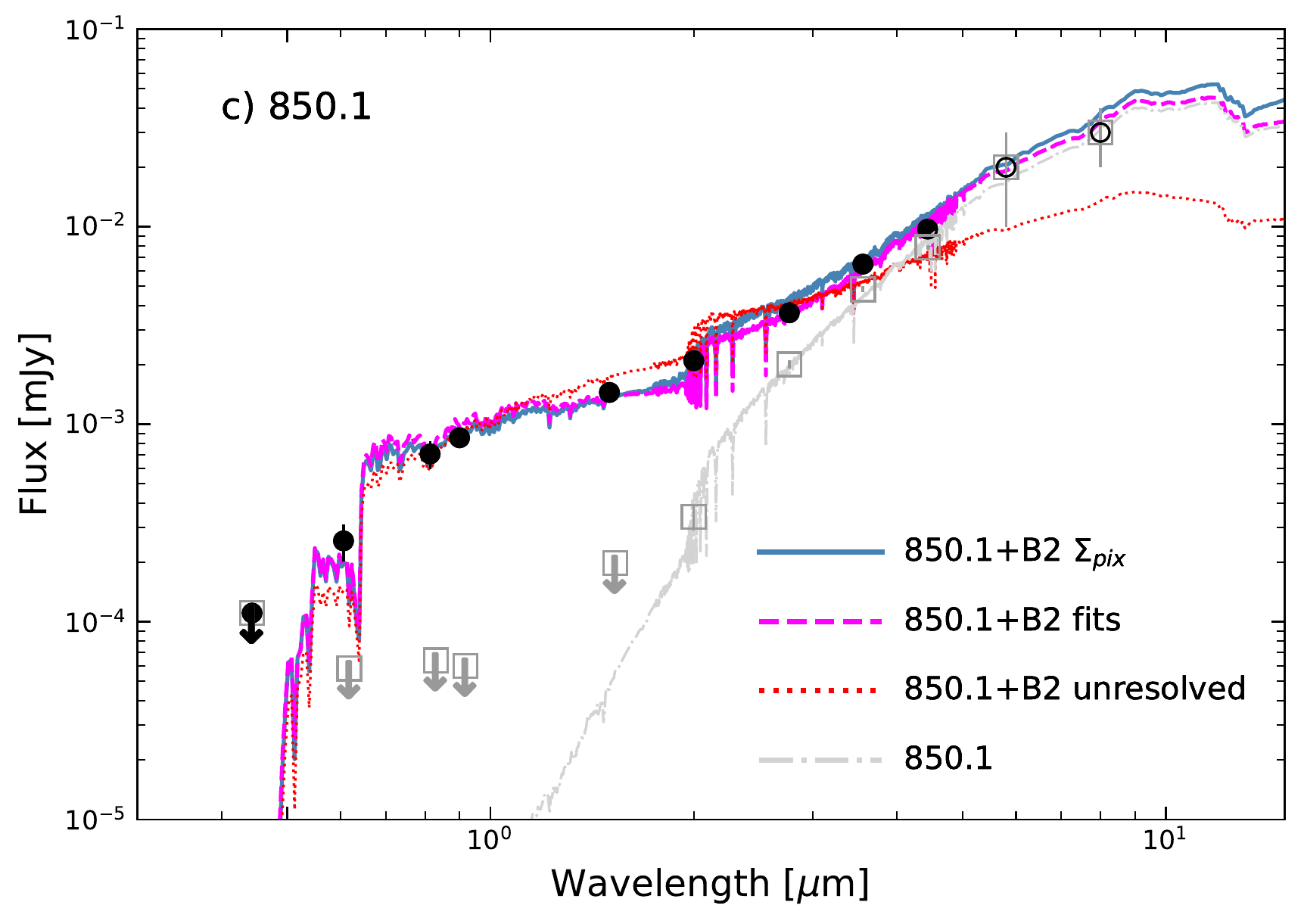}\hspace{0.2in}\includegraphics[width=3.3in]{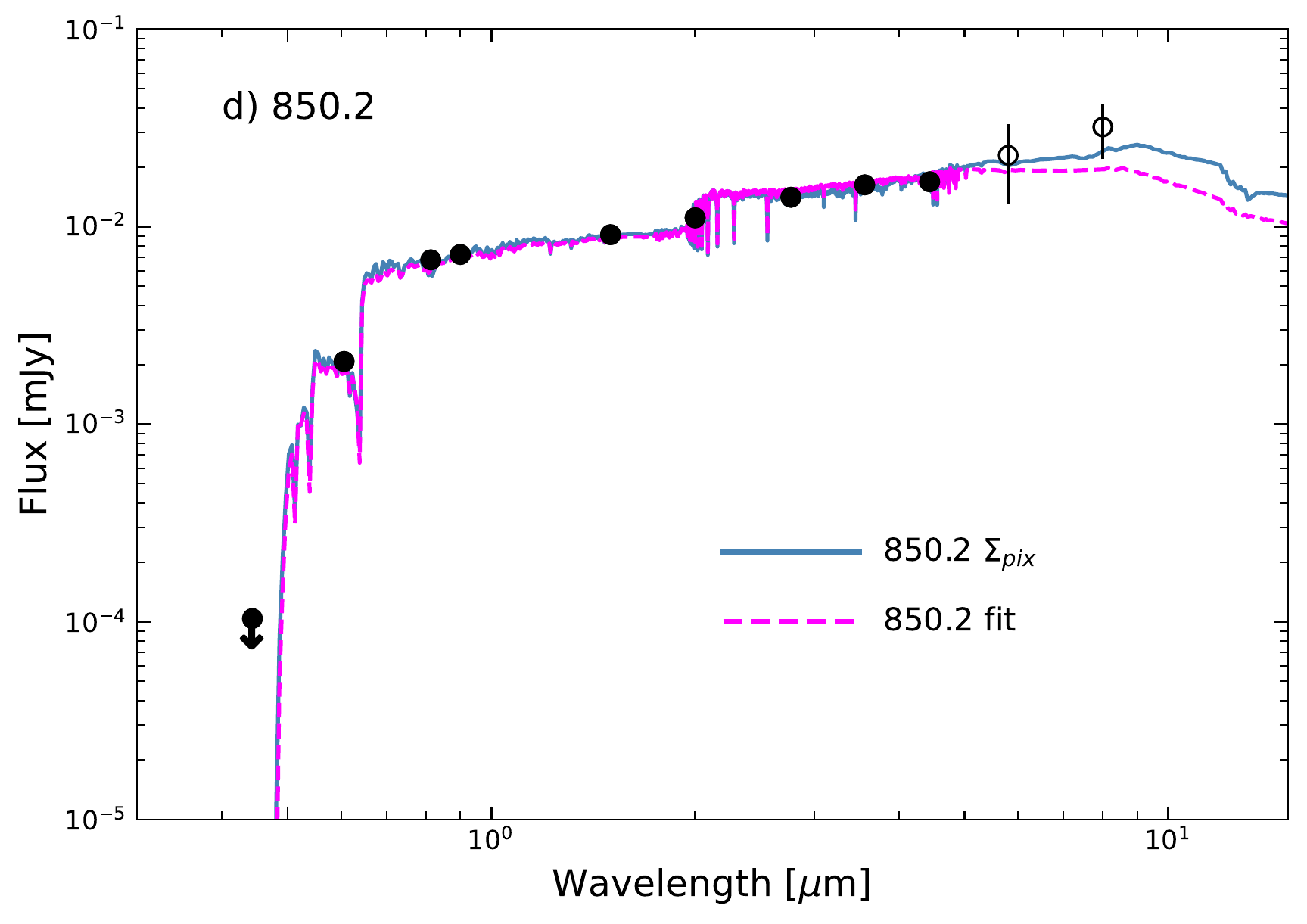} }
 \caption{\small  a) The spectral energy distributions for 850.1 (including B2) and b) 850.2, showing the measured photometry (Tables~1 \& 4) and the best-fit {\sc magphys} model at the redshifts determined from the NOEMA CO observations, $z$\,$=$\,4.26.   The lower two panels, c) and d), show an expanded view of the restframe UV--near-infrared region of the SEDs.  Two SEDs are shown for each source, one derived  from summing the resolved pixel-level SED fits   ($\Sigma_{\rm pix}$, including B2) and one from fitting to the integrated source photometry.  For the latter in 850.1, the individual fits to B2 and  850.1 were summed, although the emission longward of $\sim$\,3\,$\mu$m is dominated by 850.1. Wavebands  used in the resolved pixel fits are shown as solid points. For  the source-integrated fits these bands were used as well as the unresolved photometry in the wavebands marked with open symbols.
Also shown for comparison are the composite SED from the {\sc magphys} analysis of $z$\,$\geq$\,3 SMGs from the AS2UDS  survey by \cite{Dudzeviciute20} and  the SED of Arp\,220, both normalised to the SMA 880\,$\mu$m flux density.  These illustrate that the combined SED of 850.1+B2 is similar to  the AS2UDS composite between restframe $B$-band (observed $\sim$\,2$\mu$m) and  near-infrared, but due to the contribution from B2 it becomes much bluer into the restframe UV, better matching Arp\,220. 850.1 alone is redder than the AS2UDS composite shortward of 2\,$\mu$m.     850.2 has an SED comparable to Arp\,220 longward of $\sim$\,3\,$\mu$m (restframe $\sim$\,6000\AA) but bluer shortward of this and considerably bluer than  the AS2UDS composite.
 In c) the results of fitting  to the photometry and limits for just 850.1 are shown (gray open squares), as well as a single
 {\sc magphys} fit to the combined (``unresolved'') photometry from 850.1 and B2, illustrating that this provides a poorer description of the observations, especially longward of 3\,$\mu$m.  In all  panels, the resolved-pixel modelling, $\Sigma_{\rm pix}$, had only a single constraint beyond 4.4\,$\mu$m from the SMA at 880\,$\mu$m, yet it does at least as well out to 8\,$\mu$m as the fits to the integrated source photometry (that included the IRAC 5.8 and 8.0\,$\mu$m fluxes in the fit), although the absence of spatially resolved constraints at 450\,$\mu$m resulted in poorer agreement in the restframe far-infrared region.
}
\end{figure*}

Figure~1  shows a  {JWST}/{HST} view of the wider environment of 850.1/850.2.  More detailed zooms are shown in Figure~2, including the SMA 880$\mu$m maps that precisely locate the counterparts to the sub-millimeter sources.  Figure~4 illustrates the variation in the brightness of the two counterparts with wavelength across the nine {JWST}/{HST} bands.  Sources  were identified  in a sequence starting with ``A'' for distinct sources,  ``B'' for potential subcomponents, and ``C'' for likely emission-line features. These are labelled on the F090W images.   

To extend the wavelength coverage of these dusty, high-redshift galaxies, the pipeline reductions of the {Spitzer} IRAC and MIPS 24\,$\mu$m images of A\,1489 were retrieved from the {SST} archive.  This was undertaken prior to both the launch of {JWST} and the SMA observations, finding potential 8-$\mu$m bright counterparts within the SCUBA-2 error circles for both 850.1 and 850.2, that turned out to be the correct identifications.  These data were included in our SED analysis to extend the wavelength coverage in the gap between {JWST} and SCUBA-2.  Neither source was detected at 24\,$\mu$m with MIPS. This was unsurprising given their high redshifts and the modest depth of the data, and so a 3-$\sigma$ limit was adopted at that wavelength (Table~1). The emission in the IRAC bands was blended with the nearby galaxies for both 850.1 and 850.2, and therefore the flux estimates were based on small 2\farcs4--3\farcs6 diameter apertures.  However, the sub-millimeter galaxies were increasingly dominant at the longer wavelengths and 850.2 was effectively uncontaminated by 8\,$\mu$m.   For both sources, excluding the IRAC and MIPS data points, or adopting limits instead, did not alter the solutions for the best-fit SEDs obtained below.   

To measure integrated photometry for 850.1/850.2 and the other sources in close proximity to them in the {JWST} and {HST} data,  large  circular  apertures were used (except for 850.1) with sizes listed in Table~4. These measurements used the  photometric zero points from \citet{Windhorst23} for NIRCam and from the ACS pipeline (as given in the FITS file headers) for the ACS imaging.  These provided close-to-total fluxes for most of the sources.   PSF matching employed model PSFs created with {\sc WebbPSF} v1.1.0 or {\sc TinyTim} for {HST}/ACS, utilizing the {\sc odt} files closest to the observation date, and convolved to match the F444W PSF.   Due to the large aperture sizes, PSF corrections for  850.1/850.2 were modest in  all cases.   Sky corrections were derived from large annuli around the targets.  An additional correction was applied for  the contribution from the disk galaxy,  A4,  to those sources around 850.1.   This was modelled in each filter assuming rotational symmetry and rotating the images by 180\,degrees around the  center of A4, then smoothing slightly, before  subtracting this image from the original.   The resulting photometry is listed in Table~4 with 3-$\sigma$ limits for non-detections.

To determine the sizes and profiles of the two sources, the {\sc galfit} code was applied to the different {JWST} wavebands assuming single S\'ersic profiles and correcting for the varying resolution between the bands using an empirical PSF taken from stars in the images.  This process was challenging in the crowded environment of 850.1, especially given the fine structures visible in the surrounding galaxies as a result of the superlative resolution and depth of the data.  However, aided by the fact that the bulk of 850.1 was undetected in the F090W image,  a model was constructed in this band of the surrounding galaxies and used as the  basis for constrained fitting to the redder passbands, where  a further component was included to represent 850.1.   The structural parameters of both sources from this analysis are reported in Tables~2 \& 5.

\subsection{Integrated SED modelling}

The SED modelling of the two sub-millimeter sources was iterated with the aim of understanding the nature of the various subcomponents in these systems and so derive robust physical properties for the two galaxies.   The analysis employed both the photometric-redshift version of the {\sc magphys} code \citep{Battisti19} as well as the high-redshift version \citep{daCunha15} to derive parameters at the  known CO redshifts, both using the standard priors (including a star-formation history that rises at early times, before declining exponentially, with superimposed random bursts).    The selected codes were applied to  the integrated photometry of the  sources (including potentially unrelated subcomponents), to the integrated photometry of the individual subcomponents, and also to the resolved ``pixel-level'' photometry (described in the next subsection).   In principle, the latter resolved SED fitting should be most robust, but the modest resolution of some of the longer-wavelength observations ({Spitzer} IRAC, SCUBA-2, and NOEMA) means that there were fewer constraints available on the form of the SEDs.   For this reason the resolved SED fitting is presented as a proof-of-concept, and for most of the quantitative analysis, the fitting to the integrated photometry was used to derive the physical properties.

The photometric analysis of 850.2 was  straightforward owing to the relative isolation of the galaxy from other sources in the field, as well as the relatively good agreement between the redshifts derived from the CO/[C{\sc i}] emission lines and the archival restframe UV spectroscopy.  The latter confirmed that the emission seen in both sub-millimeter and the restframe UV--optical  all arises from the same physical system, although not necessarily precisely the same components within this system.

In contrast, 850.1 has a  complex environment and  the analysis of this system required  iteration.  Based on the source morphology and its variation with wavelength shown in Figure~4, as well as the results from the resolved SED modelling discussed in the next sub-section, the emission around 850.1 was divided into three components:  two blue ``knots''  B1 and B2, and the bulk of the remainder of the source, 850.1, which shows much redder colors and corresponds to the peak of the SMA map (Figure~2).   The analysis started   by estimating likely redshifts for these three components as well as the other sources seen close to 850.1/850.2 that lacked spectroscopic redshifts (those marked in Figure~2 and listed in Table~4 \& 5),  with the photometric redshift version of {\sc magphys}  \citep{Battisti19,daCunha15}.  These photometric redshifts are reported in Table~5.  These were then used to assess the level of agreement of the spectroscopic redshift measurements for 850.1 and 850.2 with their photometric properties.   

For 850.2, the photometric redshift, $z_{\rm ph}$\,$=$\,4.28$^{+0.08}_{-0.10}$, driven by the strong Lyman break in the broadband SED (Figures~4 \& 5) was in excellent agreement with the spectroscopic redshift  $z$\,$=$\,4.26 derived from the multiple lines seen in the NOEMA spectrum (Figure~3), and also the Gemini-N/GMOS spectrum (Nonino, Zitrin, priv.\ comm.).    The latter spectrum showed a pronounced decline in flux blueward of $\sim$\,1215\AA, no Lyman-$\alpha$ emission but a series of strong absorption features including   Si{\sc ii}\,$\lambda$1260, O{\sc i}+Si{\sc ii}\,$\lambda$1303 (blend), C{\sc ii}\,$\lambda$1334, Si{\sc ii}\,$\lambda$1526 and C{\sc iv}\,$\lambda$1548, Fe{\sc ii}\,$\lambda$1608 and Al{\sc ii}\,$\lambda$1670.  The Si{\sc  ii}\,$\lambda$1265 and C{\sc iv}\,$\lambda$1548 lines displayed  P-Cygni-like profiles.   These  low-ionisation interstellar absorption features are commonly seen in Lyman-break galaxies \citep{Shapley03}.   In 850.2, their measured line centers displayed a median blueshift of $\sim -$190\,km\,s$^{-1}$ relative to the CO-derived systemic redshift.  This velocity offset may be a signature of a wind, or  it indicates that the UV-bright element of the system corresponds to the higher frequency peak in the CO spectrum of the source (Figure~3), with the redshifted CO component then potentially arising in the redder part of the system.

A similar photometric analysis for 850.1, used a photometric aperture  that includes the emission from the two knots B1 and B2 (Figures~2 \& 4, Table~4).   This resulted in a  photometric redshift that  was significantly lower than the  CO redshift: $z_{\rm ph}$\,$=$\,2.63$^{+0.46}_{-0.23}$ (Table~5).   Figure~4  shows clearly that B1 and B2 are considerably bluer than the bulk of 850.1, both dominating the emission shortward of $\sim$\,2\,$\mu$m, and B1 is visible all the way down to F435W,  where the wavelength coverage of the filter  falls almost completely blueward of 912\AA\ in the restframe at $z$\,$=$\,4.26.     Photometric redshifts
measured for the three components individually yielded estimates of  $z_{\rm ph}$\,$=$\,1.12$^{+0.11}_{-0.12}$ and $z_{\rm ph}$\,$=$\,4.17$^{+0.16}_{-3.28}$ for B1 and B2, respectively, while the very red  850.1 gave $z_{\rm ph}$\,$=$\,4.41$^{+0.77}_{-1.95}$ in reasonable agreement with the CO redshift (Table~5).

The photometric redshift estimate for B1  indicated that the source was unlikely to be associated with either 850.1 or the foreground cluster. Thus, while the alignment of B1 with features in 850.1 is very suggestive that they are related, B1's detection in the F435W filter is hard to reconcile with it lying at $z$\,$=$\,4.26, and  it is more likely to be a superimposed lower-redshift source.  The situation is more complicated for B2, where the estimated photometric redshift suggests it could be associated with 850.1 at $z$\,$=$\,4.26, but the uncertainty allows much lower redshift solutions as well.   We therefore retain B2 as a potential UV-bright component within 850.1 or a close/interacting companion and discuss  how this influences the inferred properties of the source.

Having confirmed that the measured photometry for 850.1, B2, and 850.2 were all consistent with the CO-derived redshifts,  more representative estimates of their physical properties were derived by running the high-redshift version of {\sc magphys} \citep{daCunha15} with the redshifts fixed at $z$\,$=$\,4.26 for all three components.   The main parameters  from these fits are reported in Table~5  and  the best-fit SEDs plotted in Figure~5.   For 850.1 and B2,  the two best-fit SEDs were summed and labelled ``850.1+B2 fits'', and this is plotted  compared to the integrated photometry of the system, although   850.1 dominates the emission longward of $\sim$\,3\,$\mu$m.  In addition,  Figure~5c  show the results of fitting solely to 850.1, illustrating  the very red colors of this source.

The fit to  the combined photometry of 850.1 and B2 using a {\it single} {\sc magphys} model and labelled ``850.1+B2 unresolved'' is also shown in Figure~5c.  This  was much less successful at reproducing the optical/near-infrared SED of the combined source.  This  resulted in a
statistically much poorer fit to the observed photometry that predicted a much lower  mass, star-formation rate, $A_V$ and age (see Table~5) but a comparably high dust mass  to the preferred sum of the individual fits to 850.1 and B2.   These parameters would imply a galaxy  more similar to 850.2 in its physical properties, but we believe that this poor fit was erroneous due to the distinct low and high extinction regions that were being modelled with a single dust extinction term.    Similar model biases  may also be responsible for some  apparently low-mass sub-millimeter galaxies identified from SED fitting  of systems with a wide range in $A_V$  \citep[e.g.,][]{Pope23}.  This erroneous fit also produces an apparent power-law-like excess in the observed fluxes longward of 3.6\,$\mu$m compared to the model.   This is reminiscent of similar features seen in the SEDs of $\sim$\,10\% of sub-millimeter sources analysed by \cite{Hainline11}, who attributed it to hot dust emission from  dust-obscured AGN.    If the origin of this excess is similar in those sources to 850.1, then an obscured AGN  may not be the cause  as there is no such bright compact,  red point-source  seen in 850.1.  Instead the red mid-infrared excess may be a result of fitting a single SED model to a source with a mixture of internal dust obscuration, including both weakly and highly reddened regions.

%
%
\setcounter{figure}{5}
\begin{figure*}[ht!]
  \centerline{ \includegraphics[width=6.5in]{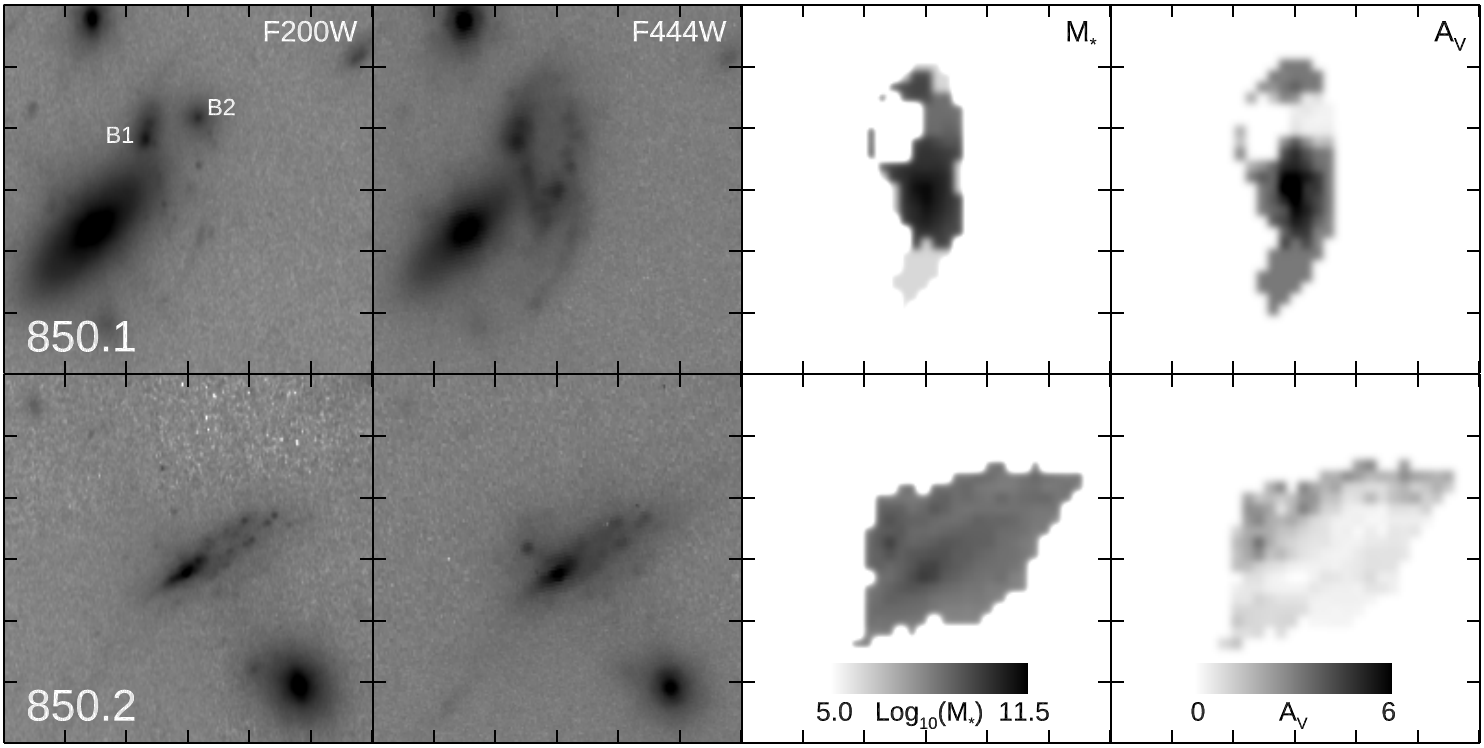} }
  
  \caption{\small 6$''\times$\,6$''$ thumbnails of 850.1 and 850.2 showing the {JWST}  F200W and F444W images along with the pixel-based {\sc magphys} estimates of the log-scaled stellar mass and $A_V$.  The stellar-mass surface density is  per 0\farcs18$\times$\,0\farcs18 pixel,  corresponding to $\sim $\,0.6\,$\times$\,0.6\,kpc in the source plane for 850.1 and $\sim $\,0.5\,$\times$\,0.5\,kpc for 850.2.    B1 was excluded from the map for 850.1 due to the poor quality of the SED fits at $z$\,$=$\,4.26, supporting its likely lower redshift.   B2 has much lower $A_V$ and stellar-mass surface density compared to the bulk of the massive and dusty remainder of 850.1.  These very distinct properties motivate the decision to fit B2 and the bulk of 850.1 separately and then sum the results.   In contrast, 850.2 exhibits a smaller variation in $A_V$ and stellar-mass surface density across the source.
}
\end{figure*}

\subsection{Resolved SED modelling}

The high resolution and depth of the {JWST}/{HST} imaging of these sources can be used to go beyond the analysis just described to undertake resolved SED modelling of the sources to investigate their internal structures. Specifically to map the distribution of stellar mass and dust reddening within the two systems \citep[see also,][]{Duncan23}.   To perform this analysis   9$''\times$\,9$''$ thumbnails were extracted from the {JWST} and {HST} imaging centered on 850.1 and 850.2.   The {JWST} thumbnails were weighted and coadded,  the combined image smoothed with a Gaussian filter with a FWHM of 0\farcs07, and then  {\sc SExtractor} \cite{Bertin96} used to construct a segmentation map.  The segmentation maps was employed to identify  regions of blank sky  that were used to estimate the pixel-to-pixel noise and also  to define areas associated with the various sources.  To aid this for 850.1 the large-scale emission from the bright galaxy, A4, was first removed by rotating the images by 180\,degrees centered on A4 and subtracting these from the original frames. 

To ensure that the final spatially resolved maps were not oversampled,   the   {JWST} and {HST}  images were  PSF-matched to the resolution of the F444W observations, $\sim$\,0\farcs15.   These images (and the associated segmentation masks) were then  rebinned from the default 0.03$''$ pixel$^{-1}$ sampling to 0.18$''$ pixel$^{-1}$, to provide nearly independent pixels.  From this binned data,  photometry was extracted in all the {JWST}/{HST} bands for each  0\farcs18 pixel, including the uncertainties measured from the pixel-to-pixel variations in the background regions.

To include information about the longer-wavelength emission, the SMA 880-$\mu$m map was used to provide a  constraint on the  long wavelength pixel-to-pixel photometry.  For each source  a model was constructed from the deconvolved 880-$\mu$m source properties measured by SMA and sampled at 0\farcs18.  For the 880-$\mu$m photometry,  liberal  uncertainties of 2\,mJy per pixel were adopted to avoid over-constraining the {\sc magphys} fits and also to reflect the fact that the resolution of the SMA map was coarser than the {JWST}/{HST} imaging.

For the  {\sc magphys} analysis of the pixel-based photometry, at least three detections were required at $>$3\,$\sigma$ in the visible or near-infrared wavebands to include the fitted parameters in the analysis.  Fits to pixels with too few photometric constraints  resulted in poorly constrained  properties.   However, the results did not change significantly if  2-$\sigma$ limits in three bands or 3-$\sigma$ in two bands were used instead, although the uncertainties became more significant.  To derive the physical properties of each source the high-redshift version of {\sc magphys} was applied to the photometry for each pixel within the segmentation mask that defines each of the sources.  A fixed redshift of $z$\,$=$\,4.26 was used, but  all other parameters were allowed to  vary according to the standard priors used in the previous section, and also adopted in previous studies \citep[e.g.,][]{Dudzeviciute20}.   For the analysis of 850.1,  the pixels that were associated with B1 were included, but then removed as this appears to be an unrelated foreground source. This conclusion  was also supported  by the $\chi^2$ of the  {\sc magphys} fits for B1  at $z$\,$=$\,4.26 being significantly worse than  those in the remainder of the source ($\gg$\,20, compared to $\sim$\,1).

Two further requirements were applied in the resolved analysis for a fit to a pixel to be included in the final maps. Firstly,   the  {\sc magphys} fit had to provide a reasonable description of the photometric observations, yielding a $\chi^2 <$\,10. This removed seven pixels in 850.2 and none in 850.1.  Secondly,   the predicted 850-$\mu$m flux density from the best-fit {\sc magphys} SED for the pixel must not exceed 10\,mJy.   This caught situations where the predicted restframe far-infrared SED for the pixel, which was only weakly constrained by the SMA observations, was strongly at odds with the expected flux distribution for the source.   This affected only  four pixels from 116 in 850.1 and twelve from 415 in 850.2.   Typically  these were in the lowest surface brightness regions on the periphery of the sources and these pixels were replaced by the local median fit.    With these cuts applied, the total sub-millimeter flux densities for the sources from summing the predicted {\sc magphys} 850-$\mu$m fluxes were 52\,mJy for 850.1, compared to 42.3$\pm$1.6\,mJy and 54.1$\pm$1.5\,mJy observed by SMA and SCUBA-2 respectively, and 14\,mJy for 850.2, compared to 22.5$\pm$1.6\,mJy and 21.2$\pm$1.5\,mJy for SMA and SCUBA-2 respectively.   Figure~5 shows the predicted full SEDs for both sources, constructed by summing the individual best-fit pixel SEDs.  These provide very good fits to the visible to near-infrared (including the longer wavelength {Spitzer} bands that were not included in the resolved analysis), and while the agreement at longer wavelengths in the restframe far-infrared to sub-millimeter was poorer, given the limited information available at these wavelengths for a spatially resolved analysis at {JWST} resolution, we view the agreement as reasonably successful.

This resolved analysis allowed the internal variation in some of the key physical properties of the sources to be mapped on $\sim$\,0\farcs18 scales, equivalent to  0.5--0.6\,kpc in the source plane after accounting for lensing magnification.    The discussion below focuses on the distribution of stellar mass and dust reddening, $A_V$, as these two quantities are most closely tied to the  photometric properties in the {JWST} and {HST} wavebands, where the observations have the highest resolution and signal-to-noise.  The absolute normalisation of these maps will depend on the SED modelling code used \citep[e.g.,][]{Carnall19,Pacifici23}, but for comparison with the earlier results and published  samples we use {\sc magphys}.
The resolved 2-D maps of the estimated stellar-mass density and dust reddening for 850.1 and 850.2 are shown in Figure~6.

\section{Results and Discussion}

Figure~2 illustrates the {JWST}/{HST}  view of the surroundings of the two sources as well as their morphologies after correcting for the image shear due to lensing, while the variation in their brightness across the nine  {JWST} and {HST} bands is shown in Figure~4.     Strikingly,  850.1 is undetected in {\it all} the {HST} ACS bands and the bluer {JWST} bands, the bulk of the galaxy only being significantly detected at 2\,$\mu$m or above,  while 850.2 displays emission down to F606W but then a strong drop between F606W and F435W  (see Figures 4 \& 5).   

There are a number of sources detected with {JWST}/{HST} lying close to 850.1 and 850.2, with photometric redshifts reported in Table~5.   These redshifts indicated that two of the galaxies near 850.1, including the bright disk galaxy A4, are probably members of the cluster A\,1489 at $z$\,$\sim$\,0.35 and that similarly, the disk galaxy A5, near 850.2, is also likely to be a cluster member.   This is unsurprising given that the two sub-millimeter sources are seen through the dense central regions of the foreground cluster.   This analysis also suggested that none of the other galaxies visible  in close proximity to 850.1 and 850.2 is likely to be physically associated with the sub-millimeter sources.

\subsection{Gravitational lens model}

A first strong-lens model for A\,1489 was presented by \citet{Zitrin20} based on multiple images identified in their {HST} observations. The analysis here uses a new lens model  based on the same set of multiple images but employing a revised and updated version of the \citet{Zitrin15} parametric code \citep{Pascale22,Furtak23}. In this model, cluster galaxies were modeled as double pseudo-isothermal ellipticals (dPIE) and dark matter haloes were modelled using pseudo-isothermal elliptical mass distributions (PIEMDs). Two such halos were used for A\,1489, centered on the two main mass clumps identified by \citet{Zitrin20} and mass components were included to represent the  cluster galaxies near to 850.1 and 850.2, including A4 and A5.

The updated lens model predicts similar lensing magnification for 850.1 and 850.2 (Table~5) of $\mu$\,$=$\,4.0$_{-2.2}^{+1.0}$ and $\mu$\,$=$\,5.6$_{-3.3}^{+1.0}$, respectively, where  the uncertainties were judged from the range of acceptable models and are expected to be conservative.   These values were employed when comparing the properties of these sources to other populations.  Table~6 lists intrinsic, lensing-corrected, properties for both sources.    The NIRCam morphologies of the sources after correcting for the lensing magnification are shown in Figure~2d and 2f.

Accounting for lensing magnification, 850.1 has an intrinsic 850-$\mu$m flux density of $S_{\rm 850\mu m}$\,$=$\,13.5\,$\pm$\,0.4\,mJy,  while 850.2 has $S_{\rm 850\mu m}$\,$=$\,3.8\,$\pm$\,0.3\,mJy. This shows that 850.1 is an example of a rare population of very luminous sub-millimeter sources with a surface density of $\sim$\,10\,$\pm$\,5 degree$^{-2}$ \citep{Geach17,Simpson19,Garratt23}, while 850.2 comes from a much more numerous population around the knee in the 850-$\mu$m counts \citep{Stach18} with a surface density $\sim$\,50\,$\times$ higher ($\sim$\,500\,degree$^{-2}$).

Corrected for lens magnification, $\Delta$\,$\sim $\,1.5\,mag,  850.1 would have $H$\,$\sim $\,29 and $K$\,$\sim$\,27 for an extended source (not a point source as frequently adopted to define survey limits).   This demonstrates that neither previous {HST} nor ground-based $K$-band surveys would have been able to detect this class of source, let alone if there are less-massive or higher-redshift examples.   

After correcting their sky positions for lensing, the relative separation of 850.1 and 850.2 was estimated to be  $\sim$\,100\,kpc ($\sim$\,15$''$) in the source plane.  Thus, without the lensing magnification of the foreground cluster, these two galaxies would  appear as a single blended bright ($S_{\rm 850\mu m}\sim$\,17\,mJy) source in a single-dish sub-millimeter survey, similar to the associated multiple $z$\,$\sim$\,4.62 components found in two SCUBA-2 sources by \citet{Mitsuhashi21}.   This small spatial separation, along with the line-of-sight velocity difference between 850.1 and 850.2 of only $\Delta v \sim$\,30\,$\pm$\,20\,km\,s$^{-1}$, indicates that  these two galaxies are part of a small group at $z$\,$\sim$\,4.26 lying behind the core of the foreground cluster.  Indeed,   given the stellar mass  for 850.1 estimated below, unless 850.2 has a very high transverse velocity, then it is very likely on a bound orbit and will merge at some point in the future.

This  group behind A\,1489 adds to a  growing list of similar systems comprising one or more sub-millimeter galaxies and other companion galaxies in groups at $z$\,$\sim$\,2--4, lensed by massive foreground clusters \citep[e.g.,][]{Ledlow02, Borys04, Kneib05, Caputi21, Frye23}.  In view of the  large volumes accessible in surveys for sub-millimeter sources due to the negative $K$-correction, the frequency of identification of such systems may reflect the similarity between the spatial size of group-sized halos with mass comparable to those of  sub-millimeter galaxies \citep[$\sim$\,10$^{13}$\,M$_\odot$, e.g.,][]{Stach21} and the source-plane extent of the high-magnification region of the most massive clusters at $z$\,$\ls$\,1 \citep[see][]{Frye23}.

\subsection{SEDs and stellar properties}

The spectral energy distributions of the combined 850.1+B2 and 850.2 shown in Figure~5 are  strikingly different.  Shortward of 2\,$\mu$m (restframe $U$-band) the only emission from 850.1+B2 comes from the component/companion B2, with 850.1 itself undetected  (see Figure~5c).  In contrast, both the disk and the high surface brightness feature in 850.2 display a blue continuum  that is detected down to the restframe far-UV (Figure~4).

The {\sc magphys} analysis provided a lensing-corrected stellar mass from the sum of the fits to 850.1 and B2   $M_\ast$\,$=$\,5.5$^{+1.3}_{-0.8}\times$\,10$^{11}$\,M$_\odot$ and a star-formation rate (SFR) of SFR\,$=$\,900$^{+400}_{-300}$\,M$_\odot$\,yr$^{-1}$ (Tables~5 \& 6).   These estimates are in reasonable agreement with the sum of  the fits to the individual resolved pixels  $M_\ast$\,$=$\,4.0$^{+1.0}_{-0.3}\times$\,10$^{11}$\,M$_\odot$ and  SFR\,$=$\,1200$^{+300}_{-100}$\,M$_\odot$\,yr$^{-1}$, even though the latter fit lacked several of the mid-infrared and sub-/millimeter photometric constraints used in the former.   If B2 is taken to be a separate galaxy from 850.1,  the estimates for the system do not change appreciably as the latter contains the majority of the mass and SFR.  For 850.2 the estimates were: $M_\ast$\,$=$\,0.18\,$\pm$0.07\,$\times$\,10$^{11}$\,M$_\odot$ and a  SFR\,$=$\,210\,$\pm$\,20\,M$_\odot$\,yr$^{-1}$, with the resolved pixel-fits  again in reasonable agreement:  $M_\ast$\,$=$\,0.40$^{+0.4}_{-0.1}\times$\,10$^{11}$\,M$_\odot$ and  SFR\,$=$\,200$^{+160}_{-70}$\,M$_\odot$\,yr$^{-1}$.    AGN contamination is not biasing the high stellar mass derived for 850.1 (or indeed 850.2) as the F444W morphology indicates only a small fraction of the light in the restframe near-infrared could be  from a central point source (see Figures~2d and 2f). Similarly, removing the  IRAC 5.8- and 8.0-$\mu$m photometry from the {\sc magphys} analysis did not change the best-fitting SEDs.

After  correcting for lensing magnification,  850.1  is  fainter  than  850.2  at all wavelengths shortward of 8\,$\mu$m (restframe $H$-band), yet the estimated stellar mass for 850.1  is $\gs$\,30\,$\times$ higher, while its star-formation rate and dust mass  are only $\sim$\,3--4\,$\times$ higher than those derived for 850.2.      The primary driver of this is the  differences in ages for the best-fit SEDs.  850.1 has a best-fit mass-weighted SED with an age of $\sim$\,450\,Myr, corresponding to a mass-to-light ratio for the stellar population  in the restframe $H$-band  (observed $\sim$\,8.5\,$\mu$m)  $M/L_H$\,$\sim$\,4.5.  In contrast,    850.2 has a much younger best-fit SED with an age of $\sim$\,50\,Myrs and  $M/L_H$\,$\sim$\,0.3.  This difference in age accounts for a factor of $\sim$\,15 in the predicted masses,  with the remainder due to the difference in extinction between the two galaxies, $A_V$\,$\sim$\,5\,mag and $\sim$\,1\,mag, equivalent to $\Delta A_H$\,$\sim $\,1.   
The  time needed to form the  stellar masses of 850.1 and 850.2, assuming the current SFRs (averaged over a 100\,Myr timescale), was $\sim$\,600\,Myrs and $\sim$\,90\,Myrs, respectively.  These are comparable to the {\sc magphys} mass-weighted ages for both galaxies,  indicating that it is possible that the bulk of the stellar mass in both  systems was formed in the on-going star-formation events, with 850.1 potentially forming at $z_{\rm form}$\,$\sim$\,6.   

\subsection{Morphologies and structures}

The extended emission of the two galaxies is well detected in  the redder {JWST} bands and shows that 850.1 is  morphologically more complex than 850.2 (Figure~2).  The F444W morphology of 850.1 in Figure~4 suggests a nearly face-on system with two arm-like features extending north and south from the ends of a linear structure of clumps.  This bar-like feature has a source plane radius of $\sim$\,2--2.5\,kpc, and spans a slightly brighter clump at the center of the galaxy.  The 880-$\mu$m emission detected by SMA in 850.1 peaks on the  central clump (Figure~2), and the resolved  SED analysis shown in Figure~6 indicates that the stellar-mass surface density and the dust reddening also both peak around this region.
The map shows a maximum stellar-mass density  of 14\,$\times$\,10$^{10}$\,M$_\odot$\,kpc$^{-2}$ and a mean density within $R^{\rm mass}_{\rm e}$ of (6.4$\pm$1.4)\,$\times$\,10$^{10}$\,M$_\odot$\,kpc$^{-2}$. As surface densities these quantities are magnification invariant.   The  resolved SED maps  also indicate that  B2,   seen in projection against the northern part of the disk of 850.1, has  a lower stellar mass surface density, $\sim$\,10$^{9}$\,M$_\odot$\,kpc$^{-2}$, than the bulk of 850.1 and also a much lower  reddening, $A_V$\,$\sim$\,0.3--1.

The variation as a function of scale of the orientation of the main structural features in 850.1 is illustrated in Figure~3.
Fitted ellipses are  overlaid at three surface brightness levels onto the F356W frame showing a change in orientation and ellipticity in the inner regions, compared to the outskirts.   This variation is similar to the characteristic change in orientation and ellipticity seen  from a bar.    More qualitatively, the morphology of this source (seen in the restframe $I$-band) is  reminiscent of the bar and arms in NGC\,1365, the dominant galaxy the Fornax group, as well as showing features similar to the  arms and bars seen in dust continuum in high-redshift sub-millimeter galaxies  \citep{Hodge19}.

Although the internal colors of 850.1 appear uniformly red in Figure~2a,  this is  misleading as the source is only well detected in the three filters longward of 2.8\,$\mu$m, two of which were combined into the red channel.   Figure~2c (and 2d),  uses the three reddest {JWST} bands and shows  variation in color within the system, with the reddest region coinciding with the central clump.   The map of $A_V$ from the resolved pixel-level SED fits in Figure~6  also indicate a  reddening gradient  in the 850.1  that ranges from $A_V$\,$=$\,6.8 in the center to $A_V$\,$\sim$\,3 in the outskirts. However, these  reddening estimates are for light detected in F444W, and will be lower than what would be measured if  longer-wavelength resolved photometry was included.   This may explain why the median F444W-weighted $A_V$ from the pixel-level SEDs was $A_V$\,$=$\,3.6$^{+1.5}_{-1.2}$ compared to $A_V$\,$\sim$\,5 measured from the integrated photometry including longer wavelength constraints.   

While the map of $A_V$ in 850.1 indicates that   the central regions are highly obscured (probably even more obscured than shown in the map as this corresponds to light that was still detectable at restframe $I$-band, see \citealt{Simpson17}), the outer parts of the galaxy are also surprisingly highly and uniformly reddened out to radii of $\sim$\,3--4\,kpc. The simplest explanation may be that this galaxy has a very dusty disk (even if the gas fraction for the system is low, see \S3.5), as well as an even more obscured nuclear region.   Such extended dust disks have been seen in similarly dust-rich galaxies \citep[e.g.,][]{Hodge16,Gullberg19}, in conjunction with more compact dust continuum structures that have been interpreted as bars and rings \citep{Hodge19}.    Another, perhaps more speculative explanation  for this wide-spread (and relatively uniform) dust reddening in this (modestly-inclined, in contrast to  \citealt{Nelson23}) galaxy is a dusty wind, as is seen on similar scales in M\,82 \citep[e.g.,][]{Yamagishi12,Beirao15}, Mg{\sc ii} absorbers at higher redshifts \citep{Menard12}, and also theoretically expected \citep[e.g.,][]{Krumholz13}.   To illustrate the plausibility of this explanation, we assume a Milky Way gas-to-dust ratio, for which an $A_V$\,$\sim$\,3 corresponds to a hydrogen column density of $\sim$\,10$^{22}$\,cm$^{-2}$ \citep{Bouchet85}.   Integrating over the extent of 850.1  gives  a  gas mass of $\sim$\,10$^{10}$\,$M_\odot$.   If this gas (and associated dust) is being expelled in a wind with an outflow rate comparable to the current SFR, then this mass  represents just $\sim$\,10\,Myrs of activity, during which time the material would cover the extent of 850.1 if it was travelling at the escape velocity.   Thus it is possible that the relatively uniform red colors of 850.1 may be due to a foreground screen of dust expelled by a starburst-driven wind.

Turning now to 850.2, this has an asymmetric structure with a bright elongated region (sheared by  lensing) and a western extension.
The overall appearance is of a dusty, disk-like galaxy (Figure~2).  In contrast to the very red colors of the bulk of 850.1, both the high-surface brightness  region of 850.2 and the western extension  are  detected in F606W, corresponding to restframe $\sim$\,1100\AA, with neither region  detected in F435W, restframe $\sim$\,830\AA, indicating the presence of a strong Lyman break in the SED  (Figure~4).  However,  in addition to these UV-bright components there is also faint red emission to the north that shows up as a band of higher-$A_V$ in Figure~6.
This northern region includes a compact red component, seen in  F277W, F356W and F444W (Figures~2 \& 4), that appears as a  distinct entity in the source plane (Figure~2f).   The peak of the SMA 880-$\mu$m emission lies between this red clump and the UV-bright region, suggesting that this may be an interacting system with both obscured and unobscured components, and with the unobscured part corresponding to the higher-frequency peak  seen in the CO spectrum of this source.

The pixel-level SED fits for 850.2 give  $A_V$\,$\sim$\,0.3--1.0 over most of the UV-detected region but rising to $A_V\sim$\,3 at the red clump on the northern side.  This behaviour contrasts with that seen  in local galaxies, where extinction typically rises towards the
galaxy center \citep[e.g.,][]{Holwerda05,Keel23}.  From the stellar mass map the median stellar-mass density within the effective radius  in 850.2 was (0.10$\pm$0.01)\,$\times$\,10$^{10}$\,M$_\odot$\,kpc$^{-2}$ with a peak of 0.7\,$\times$\,10$^{10}$\,M$_\odot$\,kpc$^{-2}$.   Thus 850.2 has a more uniform distribution of stellar mass density and $A_V$ than  850.1.

%
%
\setcounter{figure}{6}
\begin{figure*}[ht!]
 
  \centerline{\includegraphics[height=2.3in]{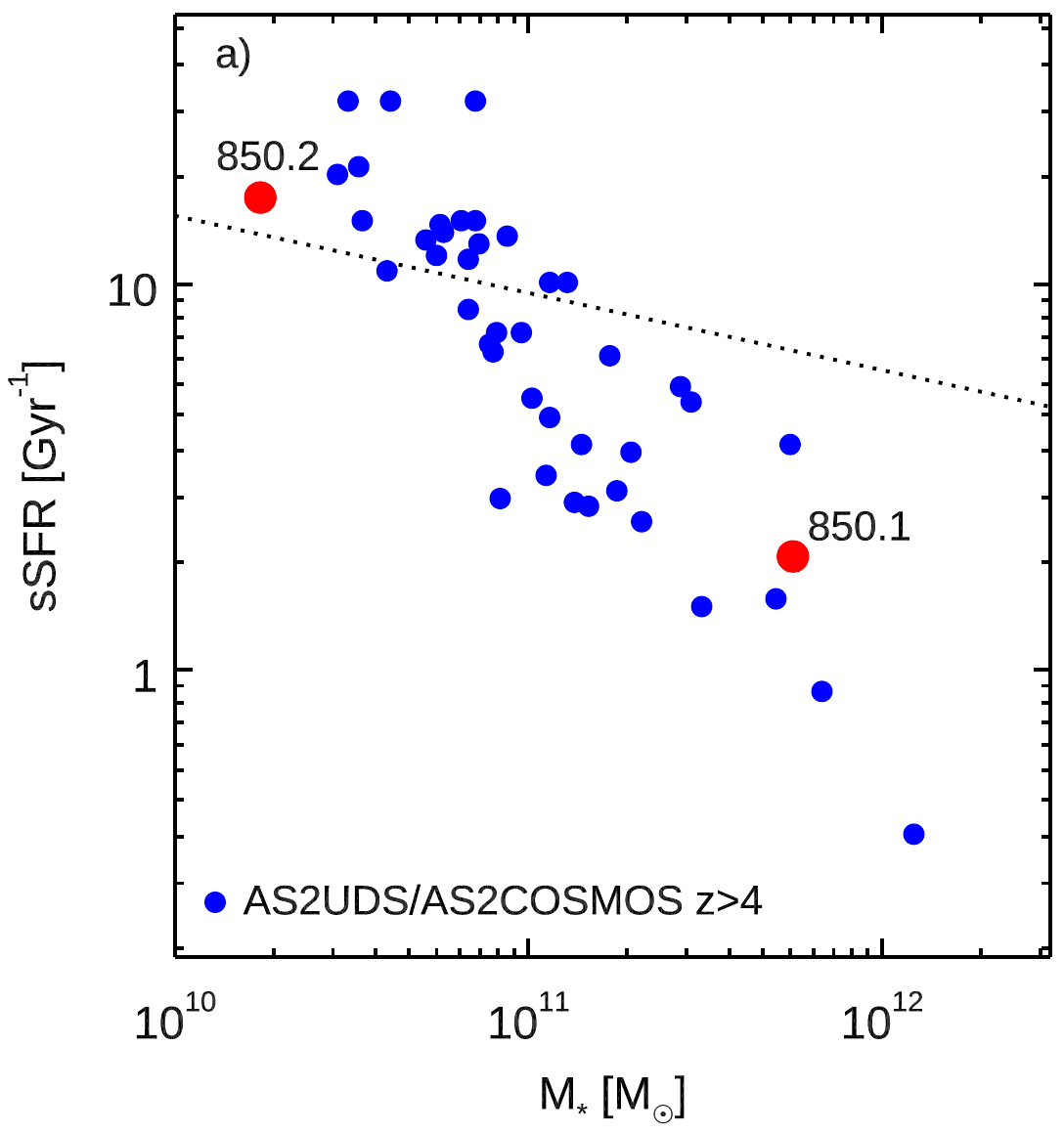} \hspace*{0.13in} \includegraphics[height=2.3in]{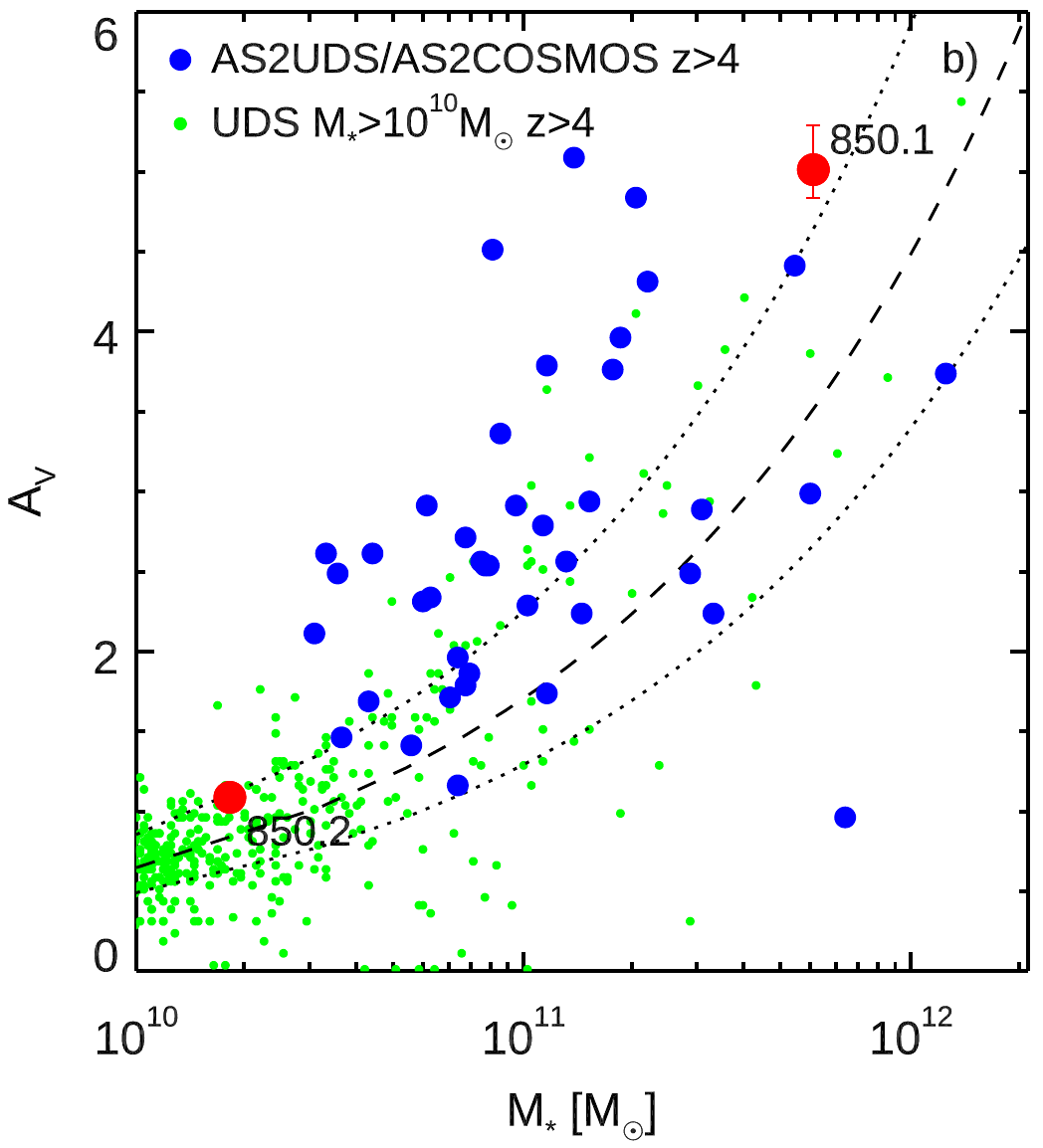}  \hspace*{0.13in} \includegraphics[height=2.3in]{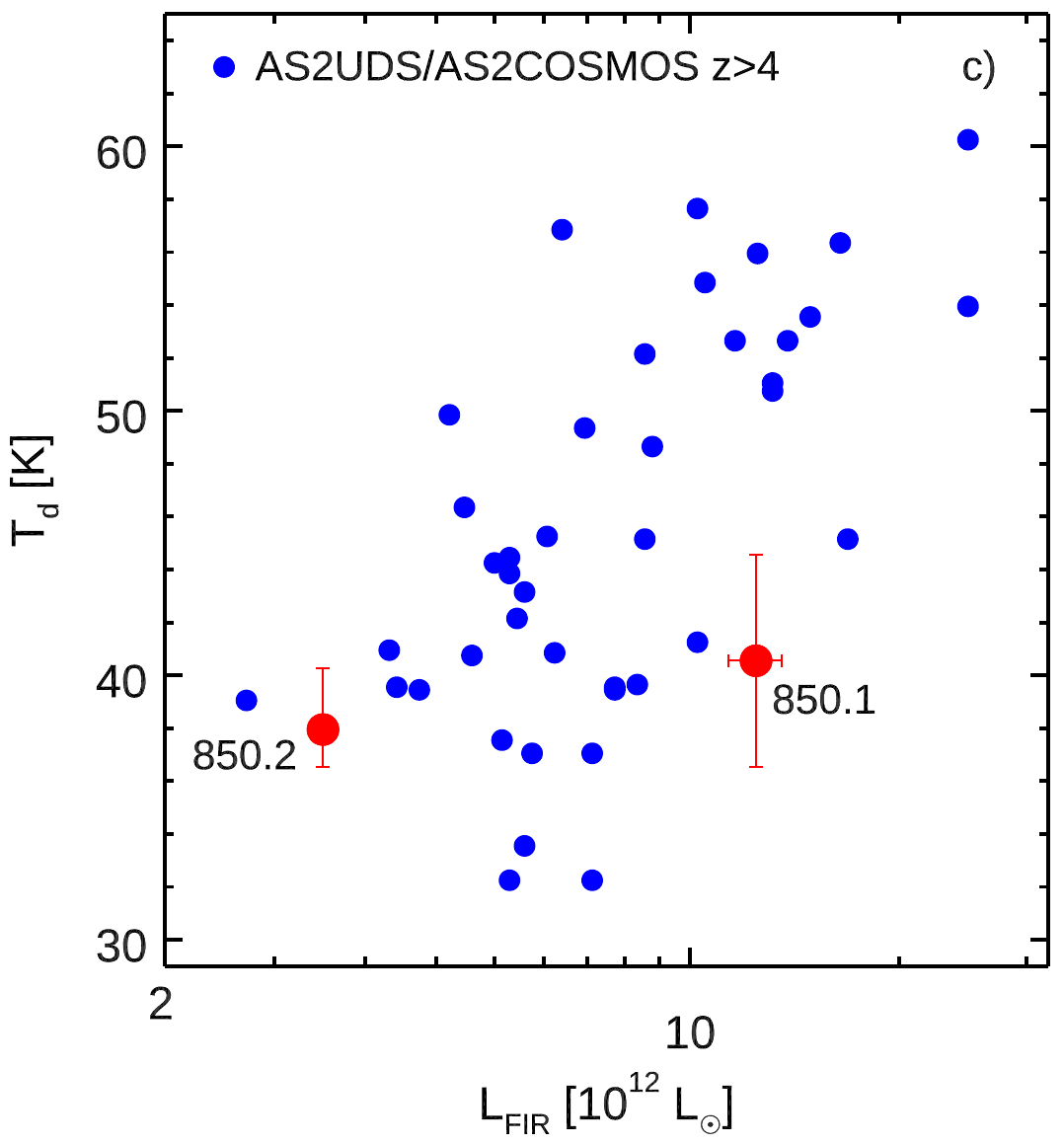}}
   \centerline{     \includegraphics[height=2.3in]{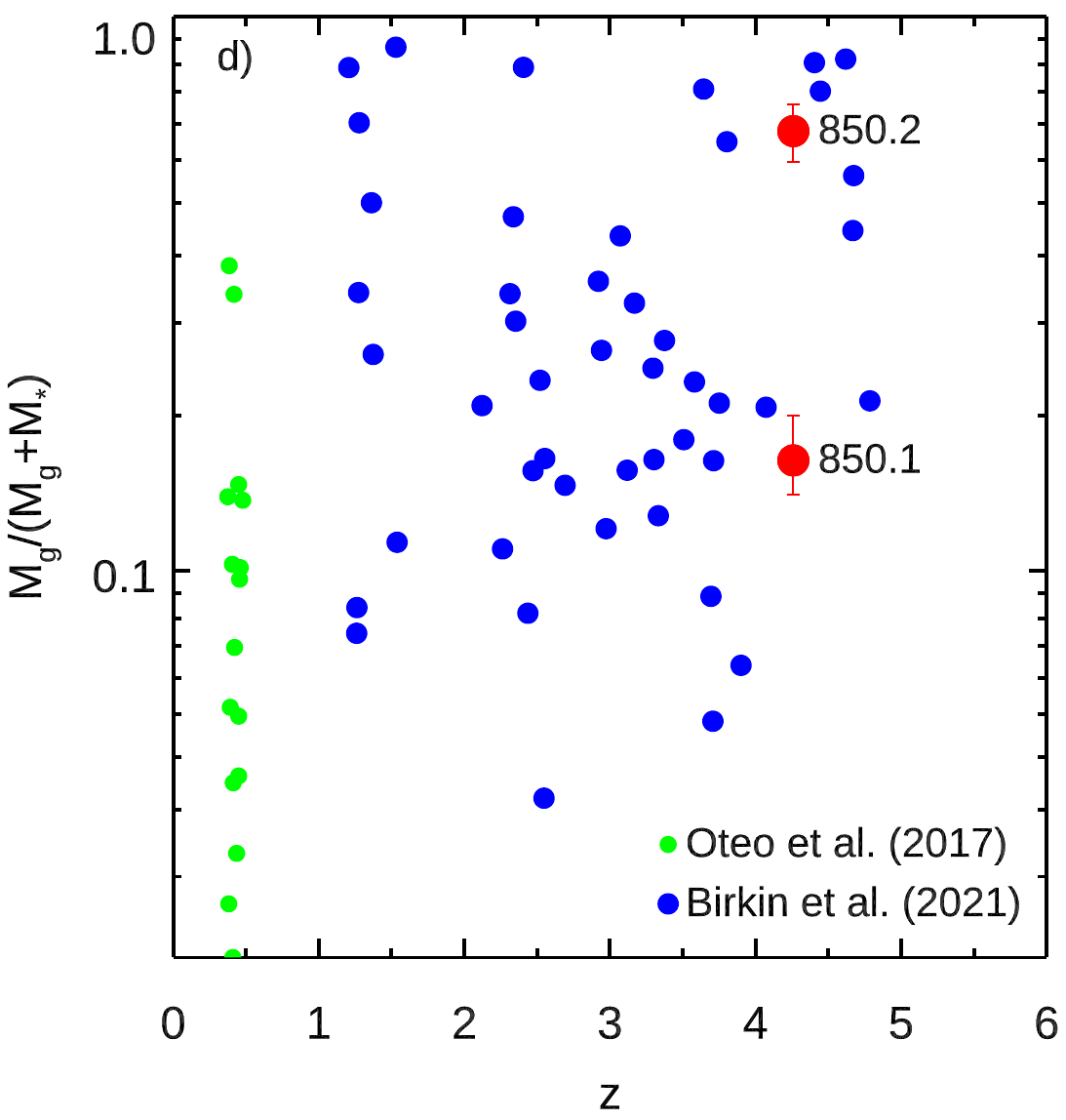} \includegraphics[height=2.3in]{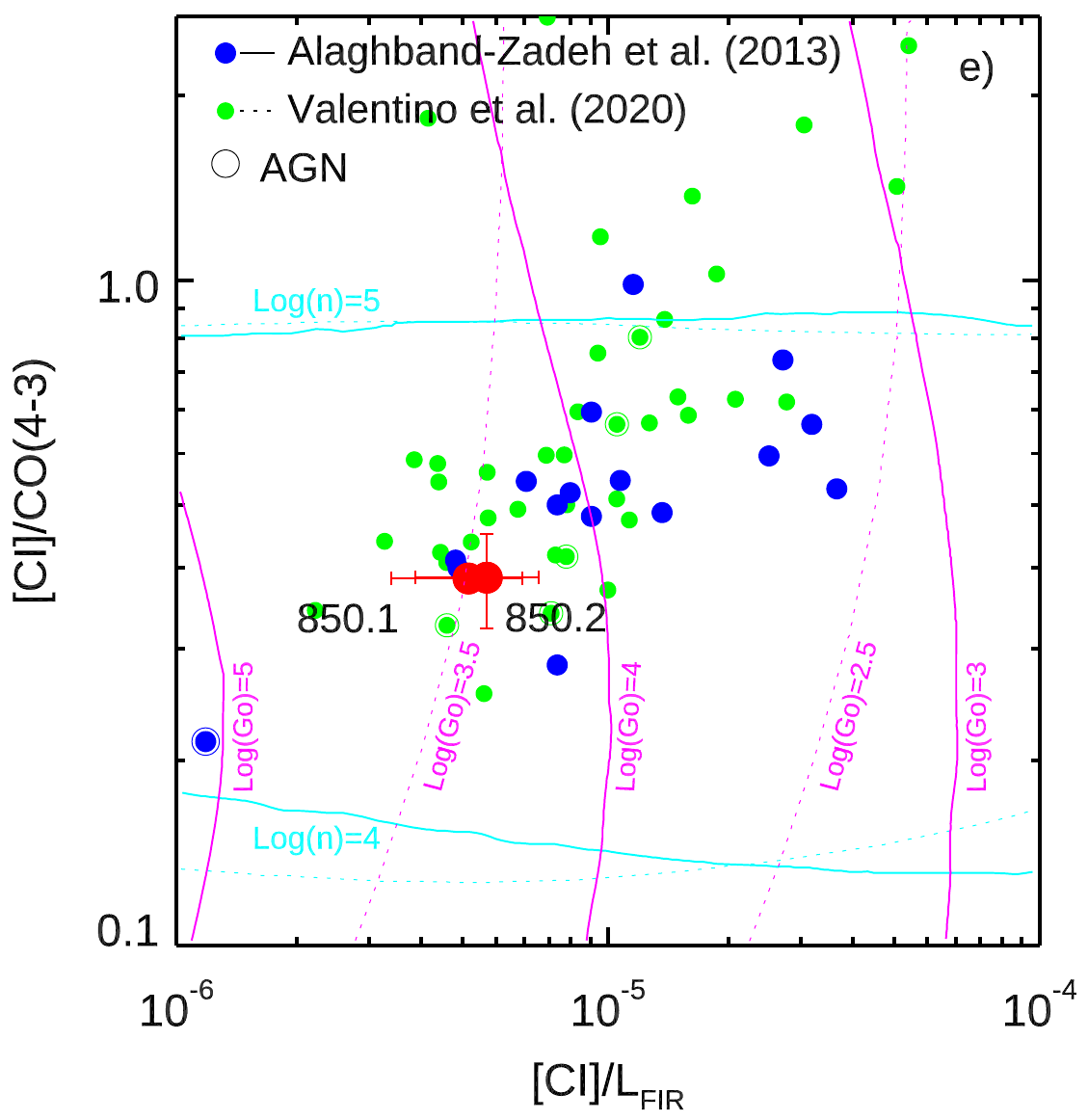} \includegraphics[height=2.3in]{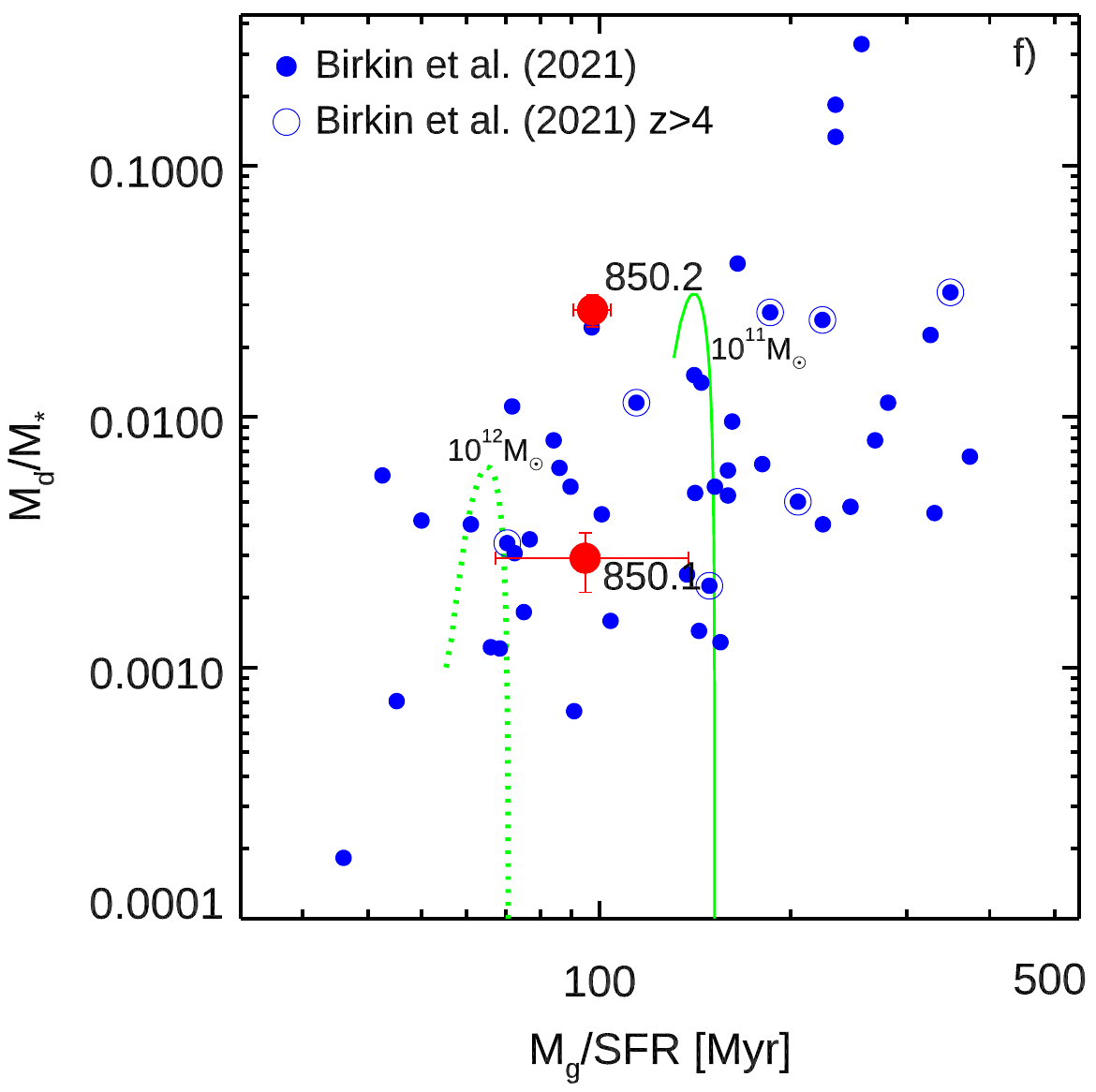}}

  
  \caption{\small The stellar and interstellar medium (ISM) properties of 850.1 and  850.2.  For comparison  several panels  show samples of 870-$\mu$m selected ALMA-identified galaxies from  the AS2UDS \citep{Stach19,Dudzeviciute20} and AS2COSMOS \citep{Simpson19,Ikarashi23} surveys and the CO properties of a sample of sub-millimeter-continuum selected galaxies from \cite{Birkin21}.  
  {\it a)} The trend of specific SFR with stellar mass (SFR/$M_{\ast}$--$M_{\ast}$) where the dotted line shows the starburst boundary at $z$\,$=$\,4 from \cite{Speagle14}.
  {\it b)} $A_V$ versus  $M_{\ast}$, where the dashed and dotted lines show the best fit relation and  1-$\sigma$ scatter for a  $K$-selected $z>$\,4 galaxies sample with $M_\ast$\,$>$\,10$^{10}$\,M$_\odot$ drawn from the UKIDSS UDS survey \citep{Lawrence07}  analysed by \cite{Dudzeviciute20}.
  {\it c)} The dust temperature versus far-infrared luminosity ($L_{\rm IR}$--$T_d$) relation.
  {\it d)}  CO-derived gas fraction, $M_{\rm g}/(M_{\rm g}+M_{\ast})$, as a function of redshift, including low-redshift SMG-analogs from \cite{Oteo17}.
  {\it e)} The ratio of the [C{\sc i}]/CO(4--3)  line luminosities versus the [C{\sc i}] to far-infrared luminosity ratio, which also include  sources and PDR model tracks from \cite{AZ13} and \cite{Valentino20}.  The literature sources are flagged if they are estimated to have $\geq$\,20\%  contributions  from an AGN to their far-infrared luminosities \citep{Valentino20}.
  {\it f)}  Dust-to-stellar mass ratio, $M_{\rm d}/M_{\ast}$, as a function of gas consumption timescale,  $M_{\rm g}/{\rm SFR}$, including two model tracks from \cite{Calura17} for halos with baryonic masses of 10$^{11}$\,M$_\odot$ and 10$^{12}$\,M$_\odot$ roughly matching 850.2 and 850.1, respectively.  The  tracks were rescaled from their assumed Larson IMF with $m_{\rm c}$\,$=$\,1.2 for comparison to the {\sc magphys} derived values.
}
\end{figure*}

Although 850.2 is very blue compared to 850.1,  the UV light traces only $\sim$\,10\% of the star formation  in this system. In addition, there is a slight excess of emission in the {Spitzer} 5.8 and 8\,$\mu$m filters above the best-fit model SED  in Figure~5 that may indicate a steeply rising SED for the dustier red northern region (but this is far from dominant in F444W).   Taken together with the strong far-infrared/sub-millimeter emission, indicating the presence of highly obscured star formation within the system and the velocity offsets seen between the optical and CO spectroscopy, this suggests that  there could be two spatially distinct components   within this source.  However, in contrast to B2 and 850.1 in 850.2 the  low-obscuration component  dominates from the optical out to at least 4.4\,$\mu$m, with  the fainter and more obscured component only  appearing at $>$\,2\,$\mu$m. 

The  {\sc galfit} analysis quantified the stellar continuum structure of the two sources.  The analysis of 850.1 in the F444W passband yielded a S\'ersic parameter  $n$\,$=$\,0.9\,$\pm$\,0.1 (with lower, but increasingly uncertain, values in bluer bands).  A fixed  $n=$\,1 was therefore adopted to simplify comparison of the derived sizes between the bands.   This showed  a constant effective radius with wavelength of $R^{\rm F444W}_{\rm e}$\,$\sim$\,3.8\,$\pm$\,0.4\,kpc after correcting for lensing \citep[see also][]{Chen22,Kamieneski23}.   The similar analysis of 850.2 yielded $n$\,$\sim$\,1.9\,$\pm$\,0.1 in F444W, with higher values in the bluer bands, although a combination of a de Vauncoleurs (S\'ersic $n=$\,4) and an $n$\,$=$\,1 S\'ersic did a better job of reproducing the asymmetric, extended emission in the source.  However, for comparison to 850.1,  $n$\,$=$\,1 was adopted, which gave sizes that modestly decrease in bluer bands, but with low significance, and  $R^{\rm F444W}_{\rm e}$\,$\sim$\,2.5\,$\pm$\,0.3\,kpc in F444W (also corrected for lensing).   These sizes are comparable to those derived in the F444W waveband for  sub-millimeter galaxies with flux densities of $S_{\rm 850\mu  m}$\,$\geq$\,1\,mJy  in the literature: $R^{\rm F444W}_{\rm e}$\,$=$\,3.9\,$\pm$\,0.8\,kpc \citep{Cheng22,Cheng23,Chen22}.   There was no evidence for a trend in restframe near-infrared size with either dust mass or redshift for the combined samples.

The dust continuum radii, $R^{\rm dust}_{\rm e}$, derived from the SMA observations (corrected for lensing magnification) were 2.8\,$\pm$\,0.7\,kpc for 850.1 and 1.9\,$\pm$\,1.0\,kpc for 850.2.  Compared to the F444W sizes, these  gave identical ratios of dust to F444W effective radii of 0.7\,$\pm$\,0.2 and 0.7\,$\pm$\,0.4 respectively, similar to previous estimates \citep[e.g.,][]{Lang19,Gullberg19,Chen22}.   These indicate the presence of compact dust-enshrouded regions in the dense cores of both galaxies, linked to the high $A_V$ regions  \citep{Simpson17}.

Finally, {\sc galfit} was applied to the stellar mass distributions from the resolved, pixel-level SED modelling to attempt to derive stellar-mass radii with a  S\'ersic profile with $n$\,$=$\,1.  However, this proved very unstable and so instead  simple numerical integration was used to derive the half-mass radii for the two sources:   $R^{\rm mass}_{\rm e}$\,$=$\,1.0\,$\pm$\,0.2\,kpc for 850.1 and 2.2\,$\pm$\,0.4\,kpc for 850.2, both corrected for lensing magnification.   While the stellar mass radius  determined for 850.2 was in  agreement with that measured at F444W, for 850.1 the stellar-mass size was around $\sim$\,4\,$\times$ smaller than seen at F444W.   This demonstrates that a significant fraction of the stellar mass in this massive system lies in a compact and very highly obscured nuclear region with  the ratios of dust to stellar mass sizes:  2.8\,$\pm$\,0.9 and 0.9\,$\pm$\,0.5 for 850.1 and 850.2, respectively.

\subsection{Comparisons to dust-mass-selected populations}

Figure~7  compares the derived properties of 850.1 and 850.2 with a representative sample of dust-mass-selected galaxies derived from ALMA interferometric identifications of $\sim$\,1,000 sources found in $\sim$\,3\,degrees$^2$ of wide-field 850-$\mu$m SCUBA-2 surveys in the UDS and COSMOS fields:   AS2UDS \citep{Stach19} and AS2COSMOS \citep{Simpson20}.   These samples have been analysed in an identical manner using {\sc magphys}, allowing us to perform a simple comparison of their derived properties \citep{Dudzeviciute20,Ikarashi23}.   After accounting for lensing magnification, Figure~7a shows that 850.1 and 850.2 lie at either end of the mass range seen in 850-$\mu$m-selected galaxies at $z$\,$>$\,4: 850.1 being one of the more massive galaxies found (from dust-selected samples or other methods) at these redshifts, while 850.2 is one of the least massive in dust-selected samples \citep[see also,][]{Pope23}. Looking at their specific SFR,  850.1 lies below and 850.2 lies just inside the claimed ``starburst'' regime at this redshift.  The strong linear trend in Figure~7a likely reflects the fact that these samples were selected primarily on their cold dust mass, the formation of which requires an equivalent characteristic stellar mass of $\sim$\,10$^{11}$\,M$_\odot$, with an
observed dispersion  of only $\sim$\,0.4\,dex.

Figure~7b shows the variation of $A_V$ with stellar mass for 850.1 and 850.2 compared with  the $\sim$\,40 ALMA-identified $z$\,$>$\,4 dusty star-forming galaxies selected from  SCUBA-2 mapping in the UDS and COSMOS fields as well as less active $K$-selected star-forming galaxies at $z$\,$>$\,4.   This figure demonstrates one of the more strikingly differences between 850.1 and 850.2: they  lie at diametrically opposite ends of the trend of $A_V$ with stellar mass for dusty galaxies at $z$\,$\sim$\,4.  Thus these two closely-associated sources span the full range in reddening seen in the high-redshift sub-millimeter population, underlining the diversity in the restframe UV-to-visible properties for dusty star-forming galaxies,   which was already present in  the population, even in the same environment, at  $z$\,$\gs$\,4.  The two sources broadly follow the trend seen in the $z$\,$>$\,4 $K$-selected galaxy population, which has the form  $\log_{10}(A_V)$\,$=$\,0.42\,$\log_{10}(M_\ast/10^{11})+$\,0.23 with a median absolute deviation of $\pm$\,0.32,  although this has a steeper slope than seen at $z$\,$\sim$\,0 (\citealt{Garn10}, but see also \citealt{GG23,Shapley22}).   

To test if there were other observed properties that would provide a better predictor of $A_V$ for this sample of $z$\,$>$\,4 SMGs,  
the correlation of various  properties ($M_\ast$, $M_{\rm d}$, $L_{\rm FIR}$, SFR, $T_{\rm d}$, Age, $\Sigma_{\rm d}$) with $A_V$ was analysed using maximal information-based non-parametric exploration ({\sc mine}, \citealt{Reshef11}).   This showed that the most significant correlation with $A_V$ for the SMG sample was indeed with stellar mass,  with a  0.3\% chance of random correlation using a jackknife test on the sample and  the MINE likelihoods were calibrated with a Monte Carlo simulation. There was only a minor reduction in scatter when  additional corrections were included for either $L_{\rm FIR}$ or $M_{\rm d}$  \citep{GG23}, and so it appeared that stellar mass was the most likely driver of the differences in dust reddening between 850.1 and 850.2 (although $L_{\rm FIR}$ or SFR did nearly as well).    

The final panel in Figure 7 showing purely SED properties is Figure~7c, which compares the dust temperature and far-infrared luminosities from the {\sc magphys} analysis of  850.1/850.2 to the wider dusty $z>$\,4 population in UDS and COSMOS, where $T_{\rm d}$ and $M_{\rm d}$ were derived in very similar ways with {\sc magphys}.    The two galaxies lie within the scatter of the $T_{\rm d}$--$L_{\rm FIR}$ trend at $z$\,$\gs$\,4, with 850.1 having a higher far-infrared luminosity and a marginally higher dust temperature (although with considerable uncertainties, even with the inclusion of the 450\,$\mu$m SCUBA-2 photometry).   The similarity in the dust temperatures
is due to the similar ratios of far-infrared luminosity to dust mass (or equivalently ISM mass, as 850.1 and 850.2 have very similar dust-to-gas ratios, see below)  for the two sources:  (8$\pm$4)\,$\times$\,10$^3$\,L$_\odot$\,M$^{-1}_\odot$ and (7$\pm$1)\,$\times$\,10$^3$\,L$_\odot$\,M$^{-1}_\odot$  for 850.1 and 850.2 respectively,  suggesting their far-infrared  luminosities  predominantly arise in similarly optically-thick starburst regions \citep{Scoville13}.

\subsection{Gas and dynamical masses}

The lines detected in the NOEMA 3-mm spectra of the sources, along with their dust continuum properties from SMA and SCUBA-2, provide a wealth of information about the properties of the galaxies including both their gas dynamics and the characteristics of their interstellar medium (ISM).   Looking first at their ISM properties, the CO line luminosity ratios for the two galaxies are very similar: $r_{54} =$\,0.54\,$\pm$\,0.01 and  [C{\sc i}]/CO(4--3)\,$=$\,0.29\,$\pm$\,0.01 for 850.1;  $r_{54} =$\,0.61\,$\pm$\,0.02  and  [C{\sc i}]/CO(4--3)\,$=$\,0.29\,$\pm$\,0.04 for 850.2.   These compare to $r_{54} =$\,0.69\,$\pm$\,0.06 and [C{\sc i}]/CO(4--3)\,$=$\,0.82\,$\pm$\,0.06 for the well-studied $z$\,$=$\,2.3 SMG the Cosmic Eyelash \citep{Swinbank10,Danielson11}, indicating that 850.1 and 850.2 have marginally less excited CO spectral line energy distributions (SLEDs) compared to the Eyelash (i.e.\  suggesting  a flattening/turnover of the SLED around $J_{\rm up}$\,$=$\,4--5).

Gas masses were estimated by converting the CO(4--3) line luminosities to CO(1--0) using the Eyelash ratio of $r_{41}=$\,0.50\,$\pm$\,0.04, noting that its CO SLED appears slightly more excited than  either 850.1/850.2 and hence this may underestimate the CO(1--0) line luminosity.   Adopting $\alpha_{\rm CO}$\,$=$\,1,  consistent with the available constraints  on dust-mass-selected samples and previous studies \citep{Birkin21},  gives cold gas masses, corrected for lensing,  $M_{\rm g}$\,$=$\,(1.05\,$\pm$\,0.03)\,$\times$\,10$^{11}$\,$\alpha_{\rm CO}$\,M$_\odot$ for 850.1 and $M_{\rm g}$\,$=$\,(0.29\,$\pm$\,0.02)\,$\times$\,10$^{11}$\,$\alpha_{\rm CO}$\,M$_\odot$ for 850.2 (Table~6).  

The NOEMA spectra also provide kinematic information about the motion of cool gas in the galaxies.
The Gaussian FWHM of the CO lines is the best measure of the kinematics of the gas in 850.1.  For 850.2, both CO transitions show evidence for double-peaked lines and so  the mean velocity difference between the peaks, 330\,$\pm$\,70\,km\,s$^{-1}$, was used as  the measure of  the projected circular velocity.  Inclination was estimated from the apparent axial ratio of the source measured in the F444W band and corrected for lensing shear (Table~2).   To calculate the dynamical masses   a fixed radius of 10\,kpc was chosen as this should be sufficient to contain the majority of the stellar and cold-gas mass in these systems.\footnote{When estimating the mass of the cold H$_2$ reservoir using a CO(1--0)  luminosity extrapolated from  moderate-$J$ CO lines using  a galaxy-integrated conversion factor, $r_{ij}$,  it is critical to appreciate that this does not mean that the extrapolated H$_2$ (or CO(1--0)) reservoir has the same physical extent as the moderate-$J$ CO emission.   Gradients in the excitation or optical depth will lead to different physical extents for the different tracers \citep{Weiss05}.} Using $r$\,$=$\,10\,kpc and $M_{\rm dyn}(<r) = r \times (V_{\rm c} / \sin(i))^2 / G$ gives estimated dynamical masses of 3.8$^{+5.1}_{-1.5}\times$\,10$^{11}$\,M$_\odot$ and  0.8$^{+0.3}_{-0.3}\times$\,10$^{11}$\,M$_\odot$ for 850.1 and 850.2, respectively.   These compare to the total baryonic masses of 6.6$^{+1.3}_{-0.5}\times$\,10$^{11}$\,M$_\odot$ and 0.50$^{+0.03}_{-0.03}\times$\,10$^{11}$\,M$_\odot$, confirming that the  baryonic masses are broadly consistent with the galaxy kinematics derived from the CO line width, given the uncertainties on the inclination corrections.   If the calculation had instead used  a Milky-Way-like $\alpha_{\rm CO}$\,$=$\,4.3 to estimate the cold gas mass, then the  baryonic masses  would rise to  10.0$^{+1.3}_{-0.5}\times$\,10$^{11}$\,M$_\odot$ and 1.4$^{+0.7}_{-0.7}\times$\,10$^{11}$\,M$_\odot$, which  both exceed the dynamical estimates, although they are still formally in agreement due to the large uncertainties on the latter from the inclination corrections and adopted sizes.   

Using the estimated gas masses assuming $\alpha_{\rm CO}$\,$=$\,1, Figure~7d illustrates the variation in gas fraction of  galaxy populations as a function of redshift for dust-mass-selected samples. The gas fraction of 0.16\,$\pm$\,0.03 for 850.1 is at the low end of the distribution for dust-selected systems at  $z$\,$\sim$\,4, while the estimate for 850.2 of 0.62\,$\pm$\,0.07 is at the upper end  and  so more consistent with the rough trend for higher gas fractions at higher redshifts.

\subsection{ISM properties}

Both 850.1 and 850.2  show  fairly lower [C{\sc i}] luminosities relative to either their CO(4--3) or far-infrared luminosities compared to many of the high-redshift dust-mass selected galaxies or other [C{\sc i}]-detected star-forming galaxies shown in Figure~7e.  
Following \citet{AZ13},  a simple PDR model can be  used assess the physical origin of these differences.    The ratio  [C{\sc i}]/CO(4--3) traces the characteristic density of the ISM and [C{\sc i}]/$L_{\rm FIR}$ indicates the strength of the radiation field, although the absolute values  are  sensitive to the chosen model, compare the tracks from  \citet{AZ13} and \citet{Valentino20} in Figure~7e.  The \citet{Kaufman99} PDR model grids indicate well-constrained   ISM densities for 850.1 and 850.2, $\log_{10}(n)$\,$=$\,4.5\,$\pm$\,0.1\,cm$^{-3}$, and  radiation fields of $\log_{10}(G_0)$\,$\sim$\,3.5--4.3 (in Habing units), depending upon the adopted tracks \citep[see the discussion in][]{Valentino20}.  These ISM properties are within the ranges previously estimated for sub-millimeter galaxies and local ULIRGs \cite[see][]{AZ13,Valentino20}.   

Assuming a simple  uniform density sphere  for the geometry of the ISM in these sources, then the ISM densities and gas masses indicate  a radius of $\sim$\,0.4\,kpc  for 850.1 and $\sim$\,0.25\,kpc  for  850.2  \citep[see][]{Yan16}.   Adopting a more plausible thin disk geometry would result in a larger radius  by a factor of 1.9--2.5\,$\times$ for disks with thickness of $\sim$\,0.1--0.05\,$\times$ their diameter.   This would correspond to diameters of $\sim$\,1.8\,kpc and $\sim$\,1.1\,kpc, a factor of $\sim$\,1.6\,$\times$ smaller than the measured dust continuum sizes in both sources (Table~6).    However, a more realistic model would involve a multi-phase ISM, with much of the cool gas in a component that is potentially distinct from the star-forming material, thus arguing for using a lower gas mass (relevant for the star-forming phase) in these calculations and hence a  smaller volume. Even without this extra complication, it is clear that the dense ISM traced by the moderate-$J$  CO lines in these galaxies is  limited to a relatively small volume, either  distributed as smaller clouds spread throughout the disks or more likely mostly restricted to a compact region in the galaxy centers.

A similar calculation can be undertaken using the ISM radiation fields.  For solar metallicities, a Salpeter IMF, and constant-SFR bursts with ages of 50--500\,Myr,  $\sim$\,10\% of the bolometric luminosity arises in ionizing radiation with 6--13.6\,eV.  Assuming a centrally illuminated spherical shell geometry then  the estimated radiation field from the PDR models implies sheel radii   $r$\,$\sim$\,1.1--2.9\,kpc for 850.1 and $r$\,$\sim$\,0.6--1.4\,kpc for 850.2.   These sizes were smaller than those measured at F444W  (Table~6) but are closer to those derived for the total stellar mass (and dust continuum) in \S3.2 as well as by previous dust continuum studies \citep{Simpson15,Gullberg19}, although these assume different geometries.

\subsection{Gas masses and fractions}

Figure~7f  shows the variation of dust-to-stellar mass ratio with the gas consumption timescale at the current SFR for the two galaxies.   The trend in Figure~7f reflects the fact that galaxy-integrated dust mass provides a crude proxy for gas mass, hence $M_{\rm d}\propto M_{\rm g}$, with similar dust-to-gas ratios of  69\,$\pm$\,16 and 61\,$\pm$\,8   for 850.1 and 850.2, slightly lower than the median of 120\,$\pm$\,50 for the $z$\,$>$\,4 subset from \citet{Birkin21}, with  60\,$\pm$\,37 for their whole sample.   The gas consumption timescales of both galaxies are $\sim$\,100\,Myr, so they will consume their current gas reservoirs by $z$\,$\sim$\,4, unless further material is accreted.   While the dust-to-stellar mass ratio for 850.2 is  an order of magnitude higher than for 850.1, at $M_{\rm d}/M_\ast$\,$=$\,0.026\,$\pm$\,0.004 compared to 0.0027\,$\pm$\,0.0009, reflecting the very different stellar masses of the two galaxies.     To interpret this difference  Figure~7f shows two  tracks for proto-spheroid galaxy models (top-hat collapse) from \citet{Calura17}.  These show the variation of dust-to-stellar mass ratio with gas consumption timescale for halos with baryonic masses of 10$^{12}$\,M$_\odot$ and 10$^{11}$\,M$_\odot$, which  roughly bracket the estimates for 850.1 and 850.2.  
These evolutionary tracks start at high $M_{\rm g}/{\rm SFR}$ and initially show rapidly increasing  $M_{\rm d}/M_\ast$ until a peak at $\sim$\,150--200\,Myrs before declining, the tracks then terminate at the point that an outflow is expected to remove the remainder of the cold gas in the system.  The higher mass model peaks at a lower dust-to-stellar mass ratio owing to the shorter star-formation timescale leading to increased dust destruction, before the galactic wind terminates all star formation.   These theoretical models demonstrate that the observed dust-to-stellar ratios for 850.1 and 850.2  can be achieved by models  at typical ages of $\sim$\,100--300\,Myrs, with their relative dust-to-stellar mass ratios being consistent with 850.1 being older than 850.2.

The ISM properties of 850.1 and 850.2 are surprisingly similar in terms of dust temperature, ionization field, and density, given the very different stellar masses and restframe UV--visible obscuration of the stellar populations in the two galaxies.  This similarity has its basis in the comparable ratios of dust mass to far-infrared luminosity (or star-formation rate) for the two systems, which indicates similar relative amounts of ISM mass available to absorb and then reradiate  the luminosity arising from the on going star formation \citep{Scoville13}.    In contrast, the dust-to-stellar mass ratios of the two galaxies are very different, reflecting their different evolutionary stages.    850.1 is a much more massive system and is relatively gas poor, suggesting that it may be in the process of quenching its activity.  While 850.2 is a  young and gas-rich system that is rapidly growing its stellar and dust mass.

%
%
\setcounter{figure}{7}
\begin{figure*}[ht!]
 
  \centerline{\includegraphics[height=3.0in]{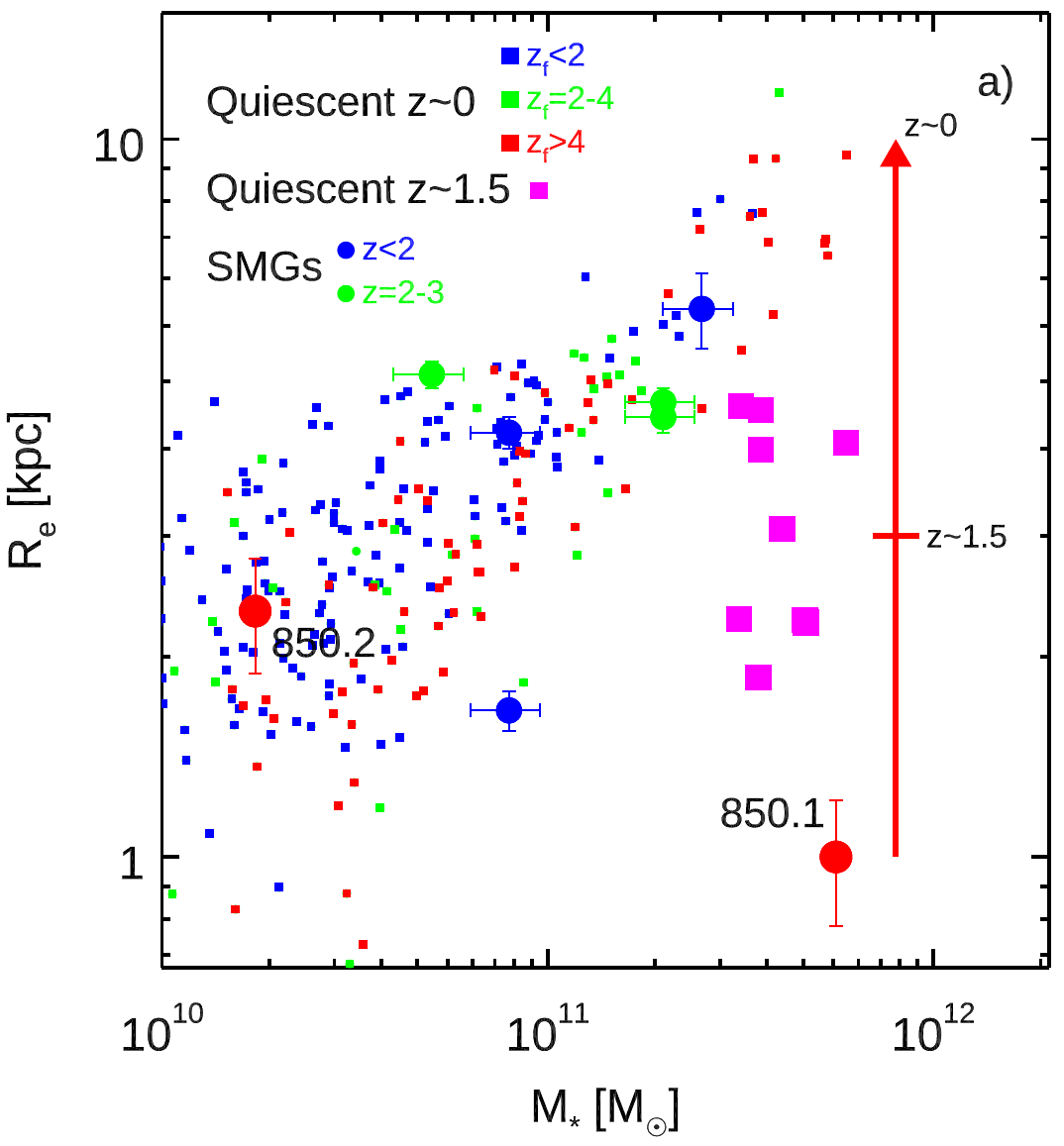} \hspace*{0.2in} \includegraphics[height=3.0in]{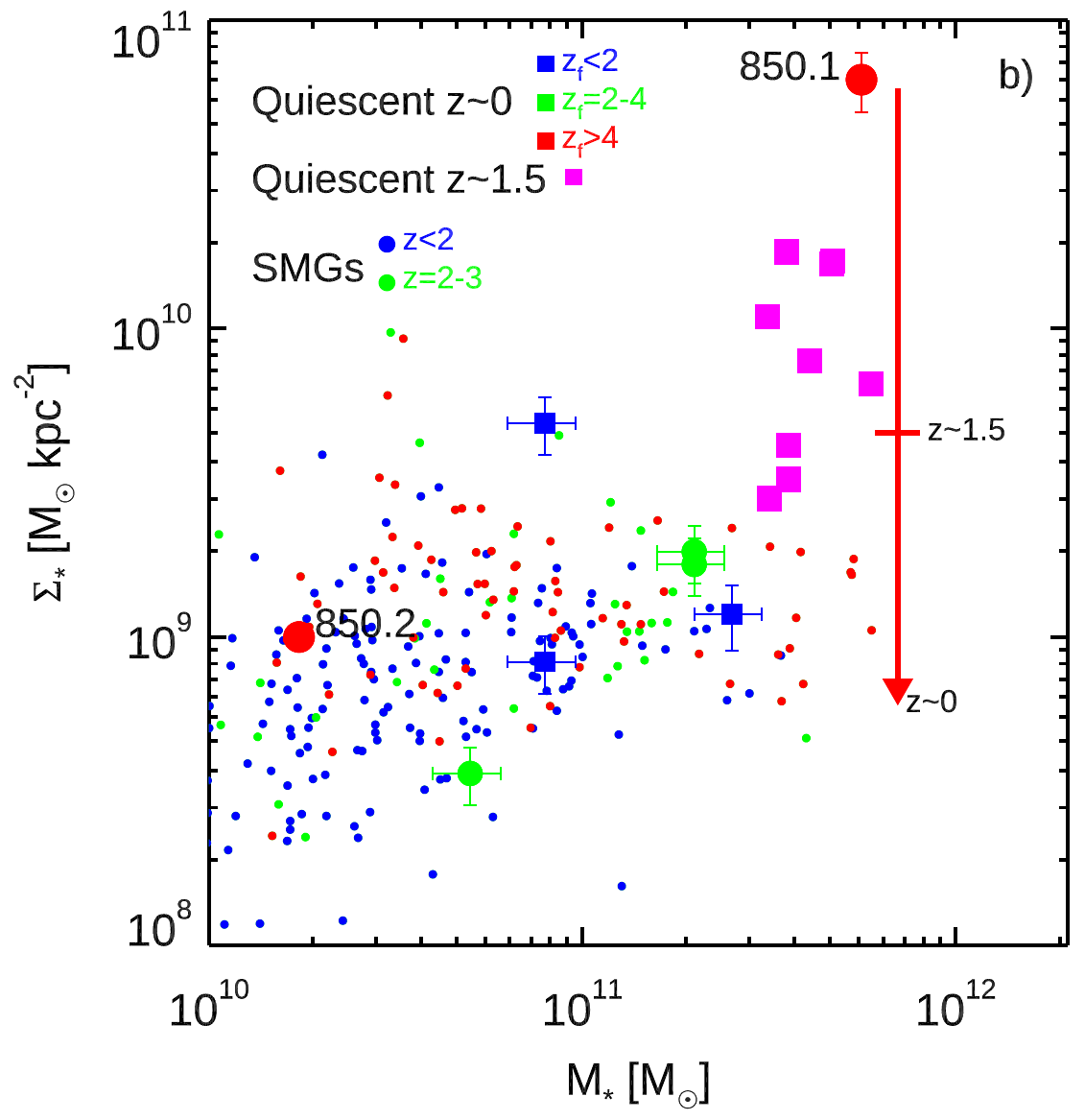}}
  
  \caption{\small 
    Comparisons between the {\it a)} stellar-mass effective radii and {\it b)}  stellar-mass surface densities of 850.1 and 850.2  with $z$\,$\sim$\,0 quiescent early-type galaxies from the ATLAS-3D survey \citep{Cappellari11,Cappellari13}, color-coded by their formation redshift as determined
    from their SSP ages in \citet{McDermid15}.  In addition, a sample of quiescent galaxies at $z$\,$\sim$\,1.5 with F160W-derived sizes and high stellar masses from \citet{Longhetti07} are shown.
    Also plotted are the F444W-derived sizes and stellar densities for $z$\,$\sim$\,1--3  SMGs from \citet{Chen22}.   The red arrow on both panels shows the effect of the expected (empirically and theoretically) size evolution from $z$\,$=$\,4.26 to $z$\,$\sim$\,0 and the tick indicates the evolution to $z$\,$=$\,1.5 for a massive galaxy such as 850.1 \citep{Cimatti12,Oser12}.  This demonstrates that 850.1 will likely evolve to have a stellar size and surface mass density similar to those of massive, compact quiescent galaxies at $z$\,$\sim$\,1.5, and then go on to match
    the properties of the most massive, and oldest, early-type galaxies at the present day.  
}
\end{figure*}

\subsection{Evolutionary links}

850.1 is one of the most massive galaxies known at $z$\,$>$\,4 in terms of stellar or total baryonic mass at $\sim$\,10$^{11.8}$\,M$_\odot$ \citep[e.g.,][]{Caputi11,Stefanon22,Marsan22},   with a space density of $\sim$\,3\,$\times$\,10$^{-7}$\,Mpc$^{-3}$ estimated from the surface density of comparably bright 850-$\mu$m sources
and assuming a redshift range of $z$\,$=$\,3--6 for the population. This is consistent with expected limits for such massive systems from $\Lambda$CDM \citep[$<$\,5\,$\times$\,10$^{-6}$\,Mpc$^{-3}$,][]{Behroozi18}.  Unlike most previous studies of high-redshift massive galaxies, the  high-resolution {JWST} imaging used here tightly constrains the restframe optical/infrared SED, and also excludes significant contamination from an obscured AGN in our estimates of the stellar mass.  Likewise, the  sub-millimeter imaging from SMA provided both the identification of this massive (but optically undetected) high-redshift galaxy and also constrained the influence of dust on 850.1's SED necessary to derive a robust stellar mass \citep[see also][]{Ikarashi22}.

This analysis also benefited from the high-spatial resolution and long-wavelength coverage of {JWST} to go beyond what has been previously possible in resolved studies of the stellar content of dusty galaxies at high redshifts \citep[e.g.,][]{Lang19}.    
The resolved stellar-mass maps from the pixel-level SED fitting indicated   stellar mass surface density  estimates within the central regions of  850.1 and 850.2  that are comparable   to the stellar density of early-type galaxies at the present day, long believed to be the potential descendants of these systems.   Moreover, the small stellar-mass  sizes for the two sources, 1--2\,kpc, are  comparable to those claimed for the most compact, massive ``quiescent/evolved'' galaxies at $z$\,$\sim$\,1--3,  $\sim$\,1--5\,kpc \citep[e.g.,][]{Waddington02,Trujillo06,Longhetti07,Buitrago08,Damjanov09,Mowla19,Lustig21}.  The stellar-mass sizes for 850.1 and 850.2 are also consistent with the expectations from  hydrodynamic simulations \citep[e.g.,][]{Wellons15}, which indicate sizes of $\sim$\,1\,kpc for systems with masses of $\sim$\,10$^{11}$\,M$_\odot$.

Figure~8 shows the intrinsic sizes and stellar-mass densities for 850.1 and 850.2 compared to their lensing-corrected stellar masses.
These are plotted alongside estimates from integrated SED fitting and {JWST} NIRCam F444W sizes for $z$\,$=$\,1--3 SMGs with $S_{\rm 850\mu m}$\,$\sim$\,2--4\,mJy from \citet{Chen22}, although  the F444W-based sizes may overestimate the true stellar-mass sizes for these dust-obscured systems.   The figure also includes a sample of high mass spectroscopically-classified quiescent galaxies at somewhat lower redshifts, $z$\,$\sim$\,1.2--1.7, with estimated stellar masses above 3\,$\times$\,10$^{11}$\,M$_\odot$, and sizes derived from {HST} F160W imaging \citep{Longhetti07}.    These comparisons show that 850.2 has a stellar-mass size and stellar-mass surface density consistent with those seen in early-type galaxies at the present day, as do the other lower redshift sub-millimeter galaxies with comparable 850-$\mu$m flux densities from \citet{Chen22}.  However, 850.1 is an outlier in both panels, with a size that is a factor of $\sim$\,10 smaller than comparably massive early-type galaxies at  $z$\,$\sim$\,0.   Nevertheless, the peak stellar-mass surface density  for 850.1, 1.4\,$\times$\,10$^{11}$\,M$_\odot$\,kpc$^{-2}$, is  similar to the characteristic maximum stellar-mass density estimated in the central regions local early-type galaxies  \citep[e.g.,][]{Hopkins09}, suggesting a potential link.    

Empirical estimates of the size evolution of massive, early-type galaxies out to $z$\,$\sim$\,3 indicate that these increase at lower redshifts as $(1+z)^{-1.39\pm0.13}$  \citep[e.g.,][]{Cimatti12,Trujillo07,Holwerda20}.  This  implies a  size  increase of $\sim$\,4\,$\times$ for 850.1 from $z$\,$=$\,4.26 to $z$\,$=$\,1.5 and $\sim$\,10\,$\times$  to the present day.    The  evolution in the size of massive galaxies is driven by minor mergers (such as the potential accretion of B2 and 850.2 onto 850.1)  that increase the total stellar mass of the system modestly but boost the effective radius considerably \citep{Naab09,Hopkins09,Oser12}.    The size evolution for lower mass galaxies is less extreme \citep{Genel18} and may be moot for 850.2 if it subsequently merges with 850.1.    As Figure~8 shows, if 850.1 undergoes  size evolution as suggested by simulations and observations at $z$\,$<$\,3, then
at $z$\,$\sim$\,1.5 
it will end up in the region of the $R_{\rm e}$--$M_\ast$ and $\Sigma_{\ast}$--$M_\ast$ planes that are populated by the high-mass quiescent galaxies from \citet{Longhetti07}, before evolving to match the properties of the most massive early-type galaxies at $z$\,$\sim$\,0.  These massive galaxies at $z$\,$\sim$\,0 are also those that appear to have formed the bulk of their stars at  $z$\,$>$\,4, again in good agreement with the observations of 850.1.  Overall, this appears to be strong circumstantial evidence  that 850.1 represents the formation phase of an ultra-massive early-type galaxy, which will subsequently evolve through the massive, passive galaxy population found at $z$\,$\sim$\,1--3 \citep{Trujillo06,Buitrago08,Damjanov09}.   

To summarise, 850.1, 850.2, and B2 reside in a probable group-like halo  serendipitously discovered behind the core of A\,1489. The two sub-millimeter sources represent  extremes of the dust-mass-selected population at $z$\,$>$\,4, with 850.1 being one of the most massive galaxies known at these redshifts.   The gas reservoirs in the two galaxies contribute very different fractions of their total baryonic content at this  early phase of the Universe (age of 1.4\,Gyr): gas fractions of 16\% and 62\%, suggesting different evolutionary states.  These differences were also reflected in the  different characteristic ages of their best-fit SEDs and dust/stellar ratios, although both were predicted to consume their current gas reservoirs by $z$\,$\sim$\,4.     The properties of these gas reservoirs are surprisingly similar, both indicating high densities, strong radiation fields, and comparable dust temperatures. Thus 850.1  appears to represent a massive, highly obscured, but relatively gas-poor (``quenching''?) galaxy with possibly two closaree neighbors that are both much less massive and more gas-rich, as well as younger.

If these  galaxies do reside in a single group-sized $\sim$\,10$^{13}$\,M$_\odot$ halo, as was claimed for sub-millimeter galaxies \citep[e.g.,][]{Stach21}, then 850.1 is very likely to be the central galaxy with B2 and 850.2 as satellites that will eventually be accreted through minor mergers.  These mergers will  increase the stellar-mass radius of 850.1, which at $\sim$\,1\,kpc is more compact than typical  examples of ``quiescent/passive'' galaxies at  $z$\,$\sim$\,2--3, into which it will likely evolve,  so that the stellar-mass surface density of this galaxy will eventually decline by two orders of magnitude to match those of the most massive early-type galaxies at the present day.   Such minor mergers may also be the trigger for the formation of the apparent stellar bar visible in the central regions of 850.1 (Figure~4).   Similar features have been reported from {JWST} imaging of galaxies at $z$\,$\sim$\,1--3  \citep[e.g.,][]{Guo23}, and likely indicate dynamical disturbance of disks.  Disks can form bars relatively quickly if they were cold and gas rich and the central regions of the halos were baryon dominated \citep[e.g.,][]{BH23}.     These non-axisymmetric structures then serve to remove angular momentum from gas in the disk and efficiently funnel it into the galaxy center potentially fueling the  central dusty starburst that alerted us to the existence of these systems in the first place.

\section{Conclusions}

This pair of sub-millimeter  galaxies  demonstrates the remarkable diversity of galaxies hosting massive dust and gas reservoirs at high redshifts.  These two galaxies are associated in a single structure at $z$\,$=$\,4.26, but their observed and intrinsic properties  span the full range known for this population at $z$\,$>$\,4.  The main conclusions of this work are:

1. 850.1 is an example of a rare population of very luminous sub-millimeter sources with  an intrinsic $S_{\rm 850\mu m}$\,$=$\,13.5\,$\pm$\,0.4\,mJy, corresponding to a population with a surface density of $\sim$\,10\,$\pm$\,5 degree$^{-2}$,  while 850.2 with $S_{\rm 850\mu m}$\,$=$\,3.8\,$\pm$\,0.3\,mJy  comes from the much more numerous population around the knee in the 850-$\mu$m counts,  having a surface density $\sim$\,50\,$\times$ higher.    850.1 is one of the most massive galaxies known at $z$\,$>$\,4 in terms of stellar or total baryonic mass at $\sim$\,10$^{11.8}$\,M$_\odot$  with an estimated formation redshift  $z_{\rm form}$\,$\sim$\,6 and a space density of $\sim$\,3\,$\times$\,10$^{-7}$\,Mpc$^{-3}$, consistent with expected limits from $\Lambda$CDM.  The  high-resolution {JWST} imaging can exclude  significant contamination from  obscured AGNs  in these mass estimates.

2. Without lensing 850.1  would have $H$\,$\sim $\,29 and $K$\,$\sim$\,27. Neither  {HST} nor ground-based $K$-band surveys would have been able to detect this class of source, let alone less-massive or higher-redshift examples.  This reflects the high dust reddening, $A_V$\,$\sim $\,3--6, over the full $\sim$\,7-kpc  detected extent of the disk in 850.1 (with the exception of the   UV-bright companion  B2, if it is actually part of 850.1).   This wide spread and relatively uniform dust reddening in a modestly inclined galaxy may be a result of either efficient  mixing of dust within the galaxy disk, or more speculatively a dusty wind emanating from the $\sim$\,1\,kpc stellar core.

3. The high-quality multi-band {JWST}/{HST} images available for this field also allows  a pixel-level SED analysis that provides unique information on the internal structures of the galaxies.  This resolved SED analysis shows good correspondence with the integrated SED fits, but only because it showed the need to exclude the foreground/companion  UV-bright sources, B1 and B2, from the integrated photometry fits for 850.1.   If these sources had been included in a single SED fit then that results in a significantly worse fit and biased derived properties.    Similar issues may affect  integrated photometry modelling of systems comprising a mix of low- and high-obscuration regions or components.  This effect may also explain previous reports of mid-infrared power-law excesses in $\sim$\,10\% of dusty galaxies that were previously attributed to buried AGN.  Such resolved analyses  benefit considerably from the inclusion of resolved long-wavelength constraints on the pixel-level SEDs, as provided by  sensitive sub-millimeter interferometry.

4.  850.1 and 850.2 are just $\sim$\,100\,kpc ($\sim$\,15$''$)  apart in the source plane and offset in velocity by only
$\sim$\,30\,km\,s$^{-1}$, indicating that they likely reside in a single structure, a small group at $z$\,$\sim$\,4.26.  Without the lensing effect of the foreground cluster they would  appear as a single, blended, bright ($S_{\rm 850\mu m}$\,$\sim$\,17\,mJy) source in a single-dish sub-millimeter survey.  There are a number of similar lensed groups at $z$\,$\sim$\,2--4, comprising one or more sub-millimeter galaxies and other companions,   reported in the literature.  The  frequency with which these systems are being found may reflect  the correspondence between the spatial scale of group-sized halos and  the size of high-magnification region of massive clusters at $z$\,$\ls$\,1, see also \citet{Frye23}.   This  adds  evidence that at least a subset of the high-redshift sub-millimeter population resides in group-scale halos. These are expected to be the most active environments for the accretion of gas and companion galaxies  from the surrounding intergalactic medium,  potentially providing both the fuel and the trigger needed to power their active star formation,

5. The properties of 850.1 suggest it is the central galaxy of this group.   Its stellar size and central stellar-mass surface density will match those measured for massive quiescent galaxies at $z$\,$\sim$\,1.5 and $z$\,$\sim$\,0 if it follows the expected size evolution due to the accretion, through minor mergers, of satellite galaxies such as 850.2 (which itself may be an interacting system) and potentially B2.

6.  The exquisite {JWST} imaging of the A\,1489 field, combined with the natural magnification of the foreground cluster lens, provides a striking view of the stellar  structure of these two strongly star-forming galaxies.   The structure of 850.1 is particularly noteworthy, with features that appear to correspond to arms and a central bar in a very massive galaxy at $z$\,$=$\,4.26.   It is tempting to identify these features as a response to dynamical perturbations  from the group environment, including the presence of 850.2  as a close companion.   Bars with $\sim$\,1-kpc radii have been reported from high-resolution dust continuum imaging of high-redshift sub-millimeter galaxies by \citet{Hodge19}, as well as from statistical analyses of the dust continuum shapes of the sub-millimeter galaxy population \citep{Gullberg19}.  These features have similar sizes to the dust continuum peak and the massive stellar core in 850.1,  suggesting that this galaxy is experiencing a bar-driven inflow that is fueling a central starburst in its baryon-dominated core.

\section*{Acknowledgments}

We thank Harold Pena for his help with the JCMT observations and reduction and Rob Ivison and Anna Puglisi for valuable comments and discussions.

All of the Durham co-authors acknowledge STFC through grant number ST/T000244/1 and ST/X001075/1.
AZ acknowledges support by the Ministry of Science \& Technology, Israel, and by Grant No.\ 2020750 from the United States-Israel Binational Science Foundation (BSF) and Grant No.\ 2109066 from the United States National Science Foundation (NSF).
RAW, SHC, and RAJ acknowledge support from NASA {JWST} Interdisciplinary
Scientist grants NAG5-12460, NNX14AN10G and 80NSSC18K0200 from GSFC. Work by
CJC acknowledges support from the European Research Council (ERC) Advanced
Investigator Grant EPOCHS (788113). 
MAM acknowledges the support of a National Research Council of Canada Plaskett
Fellowship, and the Australian Research Council Centre of Excellence for All
Sky Astrophysics in 3 Dimensions (ASTRO 3D), through project number CE17010001.
CNAW acknowledges funding from the {JWST}/NIRCam contract NASS-0215 to the
University of Arizona.
CC is supported by the National Natural Science Foundation of China, No. 11803044, 11933003, 12173045. This work is sponsored (in part) by the Chinese Academy of Sciences (CAS), through a grant to the CAS South America Center for Astronomy (CASSACA). We acknowledge the science research grants from the China Manned Space Project with NO.\ CMS-CSST-2021-A05.
JMD acknowledges support from PGC2018-101814-B-100.
The James Clerk Maxwell Telescope is operated by the East Asian Observatory on behalf of The National Astronomical Observatory of Japan; Academia Sinica Institute of Astronomy and Astrophysics; the Korea Astronomy and Space Science Institute; the National Astronomical Research Institute of Thailand; Center for Astronomical Mega-Science (as well as the National Key R\&D Program of China with No.\ 2017YFA0402700). Additional funding support is provided by the Science and Technology Facilities Council of the United Kingdom and participating universities and organizations in the United Kingdom and Canada. Additional funds for the construction of SCUBA-2 were provided by the Canada Foundation for Innovation. The data used in this project came from programs  M23AP006 and M15AI29.  
The Submillimeter Array is a joint project between the Smithsonian Astrophysical Observatory and the Academia Sinica Institute of Astronomy and Astrophysics and is funded by the Smithsonian Institution and the Academia Sinica.
This work is based on observations carried out under project number S22CX with the IRAM NOEMA Interferometer. IRAM is supported by INSU/CNRS (France), MPG (Germany) and IGN (Spain).
This work is based on observations made with the NASA/ESA/CSA {James Webb Space Telescope}. The data were obtained from the Mikulski Archive for Space Telescopes (MAST) at the Space Telescope Science Institute, which is operated by the Association of Universities for Research in Astronomy, Inc., under NASA contract NAS 5-03127 for {JWST}. These observations are associated with {JWST} program 1176.
This research is based on observations made with the NASA/ESA {Hubble Space Telescope} obtained from the Space Telescope Science Institute, which is operated by the Association of Universities for Research in Astronomy, Inc., under NASA contract NAS 5–26555. These observations are associated with program 15959.
This work is based in part on archival data obtained with the {Spitzer Space Telescope}, which was operated by the Jet Propulsion Laboratory, California Institute of Technology under a contract with NASA.
This research has made use of Canadian Astronomy Data Centre (CADC),  NASA's Astrophysics Data System (ADS) and the NASA Extragalactic Database (NED).
We recognize that Maunakea is a culturally important site for the indigenous Hawaiian people; we are privileged to study the cosmos from its summit.  We also acknowledge the indigenous peoples of Arizona, including the Akimel
O'odham (Pima) and Pee Posh (Maricopa) Indian Communities, whose care and
keeping of the land has enabled us to be at ASU's Tempe campus in the Salt
River Valley, where much of our work was conducted.

~\vspace{3mm}
\software{
{\sc galfit} \citep{galfit};
  {\sc gildas} \url{https://www.iram.fr/IRAMFR/GILDAS};
IDL Astronomy Library: \url{https://idlastro.gsfc.nasa.gov}
\citep{Landsman93};
{\sc magphys} \citep{Battisti19};
{\sc profound} \citep{Robotham18};
{\sc smurf} \citep{smurf};
{\sc sourcextractor} \url{https://www.astromatic.net/software/sextractor/} or
\url{https://sextractor.readthedocs.io/en/latest/}  \citep{Bertin96};
{\sc TinyTim} \citep{Krist11};
{\sc topcat} \citep{topcat}; 
{\sc WebbPSF} \citep{Perrin14}.
}

\vspace{3mm}
\facilities{SMA,  JCMT, NOEMA, {JWST}, {HST},  {SST}, CADC archive, Hubble and James Webb Space Telescope Mikulski Archive
\url{https://archive.stsci.edu} }

\bibliography{a1489}{}

%
%
\begin{longrotatetable}

  \begin{minipage}{\textheight}
  \movetabledown=20mm
\begin{deluxetable*}{lccccccccccc}
\tablenum{2}
\tablecaption{Kinematic properties of sources}
\tablewidth{0pt}
\tablehead{
  \colhead{Source} 
  & \colhead{$z$} & \colhead{$\nu_{\rm CO(4-3)}$} & \colhead{FWHM$_{\rm CO(4-3)}$} 
  & \colhead{$\nu_{\rm [CI]}$} & \colhead{FWHM$_{\rm [CI]}$} 
  & \colhead{$\nu_{\rm CO(5-4)}$} & \colhead{FWHM$_{\rm CO(5-4)}$}   
  &  \colhead{FWHM} 
  &  $b/a_{\rm corr}$$^a$ &  $i$ & $M_{\rm dyn}(10\,\rm kpc)$ \\
  &  &  [GHz] &  [km\,s$^{-1}$]  
  &  [GHz] &  [km\,s$^{-1}$]  
  &  [GHz] &  [km\,s$^{-1}$]  
  &   [km\,s$^{-1}$]
  &  & [deg]  & [10$^{11}$\,M$_\odot$]
}
\startdata
850.1 
& 4.2599\,$\pm$\,0.0001 & 87.6519 &  380\,$\pm$\,15  
& 93.5606 &  400\,$\pm$\,30 
& 109.558 &  375\,$\pm$\,15 
&  380\,$\pm$\,10 
& 0.88\,$\pm$\,0.11 & 28\,$\pm$\,12 & 3.8$^{+5.1}_{-1.5}$ \\
850.2
& 4.2593\,$\pm$\,0.0003 & 87.6607 &  620\,$\pm$\,70  
& 93.6452 &  1100\,$\pm$\,500  
& 109.570 &  670\,$\pm$\,80
&  \,\,640\,$\pm$\,50 
& 0.40\,$\pm$\,0.03 &  66\,$\pm$\,2 & 0.8\,$\pm$\,0.3$^b$ \\
\enddata

~\\

\tablecomments{$^a$ Axis ratio measured from the F444W isophote shape and corrected for lensing shear.
  $^b$ To estimate the dynamical mass, we used the velocity difference between the two components of the
  CO(4--3) and CO(5--4) lines, 330\,$\pm$\,70\,km\,s$^{-1}$, rather than the FWHM. }
\end{deluxetable*}
\end{minipage}

    \begin{deluxetable*}{lcccccccccc}
\tablenum{1}
\tablecaption{Sub-/millimeter positions and observed flux densities}
\tablewidth{0pt}
\tablehead{  \colhead{} & \colhead{} & \colhead{}  & \colhead{{Spitzer}} & \colhead{} & \multicolumn{2}{c}{SCUBA-2} & \multicolumn{2}{c}{SMA} &  \multicolumn{2}{c}{NOEMA} \\[-2mm]
  \colhead{Source (short name)} & \colhead{R.A. ~~~ Dec.}  & \colhead{5.8\,$\mu$m} & \colhead{8.0\,$\mu$m} & \colhead{24\,$\mu$m}   &
  \colhead{450\,$\mu$m} & \colhead{850\,$\mu$m} & \colhead{880\,$\mu$m} & \colhead{$\mu^{0.5} \rm FWHM_{\rm d}$} & \colhead{2.8\,mm} & \colhead{3.4\,mm} \\[-3mm]
\colhead{ } & \colhead{(J2000)} & \colhead{[$\mu$Jy]} & \colhead{[$\mu$Jy]} & \colhead{[mJy]} & \colhead{[mJy]} & \colhead{[mJy]} & \colhead{[mJy]} & \colhead{[kpc]} & \colhead{[mJy]}  &  \colhead{[mJy]}  
}
\startdata
SMM\,J121223.0+273351$^a$\,(850.1) &  12\,12\,23.050 +27\,33\,51.89 & 20\,$\pm$\,10 & 30\,$\pm$\,10  & $<$\,0.76 
& 72\,$\pm$\,21$^{b}$ [4.4] & 54.1\,$\pm$\,1.5 [30.5] & 42.3\,$\pm$\,1.6$^{c}$ & 5.6\,$\pm$\,1.4 & 1.03\,$\pm$\,0.08  & 0.49\,$\pm$\,0.04   \\
SMM\,J121220.4+273410$^a$\,(850.2)  &  12\,12\,20.484 +27\,34\,10.45 & 23\,$\pm$\,10 & 32\,$\pm$\,10 & $<$\,0.76
& 49\,$\pm$\,21$^{d}$ [3.5] & 21.2\,$\pm$\,1.5 [11.0] & 22.5\,$\pm$\,1.6$^{e}$ & 4.6\,$\pm$\,2.4 & 0.25\,$\pm$\,0.04  & 0.17\,$\pm$\,0.03 \\
SMM\,J121213.4+273517\,(850.3)  & 12\,12\,13.43  +27\,35\,17.0$^a$ & ... & ... & ...  & $<$\,60$^f$ & 16.5\,$\pm$\,2.2 [7.4] & ... & ... & ... & ... \\
\enddata
\tablecomments{The numbers in square brackets are the signal-to-noise of the flux density measurements.  $^{a}$ Position from SCUBA-2 850\,$\mu$m. The positions derived from the original M15AI29 SCUBA-2 observations of both sources were offset from those reported here by $\sim$\,4$''$ due to the use of M\,87 as a pointing source, where emission from the radio jet significantly perturbs the position at  850\,$\mu$m.  This issue was identified from the SMA observations analyzed in this work and M\,87 subsequently removed from the JCMT pointing catalog. 
 $^{b}$  850.1 un-deboosted  450-$\mu$m  flux density: 93\,mJy. $^{c}$ 850.1 peak SMA flux density: 18.5\,$\pm$\,0.8\,mJy.
 $^{d}$  850.2 un-deboosted 450-$\mu$m  flux density:  73\,mJy.  $^{e}$ 850.2 peak SMA flux density: 8.9\,$\pm$\,0.8\,mJy. $^f$ 2-$\sigma$ limit.}
\end{deluxetable*}

\end{longrotatetable}

%
%
\begin{longrotatetable}

\begin{minipage}{\textheight}
 \begin{deluxetable*}{lcccccccccl}
\tablenum{4}
\tablecaption{Observed visible and near-infrared photometry}
\tablewidth{0pt}
\tablehead{
\colhead{Source} & \colhead{F435W} & \colhead{F606W} & \colhead{F814W} & \colhead{F090W} & \colhead{F150W} & \colhead{F200W} & \colhead{F277W} & \colhead{F356W} & \colhead{F444W} & Comments 
}
\startdata
850.1 & $>$\,26.5   & 25.37  0.21  & 24.28  0.17  & 24.07  0.09  & 23.50  0.07  & 23.09  0.06  & 22.49  0.04 & 21.87  0.04  & 21.43  0.03 &  2\farcs0\diameter\ equivalent, including B2  \\
           &  $>$\,26.5  &  $>$\,27.2  & $>$\,26.6   & $>$\,27.1  & $>$\,25.8   & 25.07  0.15  & 23.14  0.05 & 22.19  0.04  & 21.66  0.03 &  Excluding B2  \\
           & 24.46  0.22  & 24.16  0.11  & 22.98  0.10  & 22.86  0.06  & 22.23  0.05  & 21.90  0.05  & 21.48  0.04 & 21.06  0.03  & 20.77  0.03 &  Including B1 and B2  \\
850.2 & $>$\,26.5      & 23.10  0.08  & 21.82  0.05  &  21.75  0.03 & 21.50  0.03  & 21.29  0.03 &  21.03  0.02 &  20.87  0.02 & 20.83  0.02 &  2\farcs4\diameter  \\
\noalign{\smallskip}
B1      & 25.02  0.27   & 24.98  0.18 & 23.62   0.12 &  23.44  0.07 & 22.92  0.05 & 22.69  0.05 &  22.43  0.04 & 22.22  0.04 &  22.18  0.04 & 0\farcs9\diameter, knot close to 850.1 \\
B2      & $>$\,26.5        & 25.37  0.21  & 24.28  0.17 & 24.07  0.09 & 23.50  0.07 & 23.28  0.06 & 23.35  0.07 & 23.36  0.07 & 23.23  0.06 & 0\farcs9\diameter, knot close to 850.1 \\
\noalign{\smallskip}
A1     & $>$\,26.5       & 26.04  0.29  & 25.31  0.26 & 25.14  0.15 &  24.63  0.12 &  24.52  0.11 &  24.75  0.13 &  25.24  0.16 &  25.65  0.19 & 0\farcs9\diameter, galaxy SE of 850.1 \\
A2     & $>$\,26.5       & 25.84  0.26 & 24.80  0.21 & 24.23  0.10 & 22.02  0.04 & 21.75  0.03 & 21.49  0.03  & 21.31  0.03 & 21.19  0.03 & 0\farcs9\diameter, galaxy N  of 850.1 \\
A3     & 25.13  0.28   & 24.97  0.18 & 24.14  0.15 & 24.17  0.10 & 24.23  0.10 & 24.12  0.10 & 24.27  0.10 & 24.61  0.12 & 24.90  0.14 & 0\farcs9\diameter, galaxy NW  of 850.1 \\
A4     & 23.34  0.12   & 21.52  0.04 & 20.40  0.03 & 20.17  0.02 & 19.33  0.01 & 19.08  0.01 &  19.18  0.01 & 19.67  0.01 & 19.85  0.01 & 1\farcs8\diameter,  bright spiral near 850.1 \\
A5     & 23.87  0.16   & 22.25  0.05 & 21.36 0.04 & 21.17  0.02 & 20.51  0.02 & 20.34  0.02 & 20.52  0.02 & 21.02  0.02 & 21.25  0.03 & 2\farcs4\diameter, galaxy S of 850.2 \\
\enddata
~\\
\tablecomments{All limits correspond to 3\,$\sigma$.}  
\end{deluxetable*}
\end{minipage}

    \begin{deluxetable*}{lccccccccc}
\tablenum{3}
\tablecaption{ISM properties of sources}
\tablewidth{0pt}
\tablehead{
  \colhead{Source}
  & \colhead{$\mu M_{\rm d}$} & \colhead{$\mu I_{\rm CO(4-3)}$} & \colhead{$\mu L'_{\rm CO(4-3)}$}
  & \colhead{$\mu I_{\rm [CI]}$} & \colhead{$\mu L'_{\rm [CI]}$}
  & \colhead{$\mu I_{\rm CO(5-4)}$} & \colhead{$\mu L'_{\rm CO(5-4)}$}
   & \colhead{$\mu M_{\rm g}^a$}
  &  \colhead{$M_{\rm g}/(M_{\rm g}+M_{\ast})$}   \\
  & [10$^{8}$\,M$_\odot$]  & [Jy\,km\,s$^{-1}$] &  [10$^{10}$\,K\,km\,s$^{-1}$\,pc$^2$] 
  & [Jy\,km\,s$^{-1}$] &  [10$^{10}$\,K\,km\,s$^{-1}$\,pc$^2$] 
  & [Jy\,km\,s$^{-1}$] &  [10$^{10}$\,K\,km\,s$^{-1}$\,pc$^2$] 
  &  [10$^{11}$\,M$_\odot$]
  &  
}
\startdata    
850.1 
 &  61$^{+14}_{-11}$ & 4.57\,$\pm$\,0.03 & 20.8\,$\pm$\,0.2 
 & 1.52\,$\pm$\,0.02 & 6.1\,$\pm$\,0.1
 & 3.85\,$\pm$\,0.03 & 11.1\,$\pm$\,0.1 
 & 4.2\,$\pm$\,0.1
 &  0.16\,$\pm$\,0.03 \\    
850.2
&  26$^{+3}_{-2}$ & 1.87\,$\pm$\,0.04 & 8.5\,$\pm$\,0.2 
& 0.62\,$\pm$\,0.09 & 2.5\,$\pm$\,0.4 
& 1.76\,$\pm$\,0.04 & 5.1\,$\pm$\,0.1
& 1.6\,$\pm$\,0.1
&  0.62\,$\pm$\,0.07\\ 
\enddata
~\\
\tablecomments{$^a$ Adopting $\alpha_{\rm CO}=$\,1.}
\end{deluxetable*}
\end{longrotatetable}

\setlength{\evensidemargin}{1.9in}
\setlength{\oddsidemargin}{1.9in}
\setlength{\textwidth}{6in}
\pagestyle{empty}
\newpage

%
%
\begin{deluxetable}{lcccccccccccl}
\tablenum{5}
\tablecaption{Observed properties of sources}
\tablewidth{0pt}
\tablehead{
  \colhead{Source} & \colhead{$z$} & \colhead{$z_{\rm ph}$} & \colhead{$z_{\rm Z20}$$^a$}  & \colhead{$\mu M_\ast$} & \colhead{$\mu$\,SFR} & \colhead{$\mu L_{\rm IR}$} &  \colhead{$A_V$} & \colhead{$T_{\rm d}$} & \colhead{$\mu^{0.5} R^{\rm F444W}_{\rm e}$} & \colhead{Comments} \\
  & & & & [10$^{11}$\,M$_\odot$] & [10$^3$ M$_\odot$\,yr$^{-1}$]  &  [10$^{12}$\,L$_\odot$]   &  & [K] &  [kpc] & 
}
\startdata
850.1$^{b}$ & 4.26 &  4.39$^{+0.73}_{-1.91}$ & ...                              & 22$^{+5}_{-2}$ & 3.7$^{+1.6}_{-1.2}$ & 49$^{+20}_{-13}$ & 5.0$^{+0.3}_{-0.2}$ & 41$^{+8}_{-6}$ &  ... & Sum of 850.1 \& B2 fits, $\mu=$\,4.0$^{+1.0}_{-2.2}$ \\
                    & 4.26 &  4.41$^{+0.77}_{-1.95}$ & ...                              & 22$^{+5}_{-2}$ & 4.0$^{+1.6}_{-1.4}$ & 54$^{+22}_{-15}$ & 5.1$^{+0.2}_{-0.2}$ & 41$^{+8}_{-5}$ &  7.5$\pm$0.8  & Excluding B2$^c$ \\
                    & 4.26 &  ... & ...                                                             & 16$^{+4}_{-1}$ & 4.8$^{+1.2}_{-0.3}$ & 60$^{+3}_{-4}$ & 3.6$^{+1.5}_{-1.2}$$^d$ & 43$^{+2}_{-2}$ &  ... & $\Sigma_{\rm pix}$, including B2 \\
                       \noalign{\smallskip}
850.2 & 4.26 & 4.28$^{+0.08}_{-0.10}$  &    4.47$^{+0.07}_{-0.09}$ & 1.0$^{+0.1}_{-0.1}$ &  1.2$^{+0.1}_{-0.1}$ & 18$^{+2}_{-2}$  & 1.1$^{+0.1}_{-0.1}$ & 38$^{+2}_{-2}$ &  5.8$\pm$0.6 & $\mu=$\,5.6$^{+1.0}_{-3.3}$ \\  
& 4.26 & ...   &    ...                                                               & 2.2$^{+2.0}_{-0.6}$ &  1.1$^{+0.9}_{-0.4}$ & 11$^{+13}_{-5}$  & 0.7$^{+0.3}_{-0.3}$ & 44$^{+2}_{-2}$ &  ... &  $\Sigma_{\rm pix}$ \\
   \noalign{\smallskip}
   B1 & 1.1 &  1.12$^{+0.11}_{-0.12}$ & 0.91$^{+0.47}_{-0.13}$              &    0.1$^{+0.1}_{-0.0}$ &   0.01$^{+0.01}_{-0.01}$ & 0.1$^{+0.1}_{-0.1}$  & 1.06$^{+0.50}_{-0.47}$ & ...  & ... & \\ 
   B2 & 4.26 & 4.17$^{+0.16}_{-3.28}$ & 0.33$^{+0.20}_{-0.10}$ & 0.04$^{+0.06}_{-0.02}$              & 0.04$^{+0.02}_{-0.03}$ &   0.53$^{+0.78}_{-0.50}$ & 0.6$^{+0.9}_{-0.3}$  &  ... & ... & $\mu\sim$\,4\\
                      \noalign{\smallskip}
   A1 & 0.35 & 0.31$^{+0.16}_{-0.04}$  & 0.53$^{+0.72}_{-0.38}$ & ... & ...  &   ...  & ... & ... & ... & Cluster member?\\   
   A2 & 1.8 & 1.77$^{+0.28}_{-0.25}$  & 0.55$^{+0.54}_{-0.25}$ & ... & ...  &   ...   & ... & ... & ... & \\
   A3 & 0.8 & 0.77$^{+0.11}_{-0.12}$  & 0.54$^{+1.57}_{-0.39}$ & ... & ...  &   ...  & ... & ... & ... & \\
   A4 & 0.35 &  0.46$^{+0.08}_{-0.09}$ & 0.35$^{+0.14}_{-0.05}$ & ... & ...  &   ...  & ... & ... &  ... & Cluster member? \\
   A5 & 0.35 & 0.35$^{+0.06}_{-0.03}$  & 0.29$^{+0.03}_{-0.06}$ & ... & ...  &   ...   & ... & ... & ...& Cluster member? \\
\enddata
\tablecomments{$^a$ BPZ-derived photometric redshifts from \cite{Zitrin20} using only {HST} photometry.   
  $^b$ $z_{\rm ph}=$\,2.63$^{+0.46}_{-0.23}$ when B1 and B2 are included in the photometric aperture.
  $^c$ A single {\sc magphys} fit at $z$\,$=$\,4.26 to the ``unresolved'' photometry of 850.1 including B2 increases  $\chi^2$ 
   from 1.7 to 27.1 and the fit parameters become:  $\mu M_\ast$\,$=$\,0.9$^{+0.1}_{-0.1}$\,$\times$\,10$^{11}$\,M$_\odot$,  $\mu$\,SFR\,$=$\,0.9$_{-0.1}^{+0.1}\times$\,10$^3$\,M$_\odot$\,yr$^{-1}$,   $\mu  M_{\rm d}$\,$=$\, 99$_{-20}^{+20}\times$\,10$^8$\,M$_\odot$,  $A_V$\,$= $\,2.7$_{-0.1}^{+0.1}$,     $T_{\rm d}$\,$=$\,30$^{+3}_{-1}$\,K   with an age for the best fit star-formation history of  50$^{+1}_{-1}$\,Myr.
 $^d$ F444W-weighted mean $A_V$.
}
\end{deluxetable}

%
%
\begin{deluxetable}{lccl}
\tablenum{6}
\tablecaption{Intrinsic properties of sources}
\tablewidth{0pt}
\tablehead{
  \colhead{Property$^a$} & \colhead{850.1$^b$} & \colhead{850.2} & Units 
}
\startdata
$z$ & 4.2599\,$\pm$\,0.0001 & 4.2593\,$\pm$\,0.0003 & \\
$\mu$ & 4.0$^{+1.0}_{-2.2}$ & 5.6$^{+1.0}_{-3.3}$  &  \\
$S_{\rm 850\mu m}$  & 13.5$^{+0.4}_{-0.4}$\,$^{+5.8}_{-2.7}$ & 3.8$^{+0.3}_{-0.3}$\,$^{+5.4}_{-0.6}$ & mJy \\ 
$M_\ast$  & 5.5$^{+1.3}_{-0.5}$\,$^{+2.4}_{-1.1}$ & 0.18$^{+0.02}_{-0.02}$\,$^{+0.25}_{-0.03}$ & 10$^{11}$\,M$_\odot$ \\  
$M_{\rm d}$  & 15$^{+3}_{-3}$\,$^{+7}_{-3}$ & 4.6$^{+0.5}_{-0.4}$\,$^{+6.7}_{-0.7}$ & 10$^{8}$\,M$_\odot$ \\
$M_{\rm g}$  & 1.05$^{+0.03}_{-0.03}$\,$^{+0.45}_{-0.21}$ & 0.29$^{+0.02}_{-0.02}$\,$^{+0.41}_{-0.05}$ & 10$^{11}$\,M$_\odot$ \\
$M_\ast+M_{\rm g}$  & 6.6$^{+1.3}_{-0.5}$\,$^{+2.4}_{-1.1}$ & 0.5$^{+0.03}_{-0.03}$\,$^{+0.5}_{-0.5}$ & 10$^{11}$\,M$_\odot$ \\
SFR & 900$^{+400}_{-300}$\,$^{+400}_{-150}$  & 210$^{+20}_{-20}$\,$^{+300}_{-30}$  & M$_\odot$\,yr$^{-1}$ \\
$R^{\rm F444W}_{\rm e}$ & 3.8$^{+0.4}_{-0.4}$\,$^{+0.7}_{-0.4}$ & 2.5$^{+0.3}_{-0.3}$\,$^{+1.3}_{-0.2}$  & kpc\\ 
$R^{\rm dust}_{\rm e}$ & 2.8$^{+0.7}_{-0.7}$\,$^{+0.5}_{-0.3}$ & 1.9$^{+1.0}_{-1.0}$\,$^{+1.1}_{-0.1}$ & kpc \\  
$R^{\rm mass}_{\rm e}$ & 1.0$\pm$0.2\,$^{+0.2}_{-0.1}$ & 2.2$\pm$0.4\,$^{+1.1}_{-0.2}$ &  kpc \\
\enddata
\tablecomments{$^a$ Dual error bars  give first  the random  and then the systematic uncertainties due to the lensing magnification.  $^b$ Sum of the parameter estimates for the fit to 850.1 and the fit to B2.}
\end{deluxetable}

\end{document}